\documentclass[prx,aps,floatfix,amsmath,superscriptaddress,twocolumn]{revtex4}

\usepackage{mathtools}

\usepackage{amsfonts}
\usepackage{graphicx,graphics,epsfig,times,bm,bbm,amssymb,amsmath,amsfonts,mathrsfs}
\usepackage[normalem]{ulem}
\usepackage{setspace}
\usepackage{subfigure}
\usepackage{dsfont}
\usepackage{braket}
\usepackage{upgreek }
\usepackage{tikz}
\usepackage{natbib}
\usepackage{chngcntr}

\usepackage{amssymb,enumerate}
\usepackage{graphicx}
\usepackage{graphics}
\usepackage{amsmath}
\usepackage{amsthm,bbm}
\usepackage{color}
\usepackage{dsfont}
\usepackage{hyperref}

\usepackage{lipsum}
\usepackage{lmodern}
\usepackage{tcolorbox}

\usepackage{amsfonts}
\usepackage{graphicx,graphics,epsfig,times,bm,bbm,amssymb,amsmath,amsfonts,mathrsfs}
\usepackage[normalem]{ulem}
\usepackage{setspace}
\usepackage{subfigure}
\usepackage{dsfont}
\usepackage{braket}
\usepackage{upgreek }
\usepackage{tikz}
\usepackage{natbib}
\usepackage{chngcntr}

\newtheorem{theorem}{Theorem}

\newtheorem{lemma}[theorem]{Lemma}

\newcommand{\bes} {\begin{subequations}}
\newcommand{\ees} {\end{subequations}}
\newcommand{\bea} {\begin{eqnarray}}
\newcommand{\eea} {\end{eqnarray}}
\newcommand{\be} {\begin{equation}}
\newcommand{\ee} {\end{equation}}

\def\>{\rangle}
\def\<{\langle}
\def\Tr{\textrm{Tr}}

\newcommand{\ignore}[1]{}

\usepackage{chngcntr}
\usepackage{apptools}

\newcounter{Thm}
\newcounter{Cor}
\newcounter{Lem}


\begin{document}	
	\title{Coherence distillation machines are impossible in quantum thermodynamics}
	
	\author{Iman Marvian}
\affiliation{Departments of Physics \& Electrical and Computer Engineering, Duke University, Durham, North Carolina 27708, USA}
\email{iman.marvian@duke.edu}

	\begin{abstract}
The role of coherence in quantum thermodynamics has been extensively studied in the recent years and it is now well-understood that coherence between different energy eigenstates is a resource independent of other thermodynamics resources, such as work. A fundamental remaining open question is whether the laws of quantum mechanics and thermodynamics allow the existence of a coherence distillation machine, i.e. a machine that, by possibly consuming work, obtains pure coherent states from mixed states, at a nonzero rate. This is related to another fundamental question: Starting from many copies  of noisy quantum clocks which are (approximately) synchronized with a reference clock, can one distill synchronized clocks in pure states, at a non-zero rate? Surprisingly, we find that the answer to both questions is negative for generic (full-rank) mixed states. However, at the same time, it is possible to distill a sub-linear number of pure coherent states with a vanishing error.

\end{abstract}
		
		\maketitle

\section*{Introduction}

What are the fundamental limits of nature on manipulation of quantum clocks?  Suppose we have multiple clocks, all synchronized with the same reference clock, which are  affected by noise. Then, by averaging the time read from these clocks we can obtain a more accurate estimate of the current time according to the reference clock. In other words, we can  {distill} a less noisy clock from several noisy clocks. What are the limits of this distillation process for quantum clocks? Can we distill   quantum clocks in  {pure} states from those in mixed states, at a nonzero rate?

\begin{figure} [h]
\begin{center}
\includegraphics[scale=.5]{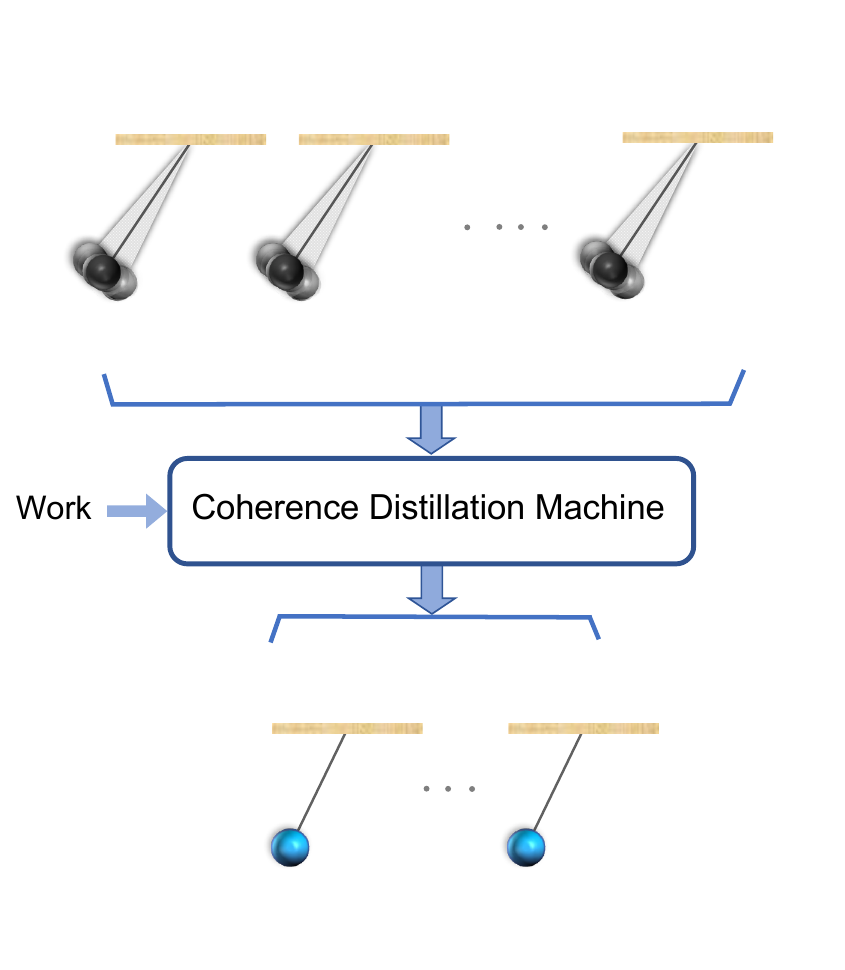}
\caption{A hypothetical  "Coherence Distillation Machine" for distilling coherence with respect to the energy eigenbasis:  It consumes work and obtains pure coherent states from  mixed states at a non-zero rate, or equivalently, purifies quantum clocks. Is this hypothetical machine consistent  with  the laws of quantum mechanics and thermodynamics?}\label{Fig1}
\end{center}
\end{figure}

Interestingly, this question is related to another fundamental question about the manipulation of coherence in quantum thermodynamics. It is now well-understood that coherence between different energy eigenstates  is a resource, independent of other thermodynamic resources such as work, and can be used to implement operations which are otherwise impossible \cite{lostaglio2015description,lostaglio2015quantumPRX,korzekwa2016extraction, narasimhachar2015low}. A fundamental open question in this context is whether the laws of quantum mechanics and thermodynamics allow the existence   a  {coherence distillation machine}, i.e. a machine that  consumes work to obtain  pure coherent states from mixed ones at a nonzero rate (See Fig.\ref{Fig1}).  The  connection between these two questions arises from the fact that the minimum requirement for a system to be  a clock is to be in a  state which contains coherence (i.e. off-diagonal terms) with respect to the energy-eigenbasis; otherwise, the system will be time-independent, and hence useless as a clock.
 
In this article, we investigate coherence distillation in the context of quantum thermodynamics,  both in the single-shot and asymptotic regimes. In particular, we settle the above questions,  which have been open heretofore \cite{winter2016operational, streltsov2017colloquium}, and show that the answer to  both of them is negative. In other words, the coherence distillation machine, depicted in Fig.\ref{Fig1}, is impossible. 
 This   is surprising,  especially when compared to the previously known results on resource distillation in the entanglement theory and other quantum resource theories (See e.g. \cite{chitambar2019quantum, winter2016operational, chitambar2018dephasing, devetak2005distillation, devetak2008resource, devetak2004relating}), and reveals important aspects of coherence in quantum thermodynamics. \color{black}  In particular, we will see that, in some precise sense, the coherence content of a single two-level system can be infinitely large. Furthermore, we find that, even though distillation with a non-zero rate is impossible, it is still possible to distill a sublinear  number of pure coherent states with a vanishing error. We also consider coherence distillation in the single-shot regime and derive a simple formula for the maximum achievable fidelity. 
\section*{Results}

\subsection*{Distillation of quantum clocks}
A quantum clock is characterized by its state and  Hamiltonian, which usually generates a periodic time evolution \cite{salecker1958quantum,  peres1980measurement, QRF_BRS_07, Lloyd_Nature_Clock, Limits_Lloyd_Clock,  Clock_Buzek, chiribella2013quantum}.  By definition, the state of a clock should be time-dependent.  Therefore, when we say a clock with Hamiltonian $H$ is in state $\rho$, we actually mean its  state is $\rho$ at a particular time, say $t=0$, with respect to a reference clock. Then, at an arbitrary time $t$ the state of clock is $e^{-i H t}\rho e^{i H t}$  (Throughout this paper we assume $\hbar=1$). 
Here, we focus on the systems with bounded Hamiltonians, with  periodic dynamics, whose period is equal to a fixed (but arbitrary) parameter $\tau$, such that $\tau=\min \{t>0: e^{-i H t}\rho e^{i H t}=\rho\}$; otherwise, the state and Hamiltonian are completely arbitrary.  In the following, when we talk about $n$  {copies} of a system with state $\rho$ and Hamiltonian $H$,  we mean $n$ non-interacting systems, with the total Hamiltonian $\sum_{i=1}^n H^{(i)}$, where $H^{(i)}=I^{\otimes (i-1)}\otimes H\otimes I^{\otimes (n-i-1)}$, and with the joint state $\rho^{\otimes n}$.

Suppose Alice is given a quantum clock with Hamiltonian $H_\text{in}$ and state $ \rho_\text{in}$, synchronized with a  standard reference clock owned by Bob.  Assume she does not  have any additional information about Bob's clock. In other words, she knows at time $t$ relative to Bob's clock, her quantum clock is in state $e^{-i H_\text{in} t} \rho_\text{in} e^{i H_\text{in} t}$; however, the parameter $t$ itself is unknown to her. 

Now suppose Alice wants to transform this  clock to a different clock, with possibly different Hamiltonian  $H_\text{out}$, which is still  synchronized with Bob's clock, such that at any time $t$ relative to his clock the new quantum clock is in state $e^{-i H_\text{out} t} \rho_\text{out} e^{i H_\text{out} t}$. For instance, the input clock with Hamiltonian $H_\text{in}$ can  be multiple copies of a noisy two-level clock in a mixed state, whereas the output clock is a single two-level system, which is more accurate than any single copy at the input, i.e. conveys more information about the parameter $t$ (This is an example of single-copy distillation of clocks, which will be discussed later).  This means that Alice wants to implement the state conversion
\be\label{eq0}
e^{-i H_\text{in} t} \rho_\text{in} e^{i H_\text{in} t}\longrightarrow e^{-i H_\text{out} t} \rho_\text{out} e^{i H_\text{out} t} \ ,\ \ \ \ \forall t\in [0,\tau)\ .
\ee
However, since parameter $t$ is unknown to her, this  conversion should be implemented by a fixed process, independent of $t$; i.e. there should exist a  physical process, described by  a completely positive trace-preserving \cite{nielsen2000quantum, wilde2013quantum}  map $\mathcal{E}$, such that $\mathcal{E}(e^{-i H_\text{in} t} \rho_\text{in} e^{i H_\text{in} t})=e^{-i H_\text{out} t} \rho_\text{out} e^{i H_\text{out} t}$, for all time  $t\in [0,\tau)$.  It turns out that this is possible if, and only if, the  {single} state conversion $\rho_\text{in} \rightarrow  \rho_\text{out} $ is possible under a Time-translation Invariant (TI) process, i.e. a process satisfying the covariance condition
\be\label{cov}
e^{-i H_\text{out} s}\  \mathcal{E}_{\text{TI}}(\sigma)\ e^{i H_\text{out} s}=\mathcal{E}_{\text{TI}}\big(e^{-i H_\text{in} s}\sigma e^{i H_\text{in} s}\big)\ ,
\ee
for  all times $s$, and input $\sigma$ \cite{ marvian2013theory, Marvian_thesis}. 
 Therefore, rather than studying the state conversions for the family of states in Eq.(\ref{eq0}), one can equivalently study state conversion for the single input-output pair $\rho_{\text{in}}$ and $\rho_{\text{out}}$ under the restricted set of TI operations.

The covariance condition in Eq.(\ref{cov}) means that TI processes are those which can be defined, and hence implemented, independent of a reference clock. Furthermore, they can be implemented without  interfering with the intrinsic time evolution generated by the system Hamiltonian.    An example of this type of processes is energy-conserving unitary transformations, i.e. those  which commute with the Hamiltonian (assuming the input and output systems have identical Hamiltonians).   There are also TI operations which are not energy-conserving, such as, preparing the system in an  {incoherent} state, i.e. any state $\rho$ commuting with the system Hamiltonian (Note that in the case of composite systems, the joint state is incoherent if it commutes with the {total} Hamiltonian).

In summary, we conclude that  for distillation or manipulation  of quantum clocks, we can restrict our attention to the set of TI operations.   In the language of quantum resource theories  \cite{horodecki2013quantumness, brandao2015reversible, coecke2016mathematical, gour2015resource, chitambar2019quantum}, these are the  {free}  operations for the resource theory of quantum clocks, which is a special case of the resource theory of asymmetry. \color{black}

It is worth emphasizing that the notion of  {resource distillation}, which can be abstractly defined in any resource theory,  has a clear operational interpretation in this framework: it is the process in which one combines noisy clocks, affected by independent noise processes, to obtain less, but more accurate clocks in pure states. More precisely, the information content of each output clock about the unknown parameter $t$, i.e. the current time relative to the standard clock, is greater than the information content of each input clock. Hence, using a distillation protocol, one can increase the efficiency of storage and transmission of quantum clocks. 
Intuitively, one expects that to maximize the information content about parameter $t$, the state of quantum clock  should be pure. This intuition is confirmed by the fact that pure states maximize any convex measure of information (about the time parameter $t$) such as quantum mutual information (Holevo quantity) \cite{Holevo:book, nielsen2000quantum, wilde2013quantum} or quantum Fisher information \cite{Holevo:book, Helstrom:book,  paris2009quantum, braunstein1994statistical}. Similarly, from the point of view of parameter estimation, to minimize the error in the estimation of the time parameter $t\in [0,\tau)$, as quantified by  any cost function which is a linear functional of state, such as mean squared error \cite{Holevo:book, chiribella2005optimal}, the system should be prepared in a pure state.

Interestingly, as we see next,  the set of TI operations also naturally arises   in the study of coherence in quantum thermodynamics. It is worth mentioning  that, in this paper we focus on a notion of coherence which is relevant in the context of quantum clocks and quantum thermodynamics, known as  {unspeakable} coherence \cite{streltsov2017colloquium, marvian2016quantify}. This notion of coherence  is a special case of a more general  property, called asymmetry \cite{gour2008resource, marvian2014asymmetry, marvian2016quantify}.  There are other resource theoretic approaches to coherence, capturing a different notion of coherence, known as  {speakable} coherence  \cite{streltsov2017colloquium, marvian2016quantify} (In these resource theories the eigenvalues of the system Hamiltonian do not play any role). 

\subsection*{Coherence Distillation Machines}

 A coherence distillation machine, as depicted in Fig.\ref{Fig1}, receives systems in a mixed coherent state, and transforms them to pure coherent  states, at a non-zero rate. Recall that a quantum state contains coherence, or is  {coherent}, if its density operator does not commute with its Hamiltonian.  
In the following, we consider two different frameworks for describing coherence distillation machines and, interestingly, find that they are equivalent and  both lead to the notion of TI  operations. 

\begin{figure} [h]
\begin{center}
\includegraphics[scale=.65]{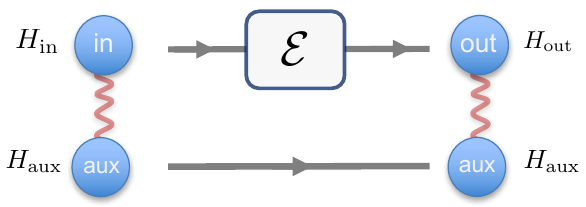}
\caption{Completely incoherence-preserving operations: Suppose the joint initial state of the input system and an auxiliary system with Hamiltonian $H_\text{aux}$  is incoherent with respect to their total Hamiltonian $H_\text{in}\otimes I_\text{aux} +I_\text{in}\otimes H_\text{aux}$.  Quantum operation $\mathcal{E}$ is called  {completely incoherence-preserving} if for any choices of $H_\text{aux}$ and the initial incoherent state,  the joint  state of the output and the auxiliary system is also incoherent with respect to their total Hamiltonian $H_\text{out}\otimes I_\text{aux} +I_\text{out}\otimes H_\text{aux}$. We show that any such operation is a TI operation and can be implemented by coupling the system to a work reservoir (battery) by an energy-conserving unitary.  } \label{Fig2}
\end{center}
\end{figure}

Our first approach is to consider the most general processes  which can be interpreted as "coherence distillation machines".  What are the constraints on  such operations?  Clearly, a distillation machine  should not generate coherence itself, i.e. should transform  incoherent states to incoherent states; otherwise,  the  coherence at the output cannot be interpreted as  {distilled} coherence. This should hold even if the input is entangled with another closed system with an arbitrary Hamiltonian; if their initial joint  state commutes with  their total Hamiltonian, then their final state should also commute, and hence be incoherent (See Fig.\ref{Fig2}).

We prove that a quantum operation  satisfies this property, or is  {completely   incoherence-preserving},  iff it is a TI operation (See Supplementary Note 1).  This means that, by proving the impossibility of coherence distillation using TI operations, we also establish its impossibility under completely  incoherence-preserving operations, which describe the most general processes relevant to coherence distillation. 

A different approach to formalizing coherence distillation is  to use  the framework of the resource theory of quantum thermodynamics (athermality) and the notion of   {thermal operations} \cite{janzing2000thermodynamic, FundLimitsNature, brandao2013resource, brandao2015reversible, aaberg2013truly, goold2016role,  gour2015resource}. Thermal operations are those which can be implemented by coupling the system to a thermal bath by  energy-conserving unitaries.  It turns out that  under these operations coherence and work are two independent resources  \cite{lostaglio2015description,lostaglio2015quantumPRX}. Therefore, to focus on coherence, one can supplement a thermal operation with an unlimited amount of work at the input (using a battery or work reservoir), which can be modeled as an auxiliary system in an energy eigenstate.  What is the set of all operations which can be implemented in this way? Interestingly, it turns out that the answer is again TI operations. In particular, any TI operation $\mathcal{E}_{\text{TI}}$ on a system $S$ with Hamiltonian $H_S$ can be implemented by coupling the system to an auxiliary system (battery) with Hamiltonian $H_\text{bat}$, such that 
\be\label{Stein}
 \mathcal{E}_{\text{TI}}(\sigma)=\Tr_\text{bat}  U(\sigma\otimes |E\rangle\langle E|_\text{bat})U^\dag\ ,
\ee
where (i) the initial state $|E\rangle_\text{bat}$ of the auxiliary system is an eigenstate of its Hamiltonian $H_\text{bat}$, and (ii) the unitary $U$ that couples it to the  system $S$ conserves the total energy $H_\text{tot}=H_S\otimes I_\text{bat}+I_S\otimes H_\text{bat}$, i.e. $[U, H_\text{tot}]=0$ (See Supplementary Note 1, Ref. \cite{keyl1999optimal}, and theorem 25 of \cite{Marvian_thesis}).   

We conclude that formalizing the notion of coherence distillation machines in the framework of  the resource theory of quantum thermodynamics (athermality), again  leads us to the notion of TI operations.

To summarize, we saw three different properties, each of which can characterize exactly  the same set of operations, namely TI operations: (\textbf{a}) invariance under time-translations, (\textbf{b})  being completely incoherence-preserving, and (\textbf{c}) being implementable with thermal operations supplemented with an arbitrary amount of work. Next, we study distillation  of coherence using these processes.

\subsection*{Main theorem: Typical states  have no distillable coherence  }
An ideal coherence distillation machine is a TI operation (or, equivalently, a completely incoherence-preserving operation) which consumes  copies of a system in a mixed state $\rho$  as the resource, to generate  copies of a system in a pure {coherent} state $\phi_\text{coh}$, at rate  $R>0$, i.e. $\rho^{\otimes n}\xrightarrow{\text{TI}} \phi_\text{coh}^{\otimes \lceil R n\rceil}$. Note that, in general, the Hamiltonians and the Hilbert spaces of the  input and output systems can be different. Also, note that  $\phi_\text{coh}$ can be any pure state of the output system, except the energy eigenstates \color{black}(For instance, one can choose a two-level system with Hamiltonian $\pi \sigma_z/\tau$, and state $|\phi_\text{coh}\rangle=(|0\rangle+|1\rangle)/\sqrt{2}$,  where $\tau$ is the period)\color{black}.

In practice, exact transformations are often impossible and physically intractable. Therefore, we can allow a small error $\epsilon$ in \color{black} infidelity  \color{black}\cite{wilde2013quantum},  provided that it vanishes in the limit of infinite copies, i.e. $\rho^{\otimes n} \xrightarrow{\text{TI}}\stackrel{\epsilon}{ \approx}   \phi_\text{coh}^{\otimes \lceil R n\rceil}\  \text{as } n\rightarrow\infty, \epsilon\rightarrow 0$ (\color{black}  Recall that infidelity is one minus fidelity, i.e. $1-\langle\psi|\sigma|\psi\rangle$ for state $\sigma$ and a pure state $\psi$. Infidelity is closely  related to the trace distance \cite{wilde2013quantum}). \color{black}  Then, by the Helstrom's theorem \cite{Helstrom:book,  wilde2013quantum}, in the limit $n\rightarrow\infty$, the actual output state is indistinguishable from the desired state $\phi_\text{coh}^{\otimes \lceil R n\rceil}$.

Consider an arbitrary  system with bounded Hamiltonian $H$ and state $\rho$. The distillable coherence  $C^{\text{TI}}_{\text{d}}(\rho)$, relative to any  {standard}  pure coherent state $\phi_\text{coh}$, is the maximum rate at which copies of  $\phi_\text{coh}$ can be obtained  from copies of this system using TI operations (or, equivalently, using completely incoherence-preserving operations),
\be
C^{\text{TI}}_{\text{d}}(\rho)\equiv \sup R: \rho^{\otimes n} \xrightarrow{\text{TI}}\stackrel{\epsilon}{ \approx}   \phi_\text{coh}^{\otimes \lceil R n\rceil}\  \text{as } n\rightarrow\infty, \epsilon\rightarrow 0\ , 
\ee
where the error $ \epsilon$ is vanishing  in infidelity (one minus fidelity). Note that this definition resembles the definition of the distillable entanglement  \cite{devetak2005distillation, devetak2008resource, devetak2004relating, bennett1996mixed,  BBP+96}, or,  more generally, distillable resource in any resource theory (See e.g. \cite{chitambar2019quantum, winter2016operational, chitambar2018dephasing}). We prove the following fundamental no-go theorem on coherence distillation:   \\

\noindent\emph{Theorem}.
 {If the  projector to the support of state $\rho$ commutes with the system Hamiltonian $H$, then  the rate of distillation of any system in a pure coherent state  $\phi_\text{coh}$ is zero, i.e. $C^{\text{TI}}_{\text{d}}(\rho)=0$. Thus, for a typical state $\rho$, which has full-rank density operator,  this rate is zero.}\\

Surprisingly, we find that the hypothetical coherence distillation machine depicted in Fig.\ref{Fig1} is impossible, i.e. starting from asymptotically many copies of a generic mixed state, using a thermal  machine we cannot distill pure coherence  at a nonzero rate, even if we spend an unlimited amount of work.  In fact, it turns out that coherence distillation remains impossible even if, in addition to copies of state $\rho$,  one is allowed to consume a finite  {helper} system in a pure state,  provided that its Hamiltonian is bounded and its Hilbert space is finite-dimensional (See Supplementary Note 5).  It is interesting to compare this result with the results of \cite{winter2016operational} and  \cite{chitambar2018dephasing}, which prove that the  rate of distillation of  {speakable} coherence is generally non-zero.

\color{black}

Finally, it is worth mentioning that although for a typical mixed state the distillable coherence is zero, there are also mixed states with non-zero distillable coherence.  The problem of classifying all such states, and determining the optimal rate of conversion remains open.
 In Supplementary Note 6  we present examples of such states, and find an achievable distillation  rate, which is closely related to a Petz-R\'enyi relative entropy. These examples rely on the previously known results on state conversions between pure states \cite{schuch2004nonlocal, schuch2004quantum, gour2008resource, marvian2018coherence}, which show that the optimal rate of conversion from a system with the pure state $\psi_1$ and Hamiltonian $H_1$ to another system with the pure state $\psi_2$ and Hamiltonian $H_2$, provided that they have the same period, is  $R=V_{H_1}(\psi_1)/V_{H_2}(\psi_2)$, where  $V_H(\psi)\equiv\langle\psi|H^2|\psi\rangle-\langle\psi|H|\psi\rangle^2$ is the energy variance for state $\psi$.
\color{black}

Next, we explain how the above no-go theorem follows from an interesting relation between two quantifiers of coherence, namely quantum Fisher information and a new quantifier, called the  {purity of coherence}.

\subsection*{Purity of coherence}

In recent years,  many quantifiers of coherence and asymmetry have been studied (See, for instance,  \cite{Marvian_thesis, marvian2014extending, girolami2014observable, yadin2016general, marvian2014asymmetry, marvian2014modes, piani2016robustness, gour2009measuring, vaccaro2008tradeoff}).  These previously known examples, however, all fail to see a simple, yet fundamental  feature of  coherence:  Given any finite copies of a generic mixed state,  it is impossible to generate a single copy of a  {pure} coherent state (with a non-zero probability), using only TI operations. Here, we introduce a new quantifier of coherence which captures the missing part of the picture and  predicts the unreachability of pure coherent states.

 For a system with state $\rho$, let the  {Purity of Coherence} with respect to the eigenbasis of an observable $H$ be 
\begin{align}
P_H(\rho)&\equiv\Tr(H \rho^2 H \rho^{-1})- \Tr(\rho H^2)\\ &=  \sum_{j,k} \frac{p^2_k-p^2_j} {p_j}\ |\langle\psi_k| H |\psi_j\rangle|^2\  \label{purity} ,
\end{align}
if $ \text{supp}(H \rho H) \subseteq \text{supp}(\rho)$, and $P_H(\rho)=\infty$ otherwise,  where $\rho=\sum_j p_j |\psi_j\rangle\langle\psi_j|$  is the spectral decomposition of $\rho$. 

As we discuss below, this function is an example of a generalized family of Fisher information introduced by Petz \cite{petz1996monotone, petz2011introduction}. Also, in Supplementary Note 2 we show that this function can be thought as the second derivative of Petz-R\'enyi relative entropy  (for $\alpha=2$)  \cite{petz1986quasi, tomamichel2015quantum}.  Using this fact  we show that purity of coherence is (i) non-negative and it becomes zero iff state is incoherent, (ii) non-increasing under any TI operation $\mathcal{E}_\text{TI}$, i.e. $P_{H_\text{out}}(\mathcal{E}_\text{TI}(\rho))\le P_{H_\text{in}}(\rho)$.  In particular, it is invariant under energy-conserving unitaries. ({iii}) Additive: for uncorrelated composite systems which are not interacting with each other, i.e. $P_{H_\text{tot}}(\rho_1\otimes \rho_2)=P_{H_1}(\rho_1)+P_{H_2}(\rho_2)$, where $H_\text{tot}=H_1\otimes  I_2+I_1\otimes  H_2$, and (iv) a convex function of $\rho$.

The above definition implies that for pure states the purity of coherence is $\infty$, unless the state is an energy eigenstate, in which  case it is zero. This unboundedness of the purity of coherence, captures  the unreachability of pure coherent states from generic mixed states: Suppose there exists a TI operation which receives   $n$ copies of a system with state $\rho_1$ and Hamiltonian $H_1$, and with probability of success $p$, transforms them  to a single copy of a system with state $\rho_2$ and Hamiltonian $H_2$. Using  properties ({i}-{iv}), in Supplementary Note 2 we show
\begin{equation}\label{lastbound}
n \ge p \times \frac{P_{H_2}(\rho_2)}{P_{H_1}(\rho_1)}  \ .
\end{equation}
Thus, to generate a single copy of a pure coherent state $\rho_2$, we need $n=\infty$ or  $P_{H_1}(\rho_1)=\infty$. 
These properties of purity of coherence make it a powerful tool to study coherence distillation, both in the asymptotic and single-shot regimes.

\subsection*{Relation with Quantum Fisher Information}

It turns out that the purity of coherence has an interesting relation with Quantum Fisher Information (QFI), and this relation plays a crucial role in the proof of our no-go theorem. Recall that for the family of states $\{e^{- i H t} \rho e^{i H t}\}_t$, QFI associated to the time parameter  $t$ is 
\be\label{Def:Fisher}
F_H(\rho)= 2 \sum_{j,k} \frac{(p_j-p_k)^2 } {p_j+p_k}\ |\langle\psi_j| H |\psi_k\rangle|^2\ .
\ee
where  $\rho=\sum_j p_j |\psi_j\rangle\langle\psi_j|$ is  the spectral decomposition of $\rho$. QFI is the central quantity of quantum metrology and estimation  theory \cite{Holevo:book, Helstrom:book,  paris2009quantum, braunstein1994statistical, petz2011introduction}, and has found extensive applications in different areas of physics (See e.g. \cite{zanardi2007information, zanardi2008quantum, zanardi2007bures, campos2007quantum, pires2016generalized,  kwon2018clock}).   QFI satisfies properties ({i}-{iv}) listed above for the purity of coherence. In particular, it is additive and monotone under TI operations.

A closer look at the properties of the purity of coherence and QFI reveals an interesting relation between them: First, comparing  Eq.(\ref{purity}) and Eq.(\ref{Def:Fisher}), one can easily show that the purity of coherence is always larger than or equal to QFI, i.e. $ P_H(\rho) \ge F_H(\rho)$, \color{black} and the equality holds iff $\rho$ is incoherent. \color{black}    Furthermore, for two-level systems, we find the nice formula
\begin{align}\label{Ex}
P_H(\rho)=\frac{F_H(\rho) }{2[1-\Tr(\rho^2)]}\ ,
\end{align}
i.e. the purity of coherence is determined by a combination of QFI and the  {purity}, $\Tr(\rho^2)$. This means that,  for states close to the maximally mixed state, $P_H(\rho)/F_H(\rho) \approx 1$, whereas for states close to a generic pure state, $P_H(\rho)$ can be arbitrarily larger than $F_H(\rho)$. We show that these properties hold beyond two-level systems: In general, if $\rho$ is $\epsilon$-close to the maximally mixed state in infidelity, then $\frac{P_H(\rho)}{ F_H(\rho)}=1+\mathcal{O}(\sqrt{\epsilon})$. In the opposite limit, where  $\rho$ is close to a pure state, we find $P_H(\rho)\ge  \frac{1}{4} F_H(\psi_\text{max}) \times [\frac{p_\text{max}^2}{1-p_\text{max}}-1]\ , $
where $p_\text{max}$ is the largest eigenvalue of $\rho$, and $\psi_\text{max}$ is the corresponding eigenvector  (See Supplementary Note 3).  Again, as $\rho$ converges to a  pure state, the purity $\Tr(\rho^2)$ and $p_\text{max}$ converge to one. In this case, $P_H(\rho)$ diverges, unless the pure state is an energy eigenstate.

We conclude that, roughly speaking, the purity of coherence $P_H(\rho)$ is lower bounded by  the ratio of QFI (for a pure state close to $\rho$) to one minus the purity of state; hence, higher $P_H(\rho)$ means more pure coherence, which justifies its name. 

It is interesting to note that the relation between the purity of coherence and QFI is analogous to the relation between the total and free energies in thermodynamics;  the latter distinguishes  ordered (low-entropy) energy and  disordered (high-entropy) energy. Similarly,  the purity of coherence, can recognize the distinction between the pure and mixed coherence. It turns out that for some operations, such as coherence distillation, the same amount of coherence quantified by QFI in states with more purity is a more useful resource. 

\subsection*{RLD and SLD Fisher Information}

It is worth mentioning that both of these quantifiers of coherence, i.e. the purity of coherence  $P_H$  and QFI $F_H$, are specials cases of  a generalized family of Fisher Information. Classically, Fisher information is the unique (up to a normalization)   stochastically monotone Riemannian metric  on the space of probability distributions    \cite{morozova1991markov}. In the quantum case, on the other hand, there is a family of monotone metrics on the space of density operators, which is fully characterized by Petz \cite{petz1996monotone, petz2011introduction}  (See also \cite{morozova1991markov}). Interestingly, functions $P_H$ and $F_H$ are extremal points in this family:  they are, respectively, the maximal and minimal  monotone metrics calculated for the one-parameter family of states $\{e^{-i H t} \rho e^{i H t}\}_t$. In quantum estimation literature, these functions are often respectively called  Right Logarithmic Derivative (RLD) and Symmetric Logarithmic Derivative (SLD) Fisher Information. Following the  physics literature convention, here we have referred to SLD Fisher information as Quantum Fisher Information (QFI).

Remarkably, these two extremal functions have also  distinguished roles in the resource theory of (unspeakable) coherence  and quantum clocks:  
it has been recently shown that QFI (SLD Fisher Information) determines the  {coherence cost}, i.e. the minimum rate of consumption of standard pure coherent states that is needed to generate the desired mixed state, using TI operations \cite{marvian2018coherence}. Also, it is well-known that QFI determines the lowest achievable mean  square error for estimating the time parameter. On the other hand, it turns out that the purity of coherence (RLD Fisher Information) plays a fundamental role in the context of coherence distillation (See Fig.\ref{Fig9}), and provides a powerful tool for proving our no-go theorem on coherence distillation.

\subsection*{Proof of the main theorem}

To prove the  impossibility of coherence distillation machines, we use the properties of the purity of coherence, namely its monotonicity and additivity, and its relation with QFI. Note that   the impossibility of distillation cannot be shown using QFI alone, because it increases linearly in $n$, for both the input and the desired output states.  As we explain in the following, the main challenge in proving this theorem is the fact that  QFI and the purity of coherence are not asymptotically continuous \cite{synak2006asymptotic}. 

In Supplementary Note 4  we prove the following result, which is of independent interest: Consider $m$ non-interacting systems, each with Hamiltonian $H$, and with the total Hamiltonian $H_\text{tot}=\sum_{i=1}^m H^{(i)}$,  in the joint state $\sigma_m$. Suppose the fidelity of $\sigma_m$ and state $|\phi\rangle^{\otimes m}$, is $\langle\phi|^{\otimes m}\sigma_m|\phi\rangle^{\otimes m}= 1-\epsilon$. Then,
 for sufficiently large $m$, e.g.  $m \ge  70 \frac{|\langle\phi|H^3|\phi\rangle|^2}{V^3_H(\phi)}$  and  sufficiently small $\epsilon$, e.g. $\epsilon\le 10^{-3}$, QFI and the purity of coherence of state $\sigma_m$ relative to the total Hamiltonian $H_\text{tot}$, are lower bounded by
\begin{align}
F_{H_\text{tot}}(\sigma_m)&\ge 4 c \times m\times F_H(\phi)\label{wfewfsss0}\ ,\\
P_{H_\text{tot}}(\sigma_m)&\ge  c\times m\times F_H(\phi) \times \frac{1}{\epsilon} \ , \label{wfewfsss}
\end{align}
where $c$ is a positive constant, e.g. $c=10^{-2}$ (Recall that for a pure state $\phi$,  QFI is $F_H(\phi)=4V_H(\phi)$). Note that similar to the case of a single qubit in Eq.(\ref{Ex}), the lower bound on the purity of coherence in Eq.(\ref{wfewfsss}) grows  linearly with $\epsilon^{-1}$.

At first glance, these bounds might seem intuitive  from our previous discussions:  For instance, Eq.(\ref{wfewfsss0}) means that to be able to have a large fidelity with state $\phi^{\otimes m}$, QFI of state $\sigma_m$ should also grow (at least)  linearly with $m$, which might be expected from the additivity of QFI.  However, a more careful analysis is needed: the Hamiltonian $H_\text{tot}$ has eigenvalues of order $m\times \|H\|$, which means relative to this Hamiltonian,  two states with infidelity $\epsilon$   can have QFI's which differ by order ${\epsilon}\times m^2 \|H\|^2$. Thus, while one state can have a large QFI, e.g. linear in $m$, the other might have a negligible QFI. This makes the proof of the above bounds non-trivial. 

Now suppose there exists a TI operation $\mathcal{E}_n$ which converts $\rho^{\otimes n}$ to state $\sigma_{m(n)}$ whose fidelity with the desired state $ \phi_\text{coh}^{\otimes m(n)}$ is $1-\epsilon_n$. To simplify the notation, we assume the Hamiltonian of each copy at the input is the same as the Hamiltonian of each copy at the output, which is denoted by $H$ (This assumption is not needed for the proof).  Then, using the additivity of the purity of coherence, the total purity of coherence of the input is $n\times P_H(\rho)$. Since  this quantity is monotone under TI operations, the purity of coherence of the output is $P_{H_\text{tot}}(\sigma_{m(n)})\le n\times P_H(\rho)$. Combined with Eq.(\ref{wfewfsss}), this leads to
\be\label{sgrrhg}
\frac{m(n)}{n} \times \le \frac{1}{c} \times \frac{P_H(\rho)}{F_H(\phi_\text{coh})} \times \epsilon_n \ .
\ee
 This interesting inequality implies that to make error $\epsilon_n$ small, the   {yield} ${m(n)}/{n}$ should also be small, unless $F_H(\phi_\text{coh})=0$, i.e. $\phi_\text{coh}$ is incoherent, or $P_H(\rho)=\infty$. Thus,  if $P_H(\rho)$ is bounded and $\phi_\text{coh}$ is coherent, then to have vanishing error $\epsilon_n\rightarrow 0$, we also need to have vanishing yield, $\lim_{n\rightarrow \infty} {m(n)}/{n}=0$, which means the distillable coherence is zero. We show that for a bounded Hamiltonian $H$, $P_H(\rho)<\infty$ iff $\Pi_\rho$, the projector to the support of $\rho$, commutes with $H$.  We conclude that if $[\Pi_\rho, H]= 0$, then  the distillable coherence is zero, which proves the theorem.

\color{black}
\subsection*{Sub-linear Coherence Distillation: Trade-off between the maximum achievable yield and fidelity}\label{Sec:meas}

Even though for states with finite purity of coherence the distillable coherence is zero, interestingly, it turns out  that  {any}  state which contains coherence can still be used to distill a sub-linear number of pure coherent states. In the above scenario, let $m_\text{opt}(n)$ be the maximum number of copies of  $\phi_{\text{coh}}$ which can be distilled with error less than $\epsilon_n$, and   $r_\text{opt}(n)=m_{opt}(n)/n$ be the maximum achievable yield.  Assuming the input  and output systems have the same period, the ratio of $r_\text{opt}(n)$ to error $\epsilon_n$ satisfies 
 \begin{align}\label{wrefqqq}
4[1-o(1)] \times \frac{F_H(\rho)}{F_H(\phi_\text{coh})} \le \frac{r_\text{opt}(n)}{ \epsilon_n}\le \frac{1}{c} \times \frac{P_H(\rho)}{F_H(\phi_\text{coh})} \ ,
\end{align}
where the upper bound on $r_\text{opt}(n)/ \epsilon_n$ follows from Eq.(\ref{sgrrhg}), and holds assuming the number of distilled copies   is  sufficiently large, e.g. $m_{opt}(n) \ge  70 \frac{|\langle\phi_\text{coh}|H^3|\phi_\text{coh}\rangle|^2}{V^3_H(\phi_\text{coh})}$, and  error $\epsilon_n$ is sufficiently small, e.g. $\epsilon_n\le  10^{-3}$. These assumptions are not required for the lower bound.

This means that  there is a trade-off between fidelity  and yield. For instance, for sufficiently large $n$, one can achieve the yield $r(n)=4 \frac{F_H(\rho)}{F_H(\phi_\text{coh})} n^{-\alpha}$, for arbitrary exponent $\alpha>0$, with infidelity $\epsilon_n=n^{-(\alpha-\delta)}$  where $\delta>0$ can be arbitrary small. Choosing smaller $\alpha>0$, means higher yield and also larger error.  This should be compared with the recent results on distillation of speakable coherence  \cite{regula2018one, zhao2018one, lami2019generic, lami2019completing} (In particular, in the case of strictly incoherent operations \cite{winter2016operational, yadin2016quantum}, there are bound states, which cannot be converted to a single copy of a pure coherent state with a vanishing error, even if one is given an arbitrary many copies of state \cite{zhao2018one, lami2019generic, lami2019completing}). This tradeoff and the linear relation  between the  yield and error, which highlights the significance of yield-to-error ratio as a fundamental quantity, are unique features of this resource theory, which have practical implications in the context of quantum clocks, and are worth further study.

In the Methods section, we also discuss an interesting corollary of this result, namely a novel operational explanation  of the violation of the monotonicity of Petz-R\'enyi relative entropy under data processing, for the parameter range $\alpha > 2$ \cite{petz1986quasi, tomamichel2015quantum}.

To establish the  lower bound on $r_\text{opt}(n)/{ \epsilon_n}$ in Eq.(\ref{wrefqqq}), we consider a  TI process  defined based on a parameter estimation task: Suppose one is given $n$ copies of state $e^{-i H t}\rho e^{i H t}$, where $t\in [0,\tau)$ is unknown (Recall that $\tau$ is the period of both the input and the desired output systems). Measuring these systems, one can obtain
 an estimate $t_\text{est}\in[0,\tau)$ of $t$, with probability density $p(t_\text{est}|t)$.  We can assume the estimator is invariant under time-translations, such that $p(t_\text{est}|t)=p(t_\text{est}-s|t-s): \forall s\in [0,
\tau)$, where  the subtraction is mod $\tau$; if this is not the case, one can always make the estimator invariant  by adding a random time translation to the input state, and then canceling it at the output of the estimator (See Supplementary Note 7). Suppose after obtaining the estimate $t_{\text{est}}$ one prepares $m(n)$ copies of state $e^{-i H t_\text{est}}|\phi_{\text{coh}}\rangle$. Then, the entire measure-and-prepare process will be described by a TI operation.   Furthermore, as we show in Supplementary Note 7, applying this TI operation on the input $\rho^{\otimes n}$, the fidelity of the resulting state with  the desired state $|\phi_{\text{coh}}\rangle^{\otimes m(n)}$ is 
\begin{align}
\int_0^\tau dt_\text{est}\ p(t_\text{est}|t=0)\ &|\langle\phi_\text{coh}|e^{i H t_\text{est}}|\phi_\text{coh}\rangle|^{2 m(n)}\\ &\ge 1-m(n) F_H(\phi_\text{coh})\times\langle\delta t^2\rangle/4\ \nonumber,
\end{align}
where $F_H(\phi_\text{coh})$ is four times the energy variance of $\phi_\text{coh}$, and  $\langle\delta t^2\rangle=\int_0^\tau dt_\text{est}\ p(t_\text{est}|t) (t-t_\text{est})^2$ is the  Mean Squared Error (MSE) of the estimator (Note that because of time-translation symmetry, MSE is independent of $t$). Therefore, the ratio of  the yield $r(n)=m(n)/n$ to infidelity $\epsilon_n$, satisfies 
  \be\label{sdmp}
  \frac{r(n)}{\epsilon_n} \ge  \frac{4}{F_H(\phi_\text{coh}) \times n \langle \delta t^2\rangle}\ .
  \ee
For any reasonable estimator the MSE $\langle \delta t^2\rangle$  scales as $1/n$. Therefore, as $n$ goes to infinity, the above lower bound remains positive. In particular, as shown in \cite{braunstein1994statistical, barndorff2000fisher}, there exists  an estimator working based on the classical Maximum Likelihood (ML) estimator, which achieves MSE $\langle\delta t^2\rangle=1/({n F_H(\rho)})+o(1/n) $, i.e. saturates the Quantum Cram\'er-Rao bound \cite{helstrom1969quantum, Holevo:book, braunstein1994statistical}. Therefore, using Eq.(\ref{sdmp}), we find that the ratio ${r(n)}/{\epsilon_n}$  for this estimator,  satisfies the lower bound in Eq.(\ref{wrefqqq}).

It is worth noting that in the high noise regime, where each input copy $\rho$ is close to the maximally mixed state, we have $P_H(\rho)/F_H(\rho)\approx 1$, and therefore the lower and upper bounds in Eq.(\ref{wrefqqq}) coincide, up to a constant factor $1/c$. Therefore, in this regime we can achieve close to optimal  distillation using a measure-and-prepare strategy. Furthermore, because asymptotically the optimal MSE can be achieved  using local adaptive measurements on  individual copies \cite{braunstein1994statistical, barndorff2000fisher}, this distillation process does not require any  entangling interactions between the input  copies. On the other hand, as we discuss in Methods section, such measure-and-prepare TI operations are, in general, sub-optimal for distillation in  the low-noise regime.

\subsection*{Single-shot Coherence Distillation: Exact formula }

 Next, we consider the problem of coherence distillation in the single-shot regime: suppose we are given $n$ copies of a system in a mixed state $\rho$  as the resource, and we want to obtain a single copy of a system in a pure coherent state $\psi$, using only TI operations? What is the maximum achievable fidelity $\max_{\mathcal{E}_\text{TI}}\  \langle\psi| \mathcal{E}_\text{TI}(\rho^{\otimes n})|\psi\rangle$, where the maximization is over all  TI operations.

Using the approach of \cite{gour2018quantum}, we find a simple  general formula for the maximum achievable fidelity:
\begin{equation}\label{hmin2}
\max_{\mathcal{E}_\text{TI}}\  \langle\psi| \mathcal{E}_\text{TI}(\rho^{\otimes n})|\psi\rangle=2^{- H_\text{min}(\text{out}|\text{in})_\Omega}\  ,
\end{equation}
where $H_\text{min}(\text{out}|\text{in})_\Omega$ is  the conditional min-entropy \cite{Konig, tomamichel2015quantum},  for the bipartite state $\Omega_{\text{in,out}}$, obtained by dephasing  state $(\rho^{\otimes n})_\text{in}\otimes |{\psi}\rangle\langle{\psi}|_\text{out}$ in the eingenbasis of Hamiltonian $H_\text{in}\otimes I_\text{out}-I_\text{in}\otimes H_\text{out}$. Here,  $I_\text{in}$ and $I_\text{out}$ are the identity operators, and  $H_\text{in}=\sum_{i=1}^n H^{(i)}$ and $H_\text{out}$ are the input and output  Hamiltonians, respectively. See Supplementary Note 9, for the proof and further discussion about  this  formula.

Although important, Eq.(\ref{hmin2}) does not clearly show  the asymptotic behavior of the maximum achievable fidelity.  On the other hand,  our  results 
on the purity of coherence and sub-linear coherence distillation yield simple general upper and lower bounds on the maximum achievable fidelity. 
Note that in Eq.(\ref{sdmp}),  the number of distilled copies $m(n)$ is arbitrary and can be independent of $n$. In fact, as we explain in Supplementary Note 7, for any (fixed) finite $m(n)=m$,  Eq.(\ref{sdmp}) is tight in the regime $n\rightarrow \infty$, and  $n\times \epsilon_n$ converges to  $m F_H(\phi_\text{coh})/4 F_H(\rho)$, where $\epsilon_n$ is the infidelity of the output with $m$ copies of $\phi_\text{coh}$.

\begin{figure} [h]
\begin{center}
\includegraphics[scale=0.44]{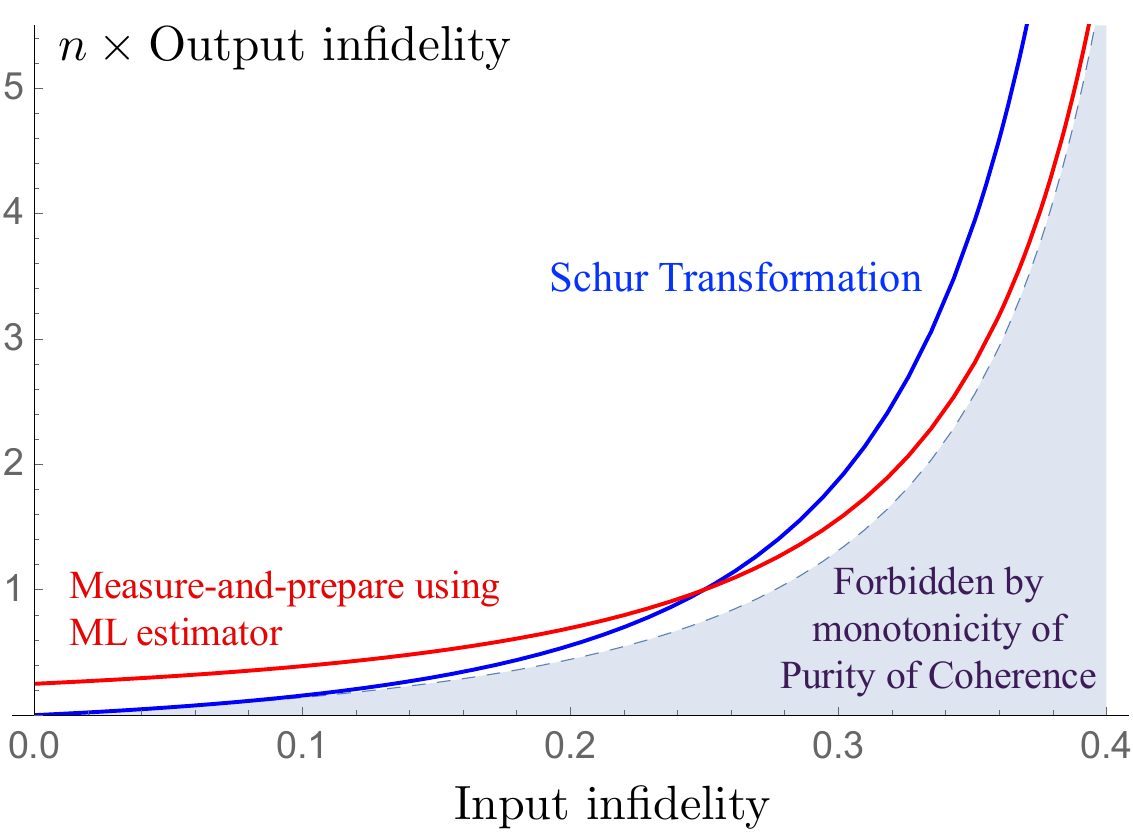}
\caption{Minimum achievable infidelity as a function of the input infidelity: We are given $n\gg 1$ two-level systems, each with Hamiltonian $\pi \sigma_z/\tau$, in state   $\rho=\lambda  |\phi_\text{coh}\rangle\langle\phi_\text{coh}| +(1-\lambda) I/2\ $,  where $0<\lambda<1$, i.e. a noisy version of state $|\phi_\text{coh}\rangle=(|0\rangle+|1\rangle)/\sqrt{2}$. The goal is to distill a single copy of $|\phi_\text{coh}\rangle$ with higher fidelity using TI operations. 
Horizontal axis is the infidelity of each input  state $\rho$ with the desired state $|\phi_\text{coh}\rangle$, which is equal to $(1-\lambda)/2$. 
For any reasonable coherence distillation process, the  infidelity at the output is in the form $e(\lambda)/n+ o(1/n)$. Vertical axis is the function $e(\lambda)$, i.e. $n$ times the output infidelity, in the limit $n\rightarrow\infty$. The dashed curve corresponds to the equation  $e(\lambda)=(1-\lambda^2)/4\lambda^2 $, dictated by the conservation of purity of coherence  (RLD Fisher information), i.e. is found by minimizing the infidelity with the desired state $|\phi_\text{coh}\rangle$, under the constraint that the purity of coherence remains conserved.  The shaded area below this curve is forbidden by the monotonicity of this quantity.  
 The blue curve is $e(\lambda)= (1-\lambda)/2\lambda^2$, achieved by a distillation process which works based on the Schur transformation \cite{cirac1999optimal}. The red curve is $e(\lambda)= 1/4\lambda^2$, achieved by a measure-and-prepare process which uses  ML estimator. Note that the lower bound imposed by the purity of coherence is tight in both 
 high-noise ($\lambda\rightarrow 0$) and low-noise ($\lambda\rightarrow 1$) regimes, but each of these TI operations achieves this lower bound only in one limit.  
} \label{Fig9}
\end{center}
\end{figure}

\subsection*{Example: Single-shot distillation of a two-level system}

The smallest  quantum clock is a system with two different energy levels. Without loss of generality we assume the  Hamiltonian of this system is $H=\pi \sigma_z/\tau$. Suppose  we want to prepare this clock in state $|\phi_\text{coh}\rangle=(|0\rangle+|1\rangle)/\sqrt{2}$, but we have access to a noisy version of this state, i.e. $\rho=\lambda |\phi_\text{coh}\rangle\langle \phi_\text{coh}|_{\text{}}+(1-\lambda) I/2\ $,  with  $0<\lambda < 1$. The goal is to use $n\gg 1$ copies of $\rho$ to obtain a state with higher fidelity with $ |\phi_\text{coh}\rangle$. What is the lowest achievable infidelity? Using the properties of the purity of coherence and, in particular, Eq.(\ref{lastbound}), in Supplementary Note 10 we show that the infidelity is lower bounded by
\be\label{Eq123}
1-\max_{\mathcal{E}_\text{TI}}\ \langle\phi_\text{coh}| \mathcal{E}_\text{TI}(\rho^{\otimes n})   |\phi_\text{coh}\rangle \ge \frac{1}{n} \frac{1-\lambda^2}{4\lambda^2}+\mathcal{O}(\frac{1}{n^2})\ .
\ee
\color{black}
 Therefore, in the limit of large $n$,  infidelity times $n$ is lower bounded by $(1-\lambda^2)/4\lambda^2$.  In Fig.(\ref{Fig9}) we compare this lower bound  with the infidelity achieved by two different TI processes: (i)  an operation  related to quantum Schur transformation,  studied previously in \cite{cirac1999optimal}, which has full SU(2) symmetry, and hence is also TI. As we discuss in Supplementary Note 10, the results of \cite{cirac1999optimal} implies that using this process we can achieve the infidelity $(1-\lambda)/2n\lambda^2$. (ii) The measure-and-prepare process based on the ML estimator, discussed in the previous section, which achieves the infidelity $n^{-1} \times F_H(\Phi)/4F_H(\rho)=1/(4n\lambda^2)$. 

Remarkably, we find that the bound imposed by the purity of coherence in Eq.(\ref{Eq123}) is tight in both high-noise ($\lambda\rightarrow 0$) and low-noise ($\lambda\rightarrow 1$) regimes. This suggests that this bound is achievable for all values of $\lambda$, and, at least in this example,  the purity of coherence determines the ultimate limit of coherence distillation in the single-shot regime. 

\subsection*{Discussion}

In recent years there has been a significant progress in understanding the concept of coherence in the context of quantum thermodynamics (See e.g. \cite{lostaglio2015description,lostaglio2015quantumPRX,korzekwa2016extraction, narasimhachar2015low, streltsov2017colloquium, goold2016role, chiribella2017optimal, halpern2016beyond}).   Nevertheless,  some aspects of coherence are still not well-understood. Here, we highlighted an important feature of quantum coherence which manifests itself, for instance, in the unreachability of pure coherent states from mixed states in both the single-shot and asymptotic regimes, and the fact that (in some precise sense) the coherence content of a single qubit can be arbitrarily large.  To quantify this feature of coherence, we introduced a new quantifier of coherence, called the purity of coherence  and showed that the monotonicity of this quantity under TI operations  gives a tight bound on the coherence distillation in the single-shot regime. The tightness of this bound supports the idea that the purity of coherence is adequately quantifying the unreachability of pure coherent states from mixed states.

In this paper, we focused on the implications of our results in the context of quantum clocks and thermodynamics. Another important area of applications is quantum metrology \cite{giovannetti2006quantum, giovannetti2011advances,  marvian2014extending, marvian2016quantify, yang2017units}, which will be discussed in future works.   \\

\section*{Methods}

\subsection*{Limited power of TI measure-and-prepare processes for distillation}

In the above example, it is interesting to note that in the high noise regime,  the optimal distillation can be achieved using a measure-and-prepare TI process. On the other hand, in the opposite limit, where the input state $\rho$ is almost pure,  measure-and-prepare TI processes are not optimal for coherence distillation. In fact, as it can be seen in Fig.(\ref{Fig9}), even if the input is  $n$ copies of a pure coherent state $\phi_\text{coh}$, the output of a measure-and-prepare distillation process can not be a pure coherent state for any finite $n$.

To understand  this fact better, in the following we derive a strong constraint on the power of measure-and-prepare TI processes  for manipulation of coherence. This constraint is a corollary of the following result: For any state $\rho$ and any Measure-and-Prepare  TI process $\mathcal{E}_\text{MP-TI}$, it holds that 
\be\label{ffsr}
P_{H_\text{out}}(\mathcal{E}_\text{MP-TI}(\rho))\le F_{H_\text{in}}(\rho)\le P_{H_\text{in}}(\rho)\ ,
\ee
i.e. the purity of coherence of the output is upper bounded by the input QFI, where $H_\text{in}$ and $H_\text{out}$ are, respectively, the input and output Hamiltonians (See below for further discussion). This means that for input $\rho^{\otimes n}$, the purity of coherence of the  output of a measure-and-prepare TI  process  is upper bounded by $n\times F_{H_\text{in}}(\rho)$. On the other hand, for a general TI process the purity of coherence of the output can be as large as $n\times P_{H_\text{in}}(\rho)$, which is much larger than $n\times F_{H_\text{in}}(\rho)$, if $\rho$ is close to a coherent pure state (For instance, in the above example, Schur transformation reaches this bound in the low noise regime). 

Combining this result with the lower bound on the purity of coherence in Eq.(\ref{wfewfsss}), we find that if one applies a  measure-and-prepare TI process to $n$ copies of $\rho$ to obtain $m(n)$ copies of a pure coherent state $\phi_\text{coh}$ with error $\epsilon_n$, then for sufficiently large $m(n)$ and small error $\epsilon_n$,  the yield $r(n)=m(n)/n$ and error $\epsilon_n$ satisfy  $r(n)/\epsilon_n\le 1/c \times F_H(\rho)/F_H(\phi_\text{coh}) $. Therefore, if QFI of state $\rho$  is finite, which is always the case for systems with bounded Hamiltonians, then using measure-and-prepare TI processes it is not possible to achieve a finite yield $r(n)>0$ with a vanishing error $\epsilon_n\rightarrow 0$, even if  $\rho$ is a pure coherent state, i.e. has an unbounded purity of coherence.

In Supplementary Note 8  we present the proof of inequality $P_{H_\text{out}}(\mathcal{E}_\text{MP-TI}(\rho))\le F_{H_\text{in}}(\rho)$ in Eq.(\ref{ffsr}). We also note that this inequality follows from the previous result of \cite{matsumoto2005reverse}. The main idea is the following: By definition any measure-and-prepare process can be realized by a measurement on the input followed by a state preparation at the output, which solely depends on the classical outcome of the measurement.  For input states $\{e^{-i H_\text{in} t}\rho e^{i H_\text{in} t}\}_t$, consider the distribution of outcomes of this measurement, as a function of parameter $t$. Then, the (classical) Fisher information corresponding to parameter $t$ is upper bounded by QFI of the input state, i.e. $F_{H_\text{in}}(\rho)$. As we show in Supplementary Note 8, this classical Fisher information, itself, is an upper bound on  $P_{H_\text{out}}(\mathcal{E}_\text{MP-TI}(\rho))$, the purity of coherence of the output (This also has been shown previously in \cite{matsumoto2005reverse}). Roughly speaking, this is true because at the classical level, the distinction between  Fisher information and the purity of coherence vanishes (This is related to \v Cencov's  theorem \cite{morozova1991markov} which asserts that, up to a normalization,  Fisher information is the unique monotone metric on the space of classical probability distributions). %

  \color{black}

\subsection*{Violation of monotonicity of Petz-R\'enyi relative entropy in the light of coherence distillation}
  
Our results on coherence distillation, and in particular Eq.(\ref{wrefqqq}) and Eq.(\ref{Eq123}),  provide a novel operational understanding  of the violation of   monotonicity of Petz-R\'enyi relative entropy under data-processing, for $\alpha>2$. Recall that for  $\alpha>1$, Petz-R\'enyi relative entropy is defined as $D_\alpha(\rho\|\sigma)=\frac{1}{\alpha-1}\log \Tr(\rho^\alpha\sigma^{1-\alpha})$ , if $\text{supp}(\rho) \subseteq \text{supp}(\sigma)$ and $D_\alpha(\rho\|\sigma)=\infty$, otherwise \cite{petz1986quasi, tomamichel2015quantum}. For $\alpha\in (1,2]$, and 
any completely positivity  trace-preserving map $\mathcal{E}$,  $D_\alpha(\mathcal{E}(\rho)\|\mathcal{E}(\sigma))\le D_\alpha(\rho\|\sigma)$,  whereas  this bound is violated for $\alpha>2$ \cite{petz1986quasi, tomamichel2015quantum}.  As we mentioned before, the purity of coherence can be derived from the second derivative of  the Petz-R\'enyi relative entropy for $\alpha=2$, and its monotonicity under TI operations follows from the monotonicity of this relative entropy (See Supplementary Note 2).    Considering the second derivative of  Petz-R\'enyi relative entropy for other values of $\alpha\in (1,\infty)$,  we can generalize the purity of coherence, and obtain the family of  functions defied by the formula $P_{H,\alpha}(\rho)\equiv\Tr(\rho^\alpha H \rho^{1-\alpha} H)-\Tr(\rho H^2)$, if the projector to the support of $\rho$ commutes with $H$, and $P_{H,\alpha}(\rho)=\infty$ otherwise. Similar to the purity of coherence, all these functions are (i) additive, (ii) non-zero iff state is coherent, and (iii) bounded if the projector to the support of $\rho$ commutes with $H$. Furthermore, for any state $\rho$ whose infidelity with a pure coherent state is $\epsilon$,  $P_{H,\alpha}(\rho)$ scales (at least) as $\epsilon^{1-\alpha}$. It follows that, if instead of the purity of coherence  we use  other monotone functions in this family, we obtain lower bounds on achievable infidelity,   which is stronger than the bounds obtained from purity of coherence. In particular, such a bound would imply that if the purity of coherence of a mixed state $\rho$ is finite, then   to distill a single copy of a pure coherent state $\phi_\text{coh}$ with error $\epsilon$,  
the  required number of copies of $\rho$ is, at least, of order $\epsilon^{1-\alpha}$, i.e. $n\in\Omega(\epsilon^{1-\alpha})$. For $\alpha>2$ this bound is asymptotically stronger than  the bound imposed by purity of coherence, which is linear in $\epsilon^{-1}$.
 
However, as we have seen in the proof of Eq.(\ref{wrefqqq}) and also in Fig.(\ref{Fig9}), there exists a TI process based on the ML estimator which achieves errors of order $\epsilon$, by consuming only order $\epsilon^{-1}$ copies of $\rho$. Therefore, if  Petz-R\'enyi relative entropy was monotone for $\alpha>2$, we had a lower bound on the number of required copies, which was violated by this coherence distillation process. This provides an operational explanation that why the Petz-R\'enyi relative entropy cannot be monotone under data-processing for $\alpha>2$: $\alpha=2$ is the largest value for which the monotonicity of Petz-R\'enyi relative entropy is not violated by  coherence distillation processes. 

\subsection*{Proofs}

All the results in the paper are rigorously proven in the Supplementary Notes 1-10.

\section*{Acknowledgments}

\noindent I am grateful to Gilad Gour and David Jennings for reading the manuscript carefully, and providing many useful comments and suggestions. 
Also, I would like to thank Anna Jen\v cov\' a,  Mil\'an Mosonyi, and Keiji Matsumoto for helpful discussions on Fisher Information.

\bibliography{Ref_2018}

$$ $$
\noindent \textbf{Contributions}

\noindent IM was the sole contributor to all the aspects of this work.\\

\noindent \textbf{Competing Interest}

\noindent The author declares no competing interest.

\newpage

\onecolumngrid

\onecolumngrid

\newpage

\maketitle
\vspace{-5in}
\begin{center}

\Large{Supplementary Material:\\ $ $ \\   Coherence distillation machines are impossible in quantum thermodynamics }
\end{center}
\appendix

	
	\title{\textbf{Supplementary Material}:\\ $ $ \\   Coherence distillation machines are impossible in quantum thermodynamics}
	
	\author{Iman Marvian}
\affiliation{Departments of Physics \& Electrical and Computer Engineering, Duke University, Durham, North Carolina 27708, USA}



\maketitle

$$  $$

$$  $$

{\Large{\textbf{Contents}}}
$$   $$
\begin{itemize}
\item \textbf{Supplementary Note 1:} Three equivalent definitions of TI operations\\
 We review some properties of TI operations and present three different
ways for characterizing this set of operations. In particular, we prove an operation is TI, if and only if, it is completely incoherence-preserving.

\item \textbf{Supplementary Note 2: Purity of Coherence}

In this section we introduce purity of coherence and study its properties. This section includes the following subsections:
\begin{itemize}
\item Connection with relative Petz-Renyi relative entropy

\item Stochastic state conversions under TI operations (Proof of  Eq.7 in the paper) 

\item{States with infinite purity of coherence}
\item Purity of coherence is lower-bounded by Quantum Fisher Information

\item Purity of coherence for Qubits (Proof of Eq.9 in the paper)

\item Purity of coherence for states close to the maximally mixed state

\end{itemize}

\item \textbf{Supplementary Note 3: Purity of coherence for a mixed state close to a pure state}

\item \textbf{Supplementary Note 4: QFI and purity of coherence in the iid regime (Proof of Eq.10 and  Eq.11 in the paper)} 

\item \textbf{Supplementary Note 5: Extension of the main theorem (Finite helper systems do not help)}

\item\textbf{Supplementary Note 6: Mixed states with distillable coherence}

\item \textbf{Supplementary Note 7: Sub-linear distillation with a measure-and-prepare TI process (Proof of Eq.14 in the paper)}

\item\textbf{Supplementary Note 8: Purity of coherence of the output of Measure-and-Prepare TI channels is upper bounded by QFI of the input}
\item \textbf{Supplementary Note 9: Distillation in the single-shot regime (Proof of  Eq.16 in the paper)}

We present a simple formula for the maximum achievable fidelity of distillation, in terms of conditional min-entropy.

\item \textbf{Supplementary Note 10: Qubit example (Proof of Eq.17 in the paper)}

\end{itemize}

\newpage
\section*{Supplementary Note 1: Three equivalent definitions of TI operations}\label{Sec:app0}

In this section we review some useful properties of Time-translationally Invariant (TI) operations (See e.g. \cite{Marvian_thesis} for further discussion). We can summarize these properties in the following theorem.\\


\refstepcounter{Thm}\label{Thm_three}
\noindent\emph{Theorem 1}
Let $\mathcal{E}$ be a Completely Positive Trace-Preserving (CPTP) linear map with arbitrary input and output spaces with Hamiltonians $H_\text{in}$ and $H_\text{out}$, respectively. 
The following three properties are equivalent:
\begin{enumerate}
\item  \textbf{Invariance under time-translations}: The map  $\mathcal{E}$ satisfies
\be\label{qwert}
\forall t\in\mathbb{R}:\ \ \  \mathcal{E}(e^{-i H_\text{in} t}\rho_\text{in}  e^{i H_\text{in} t})=e^{-i H_\text{out} t}\mathcal{E}(\rho_\text{in} ) e^{i H_\text{out} t}\ ,
\ee
for arbitrary input state $\rho_\text{in} $.
\item  \textbf{Covariant Stinespring Dilation}: The map  $\mathcal{E}$  can be implemented by coupling the input system to an auxiliary system $A$ with Hamiltonian $H_{\text{A}}$, whose initial state $|\eta\rangle$ is an eigenstate of $H_{\text{A}}$, via an energy-conserving unitary $U$, such that  
\be
\mathcal{E}(\rho_\text{in} )=\Tr_{\text{A}'}\Bigl(U[\rho_\text{in} \otimes |\eta\rangle\langle \eta|_{\text{A}}]U^\dag \Bigr)\ ,
\ee 
where (i) $A'$ denotes the discarded output auxiliary system which is a closed system with Hamiltonian $H_{\text{A}'}$, (ii) the unitary $U$ is energy-conserving, i.e.
\be
U (H_\text{in}\otimes I_{\text{A}}+I_\text{in}\otimes H_{\text{A}}) =(H_\text{out}\otimes I_{\text{A}'}+I_\text{out}\otimes H_{\text{A}'})U\ ,
\ee
  where $I_\text{in}$, $I_\text{out}$,  $I_{\text{A}}$ and $I_{\text{A}'}$ are,  the identity operators on the input and output systems, and input and output auxiliary systems, respectively,  and (iii) $|\eta\rangle$ is an eigenstate of $H_\text{A}$.   
 \item \textbf{Completely incoherence-preserving}: $\mathcal{E}$ is completely incoherence-preserving, that is for any auxiliary system B with an arbitrary Hamiltonian $H_{\text{B}}$, and any joint  state $\rho_{\text{in}, \text{B}}$ of  the input system (denoted by $in$) and the auxiliary system $B$,  if $\rho_{\text{in}, \text{B}}$ is incoherent with respect to the total Hamiltonian $H_\text{in}\otimes I_{\text{B}}+I_\text{in}\otimes H_{\text{B}}$, then the corresponding  output state  $\mathcal{E}\otimes \mathcal{I}_{\text{B}}(\rho_\text{{in}, B})$ is also incoherent with respect to the Hamiltonian $H_\text{out}\otimes I_{\text{B}}+I_\text{out}\otimes H_{\text{B}}$, i.e.
\be
[\rho_\text{{in}, B}\ , H_\text{in}\otimes I_{\text{B}}+I_\text{in}\otimes H_{\text{B}}]=0\ \  \Longrightarrow \ \ [\mathcal{E}\otimes \mathcal{I}_{\text{B}}(\rho_\text{{in}, B})\ , H_\text{out}\otimes I_{\text{B}}+I_\text{out}\otimes H_{\text{B}}]=0\ .
\ee
\end{enumerate}




\noindent \textbf{Remark.} As we explain later, to check whether a quantum operation is completely-incoherence preserving or not, one only needs to consider this condition for a system $B$ whose dimension is  equal to $d_\text{in}$, the dimension of the input  of $\mathcal{E}$. Also, one only needs to check this condition for one initial state, namely the maximally entangled state $|\Psi\rangle_{\text{in},\text{B}}= \frac{1}{\sqrt{d_\text{in}}} \sum_{i=1}^{d_\text{in}} |ii\rangle$, with the Hamiltonian of $B$ equal to $H_{\text{B}}=-H^T_{\text{A}}$, where the transpose is defined with respect to the basis $\{|i\rangle\}$.\\

The notion of completely incoherence-preserving operations can be compared with the notion of \emph{incoherence-preserving}  operations, also known as \emph{maximally incoherent} operations. These are operations which map incoherent state of the  input system $S$ to incoherent states of the output system \cite{aberg}.  It turns out this set is strictly larger than the set of  completely incoherence-preserving operations, i.e. there are operations which are incoherence-preserving, but not  completely incoherence-preserving. For instance, any unitary transformation which permutes  energy eigenstates with different energies, is an incoherence-preserving but not completely incoherence-preserving operation (The relation between the two sets in analogous to the relation between positive operations and completely positive operations).

\subsection*{Covariant Stinespring Dilation theorem (Equivalence of statements (1) and (2) in theorem \ref{Thm_three})}

Equivalence of properties (1) and (2), i.e. Invariance Under Time-Translation and Covariant Stinespring Dilation is proven before, e.g. in  \cite{Marvian_thesis}. In fact, theorem 25 of \cite{Marvian_thesis} establishes this equivalence for a general symmetry group.  For completeness,  we present the proof in the case of time-translation symmetry. 

First, it is straightforward to see that  the existence of Covariant Stinespring Dilation implies Time-translation symmetry  (Intuitively,        covariant Stinespring dilation provides a method for implementing the quantum operation. Since each step in this method respects the time-translation symmetry, the composition should also respect the symmetry).  

To prove statement (1) implies statement (2), we use a result of \cite{gour2008resource}, which shows the Kraus representation of TI operations can be written in a special canonical form. According to this lemma, by exploiting  the  unitary freedom in defining the Kraus operators of a general quantum operation \cite{nielsen2000quantum},  we can write any TI operation $\mathcal{E}_\text{TI}$ , in the form
\be
\mathcal{E}_\text{TI}(\sigma)=\sum_{E,\alpha} K_{(E,\alpha)} \sigma K^\dag_{(E,\alpha)} \ ,
\ee
where, in addition to the normalization condition  $\sum_{E,\alpha} K^\dag_{(E,\alpha)} K_{(E,\alpha)}=I_\text{out}$, Kraus operators satisfy the condition
\be\label{Kraus}
e^{-i H_\text{out} t} K_{(E,\alpha)} e^{i H_\text{in} t}= e^{-i E t} K_{(E,\alpha)}\ .
\ee
For completeness, we present the proof of this result, originally proven in \cite{gour2008resource}:  
Consider an arbitrary Kraus decomposition $\mathcal{E}_\text{TI}(\cdot)= \sum_\mu \tilde{K}_\mu(\cdot) \tilde{K}_\mu^\dag$, with linearly independent Kraus operators $\{\tilde{K}_\mu\}_\mu$. Then,  invariance under time translation  implies for arbitrary $t\in \mathbb{R}$, the set $\{e^{-i H_\text{out} t} \tilde{K}_\mu e^{i H_\text{in} t}\}_\mu$ also defines a  valid Kraus representation of  $\mathcal{E}_\text{TI}$. But,  two sets of linearly independent Kraus operators describe the same quantum operation, if and only if they are related via a unitary, i.e. there exists a unitary transformation $V(t)$ such that, 
$e^{-i H_\text{out} t} \tilde{K}_\mu e^{i H_\text{in} t}=\sum_{\mu'} V_{\mu \mu'}(t) \tilde{K}_{\mu'}  \ .
$

This equation together with the fact that $\{\tilde{K}_\mu\}$  are linearly independent implies that $V(t_1) V(t_2)=V(t_1+t_2)$  (indeed, it implies that  the unitary $V(t)$ is itself a representation of the time-translation symmetry). 

We conclude that there exists a unitary $S$ which simultaneously diagonalizes all unitaries $V(t)$ for all $t\in \mathbb{R}$ and decomposes this representation to irreducible 1-dimensional representation, as $S V(t) S^\dag=\sum_{(E,\alpha)} e^{-i E t}|E,\alpha\rangle\langle E,\alpha| $, where $\alpha$ is a multiplicity index. Define the new Kraus  operators 
\be\label{asjdlhg}
{K}_{(E,\alpha)}\equiv \sum_{\mu} S_{(E,\alpha) ,\mu}\tilde{K}_\mu\ ,
\ee
which implies
\be\label{asjdlhg2}
\tilde{K}_\mu= \sum_{(E,\alpha)}  S^\ast_{(E,\alpha) ,\mu} {K}_{(E,\alpha)}\ ,
\ee

The unitarity of $S$ guarantees that $\sum_{E,\alpha} {K}^\dag_{(E,\alpha)} {K}_{(E,\alpha)}=\sum_{\mu} {K}^\dag_{\mu} {K}_{\mu}=I_\text{out}$. Furthermore, 
\begin{align}
e^{-i H_\text{out} t} {K}_{(E,\alpha)} e^{i H_\text{in} t}&=  \sum_{\mu}  S_{(E,\alpha) ,\mu}\ e^{-i H_\text{out} t}  \tilde{K}_\mu  e^{i H_\text{in} t} \\ &= \sum_{\mu}  S_{(E,\alpha) ,\mu} \sum_{\mu'} V_{\mu \mu'}(t) \tilde{K}_{\mu'}  \\ &= \sum_{\mu}  S_{(E,\alpha) ,\mu} \sum_{\mu'} V_{\mu \mu'}(t)  \sum_{(E',\alpha')}  S^\ast_{(E',\alpha') ,\mu'} {K}_{(E',\alpha')} \\ &=\sum_{(E',\alpha')}  {K}_{(E',\alpha')}   \sum_{\mu,\mu'}  S_{(E,\alpha) ,\mu}  V_{\mu \mu'}(t)   S^\ast_{(E',\alpha') ,\mu'} \\ &=\sum_{(E',\alpha')}  {K}_{(E',\alpha')}   e ^{-i Et}\  \delta_{E,E'}  \delta_{\alpha,\alpha'}\\ &= {K}_{(E,\alpha)}   e ^{-i Et} \ ,
\end{align}
where, we have used Supplementary Eq.(\ref{asjdlhg}) to get the first line, and Supplementary Eq.(\ref{asjdlhg2}) to get the third line.  

Next, we use this result to construct an energy-conserving unitary which implements $\mathcal{E}_\text{TI}$. Define the operator
\be
W=\sum_{(E,\alpha)} K_{(E,\alpha)}\otimes |E,\alpha \rangle\langle \eta|\ ,
\ee
where $\{|E,\alpha\rangle\}$ is an arbitrary set of orthogonal states of system A$'$, and $|\eta\rangle$ is an arbitrary state of A. 
 Then, one can easily see that:
 
 \begin{itemize}
\item Using the fact that $\sum_{E,\alpha} K^\dag_{(E,\alpha)}K_{(E,\alpha)}=I_\text{in}$, we find  $W$ is an isometry, i.e.
\be
W^\dag W=I_\text{in} \otimes |\eta\rangle\langle \eta|_{\text{A}}\ ,
\ee
\item Isometry $W$ is an Stinespring dilation of $\mathcal{E}_\text{TI}$, i.e.
\be
\mathcal{E}_\text{TI}(\rho)= \Tr_{\text{A}'}(W [\rho\otimes |\eta\rangle\langle \eta|_{\text{A}} ] W^\dag)\ .
\ee
\item Suppose we define $H_{\text{A}}$, the Hamiltonian of A, to be a Hermitian operator with  state $|\eta\rangle$ as its eigenvector with eigenvalue zero,  such that $H_{\text{A}}|\eta\rangle=0$, and $H_{\text{A}'}$ as a Hermitian operator with eigenvectors $|E,\alpha\rangle$, such that   $H_{\text{A}'}|E,\alpha\rangle=-E |E,\alpha\rangle $, then  
\begin{align}
(e^{- i H_\text{out} t} \otimes e^{- i H_{\text{A}'} t}) W (e^{i H_\text{in} t} \otimes e^{i H_{\text{A}} t}) &= \sum_{(\mu,\alpha)} e^{- i H_\text{out} t} K_{(E,\alpha)} e^{i H_\text{in} t} \otimes e^{-i H_{\text{A}'} t} |E,\alpha \rangle\langle \eta|e^{i H_{\text{A}} t} \\ &=\sum_{(\mu,\alpha)}  e^{-i E t} K_{(E,\alpha)}\otimes |E,\alpha \rangle\langle \eta| e^{i Et}\\ &=  W  \ , 
\end{align}
where to get the second line we have used Supplementary Eq.(\ref{Kraus}). Equivalently, this implies 
\begin{align}
(e^{- i H_\text{out} t} \otimes e^{- i H_{\text{A}'} t}) W =W (e^{i H_\text{in} t} \otimes e^{i H_{\text{A}} t})  \ .
\end{align}
Taking the derivative with respect to $t$ this implies,
\be
W (H_\text{in}\otimes I_{\text{A}}+I_\text{in}\otimes H_{\text{A}}) =(H_\text{out}\otimes I_{\text{A}'}+I_\text{out}\otimes H_{\text{A}'})W\ .
\ee
\end{itemize}

It can be easily shown that the isometry $W$ can always be extended to a unitary which is also energy-conserving (See \cite{Marvian_thesis}). This completes the proof that (1) implies (2).

\subsection*{Completely Incoherence-Preserving operations (Equivalence of statements (1) and (3) in theorem \ref{Thm_three})}

The fact that any TI operation is completely incoherence-preserving, follows immediately from the covariance condition in Supplementary Eq.(\ref{qwert}). In the following we prove any  completely incoherence-preserving operation is TI. To prove this we consider an auxiliary system with dimension equal to the input space of $\mathcal{E}$, and with the Hamiltonian  $H_\text{B}=-H^T_\text{in} $, where $T$ denotes transpose in an orthonormal basis $\{|i\rangle\}$.  Then, consider maximally entangled state
\be
|\Psi\rangle_{\text{in},\text{B}}= \frac{1}{\sqrt{d_\text{in}}} \sum_{i=1}^{d_\text{in}} |ii\rangle\ ,
\ee 
where $d_\text{in}$ is the dimension of the input space of $\mathcal{E}$.

It can be easily seen that this state is incoherent with respect to the total Hamiltonian 
\be
H_\text{tot}=H_\text{in}\otimes I_{\text{B}}+I_{\text{in}}\otimes H_{\text{B}}=H_\text{in}\otimes I_{\text{A}}-I_{\text{in}}\otimes H^T_{\text{in}}
\ee
This can be seen, for instance, by noting that
\be
e^{-i H_\text{tot} t}|\Psi\rangle_{\text{in},\text{B}}= (e^{-i H_\text{in} t}\otimes  e^{i H^T_\text{in} t}) |\Psi\rangle_{\text{in},\text{B}}=(e^{-i H_\text{in} t}\otimes  e^{i H^T_\text{in} t}) \frac{1}{\sqrt{d_\text{in}}} \sum_{i=1}^{d_\text{in}} |ii\rangle_{\text{in},\text{B}}= (e^{i H_\text{in} t} e^{-i H_\text{in} t}\otimes I )|\Psi\rangle_{\text{in},\text{B}}= |\Psi\rangle_{\text{in},\text{B}}\ ,
\ee 
where we have used the fact that for any operator $X$, $(I\otimes X) \sum_{i=1}^{d_\text{in}} |ii\rangle=(X^T\otimes I) \sum_{i=1}^{d_\text{in}} |ii\rangle$. 

Therefore, since state $|\Psi\rangle_{\text{in},\text{B}}$  is incoherent and the quantum operation $\mathcal{E}$ is, by assumption, completely incoherence-preserving, then state $ \mathcal{E}\otimes \mathcal{I}_{\text{B}}(|\Psi\rangle\langle\Psi|_{\text{in},\text{B}})$ is also incoherent, i.e. 
\be
\left[ H_\text{out}\otimes I_{\text{B}}+I_{\text{out}}\otimes H_{\text{B}}\ ,\ \mathcal{E}\otimes \mathcal{I}_{\text{B}}(|\Psi\rangle\langle\Psi|_{\text{in},\text{B}})\right]=0\ ,
\ee
where $\mathcal{I}_{\text{B}}$ is the identity operation on system B. This implies
\bes\label{Eq4653}
\begin{align}
[e^{-i H_\text{out} t}\otimes  e^{-i H_{\text{B}} t}] \mathcal{E}\otimes \mathcal{I}_{\text{B}}(|\Psi\rangle\langle\Psi|_{\text{in},\text{B}}) [e^{i H_\text{out} t}\otimes  e^{i H_{\text{B}} t}]&=  
[e^{-i H_\text{out} t}\otimes  e^{i H^T_\text{in} t}] \mathcal{E}\otimes \mathcal{I}_{\text{B}}(|\Psi\rangle\langle\Psi|_{\text{in},\text{B}}) [e^{i H_\text{out} t}\otimes  e^{-i H^T_\text{in} t}]\\ &=\mathcal{E}\otimes \mathcal{I}_{\text{B}}(|\Psi\rangle\langle\Psi|_{\text{in},\text{B}}) \ ,
\end{align}
\ees
for all $t\in\mathbb{R}$. Recall that $\mathcal{I}_{\text{B}}$ is the identity operation on system $B$, and therefore, the left-hand side can be written as 
\be
[e^{-i H_\text{out} t}\otimes  I_{\text{B}}] \mathcal{E}\otimes \mathcal{I}_{\text{B}}\left([I_\text{in}\otimes  e^{-i H_{\text{B}} t}] |\Psi\rangle\langle\Psi|_{\text{in},\text{B}} [I_\text{in}\otimes  e^{i H_{\text{B}} t}]  \right) [e^{i H_\text{out} t}\otimes  I_{\text{B}}]\ .
\ee
Then, using the fact that $H_{\text{B}}=-H^T_\text{in}$ and the identity $(I\otimes X) \sum_{i=1}^{d_\text{in}} |ii\rangle=(X^T\otimes I) \sum_{i=1}^{d_\text{in}} |ii\rangle$, we find that the left-hand side of Supplementary Eq.(\ref{Eq4653}) can be rewritten as
\be
[e^{-i H_\text{out} t}\otimes I] \mathcal{E}\otimes \mathcal{I}\bigl([e^{i H_\text{in} t}\otimes I] |\Psi\rangle\langle\Psi|_{\text{in},\text{B}} [e^{-i H_\text{in} t}\otimes I]\bigr) [e^{i H_\text{out} t}\otimes  I]=\mathcal{E}_t\otimes \mathcal{I}_{\text{B}}(|\Psi\rangle\langle\Psi|_{\text{in},\text{B}}) \ ,
\ee
where we have defined the quantum operation $\mathcal{E}_t$ to be the time-translated version of $\mathcal{E}$, i.e.
\be
\mathcal{E}_t(\sigma)\equiv  e^{-i H_\text{out} t} \mathcal{E}\bigl(e^{i H_\text{in} t}\sigma e^{-i H_\text{in} t}) e^{i H_\text{out} t}\ .
\ee
Then, Supplementary Eq.(\ref{Eq4653}) can be rewritten as
\be
\forall t\in\mathbb{R}:\ \  \mathcal{E}_t\otimes \mathcal{I}_{\text{B}}(|\Psi\rangle\langle\Psi|_{\text{in},\text{B}})=\mathcal{E}\otimes \mathcal{I}_{\text{B}}(|\Psi\rangle\langle\Psi|_{\text{in},\text{B}})  \ .
\ee
In other words, the Choi matrix of quantum operations $\mathcal{E}_t$ is independent of $t$. It is well-known that the Choi matrix uniquely determines the quantum operation. This immediately implies that 
\be
\forall t\in\mathbb{R}:\ \  \mathcal{E}_t=\mathcal{E}\ ,
\ee
which means $\mathcal{E}$ is a TI operation. This completes the proof.

\subsection*{General symmetries: Completely symmetry-preserving operations}

In this paper, we study the notion of coherence as asymmetry with respect to time-translation symmetry. This is a specific type of asymmetry. The resource theory of asymmetry studies asymmetry with respect to am arbitrary symmetry group $G$. In this recourse theory, one studies the consequences of restriction to quantum operations which satisfy the following covariance condition 
\be
\forall g\in G:\  \ U_{\text{B}}(g)(\mathcal{E} (\cdot)U^\dag_{\text{B}}(g)=\mathcal{E} (U_{\text{A}}(g)(\cdot) U^\dag_{\text{A}}(g))\ ,
\ee
where $\mathcal{E}$ be a quantum operation from the input system A to the output system B with the unitary representation $G \ni g\rightarrow U_{\text{A,B}}(g)$ on the input A and output B.  
 
Any operation $\mathcal{E}$ which satisfies this condition is called a \emph{covariant} or \emph{symmetric} operation.  This definition is a natural generalization of  the notion of TI operations. 

Similarly, we can define a generalization of the notion of completely incoherence-preserving operations:  
 Let C be an arbitrary auxiliary system with arbitrary representation of symmetry, $G \ni g\rightarrow U_{\text{C}}(g)$. 
We say quantum operation $\mathcal{E}$ from A to B is a \emph{completely symmetry-preserving} (with respect to group $G$), if for all such auxiliary systems, and for all symmetric input states $\rho_{\text{AC}}$, it holds that    
\be
\forall g\in G :\  [\rho_{\text{AC}}, U_{\text{A}}(g)\otimes U_{\text{C}}(g)]=0, \ \ \Longrightarrow \forall g\in G :\  [\mathcal{E}\otimes \mathcal{I}_{\text{C}}(\rho_{\text{AC}}), U_{\text{B}}(g)\otimes U_{\text{C}}(g)]=0\ , 
\ee
where $\mathcal{I}_{\text{C}}$ is the identity quantum operation on system C. This condition  means that any symmetric input state of systems A and C is mapped to a symmetric state of systems B and C.\\

\noindent \emph{Proposition}
An operation is completely symmetry-preserving if and only if it is covariant.

\begin{proof}
Obviously any covariant operation is non-asymmetry-generating. To see the other direction, suppose the auxiliary system C has dimension equal to the dimension of the input system A, and the representation of symmetry on C is the complex conjugate of the representation of symmetry on A, i.e. $\forall g\in G:\ U_{\text{C}}(g)=\overline{U}_{\text{A}}(g)$. Consider the \emph{singlet} state 
\be
|\Psi\rangle_{\text{AC}}=\frac{1}{\sqrt{d_{\text{A}}}} \sum^{d_{\text{A}}}_{i=1} |ii\rangle_{\text{AC}}\ .
\ee 
This state is invariant under the action of symmetry, i.e. 
\be
[U_{\text{A}}(g)\otimes U_{\text{C}}(g)]|\Psi\rangle_{\text{AC}}=[U_{\text{A}}(g)\otimes \overline{U}_{\text{A}}(g)]|\Psi\rangle_{\text{AC}}=|\Psi\rangle_{\text{AC}}\ .
\ee
 Now for the initial state $|\Psi\rangle_{\text{AC}}$, suppose we act on system $A$ with quantum operation $\mathcal{E}$ and obtain state 
\be
\sigma_{\text{BC}}=\mathcal{E} \otimes \mathcal{I}_{\text{C}} (|\Psi\rangle\langle\Psi|_{\text{AC}})\ .
\ee 
Then, the fact that  $\mathcal{E}$ is non-asymmetry-generating, implies $\sigma_{\text{BC}}$ does not break the symmetry, i.e.
\be
\forall g\in G:\  \ [U_{\text{B}}(g)\otimes U_{\text{C}}(g)]\sigma_{\text{BC}} [U^\dag_{\text{B}}(g)\otimes U^\dag_{\text{C}}(g)]=\sigma_{\text{BC}}\  .
\ee 
This implies
\begin{align}
\forall g\in G:\ \ \mathcal{E} \otimes \mathcal{I}_{\text{C}} (|\Psi\rangle\langle\Psi|_{\text{AC}})&=[U_{\text{B}}(g)\otimes U_{\text{C}}(g)] \mathcal{E} \otimes \mathcal{I}_{\text{C}} (|\Psi\rangle\langle\Psi|_{\text{AC}})[U^\dag_{\text{B}}(g)\otimes U^\dag_{\text{C}}(g)]\\ &=[U_{\text{B}}(g)\otimes I_{\text{C}}] \mathcal{E} \otimes \mathcal{I}_{\text{C}} ([I_{\text{A}}\otimes U_{\text{C}}(g)] |\Psi\rangle\langle\Psi|_{\text{AC}} [I_{\text{A}}\otimes U^\dag_{\text{C}}(g)] )[U^\dag_{\text{B}}(g)\otimes I_{\text{C}}]\\ &=[U_{\text{B}}(g)\otimes I_{\text{C}}] \mathcal{E} \otimes \mathcal{I}_{\text{C}} ([U^\dag_{\text{A}}(g)\otimes I_{\text{C}}] |\Psi\rangle\langle\Psi|_{\text{AC}} [U_{\text{A}}(g)\otimes I_{\text{C}}] )[U^\dag_{\text{B}}(g)\otimes I_{\text{C}}] \\ &= \mathcal{E}_g \otimes \mathcal{I}_{\text{C}} (|\Psi\rangle\langle\Psi|_{\text{AC}})
\end{align}
where we have defined the map $\mathcal{E}_g$ to be the rotated the version of $\mathcal{E}$, such that
\be
\mathcal{E}_g(\rho)=U_{\text{B}}(g)\mathcal{E}(U^\dag_{\text{A}}(g)\rho U_{\text{A}}(g) ) U^\dag_{\text{B}}(g)
\ee
for all $\rho$.  
But, because $|\Psi\rangle_{\text{AC}}$ is a maximally entangled state, this implies that 
\be
\forall g\in G:\ \mathcal{E}_g=\mathcal{E} .
\ee
which implies $\mathcal{E}$ is a covariant map.

 \end{proof}

\newpage

\section*{{Supplementary Note 2: Purity of Coherence}} \label{Sec:purity}

In this paper we introduce a new measure of asymmetry, which we call it \emph{Purity of Coherence}.  The purity of coherence, with respect to the eigenbasis of an observable $H$, is defined by 
\begin{align}
P_H(\rho)&\equiv\Tr(H \rho^2 H \rho^{-1})- \Tr(\rho H^2)\
\end{align}
if $ \text{sup}(H \rho H) \subseteq \text{sup}(\rho)$, and $P_H(\rho)=\infty$ otherwise (Note that $\text{sup}(H \rho H)=\text{sup}(H \rho^2 H)$).  

Equivalently, 
\be
P_H(\rho)=f_H(\rho)-f_H(\mathcal{D}(\rho))\ ,
\ee
where for any state $\sigma$
\be
f_H(\sigma)\equiv\Tr(H\sigma^2 H \sigma^{-1})\ ,
\ee
and
\be
\mathcal{D}(\rho)=\sum_n P_n \rho P_n=\lim_{T\rightarrow \infty}\frac{1}{T} \int^{T}_0 dt\  e^{-i H t} \rho  e^{i H t}\ ,
\ee
is the map that dephases $\rho$ in the eigen-basis of $H=\sum_n E_n P_n$ (also known as the resource-destroying map \cite{liu2017resource}).

Using the spectral decomposition of state $\rho$, as $\rho=\sum_j p_j |\psi_j\rangle\langle\psi_j|$, we can rewrite  the formula for the purity of coherence as
\begin{align}
P_H(\rho) &=  \sum_{j,k} \frac{p^2_k-p^2_j} {p_j}\ |\langle\psi_k| H |\psi_j\rangle|^2\ .
\end{align}

\subsection*{Properties of purity of coherence}

The important properties of purity of coherence, such as monotonicity under TI operations and convexity,  follow from the properties of  the function 
\be
\overline{Q}_2(\rho\|\sigma)\equiv \Tr(\rho^2\sigma^{-1}) , 
\ee
if $ \text{sup}(\rho) \subseteq \text{sup}(\sigma)$, and $\overline{Q}_2(\rho\|\sigma)=\infty$ otherwise. As we will discuss later, the logarithm of this function is the Petz-R\'enyi  relative entropy for $\alpha=2$.

In particular, this function satisfies the following properties (See \cite{tomamichel2015quantum} for further discussions and proofs of these properties):
\begin{itemize}
\item Unitary invariance: It is invariant under any unitary transformation $U$, i.e. $\overline{Q}_2(U\rho U^\dag\|U\sigma U^\dag)=\overline{Q}_2(\rho\|\sigma)$.
\item Joint convexity: For any $0\le p \le 1$:
\be\label{app:joint}
p\ \overline{Q}_2(\rho_1\|\sigma_1)+ (1-p)\overline{Q}_2(\rho_2\|\sigma_2)\ge  \overline{Q}_2\bigg([p\rho_1+(1-p)\rho_2] \bigg\| [p\sigma_1+(1-p)\sigma_2] \bigg)
\ee
\item Information-processing inequality: For any completely positive trace-preserving map $\mathcal{E}$, 
\be\label{app:info}
\overline{Q}_2(\mathcal{E}(\rho)\|\mathcal{E}(\sigma))\le   \overline{Q}_2(\rho\|\sigma)\ .
\ee
This follows from the Stinespring dilation theorem, together with unitary invariance and joint convexity. 
\end{itemize}

Using these properties, it can be easily seen that for any time $t\in\mathbb{R}$, the function $B(\rho)\equiv \overline{Q}_2(\rho\| e^{-i t H}\rho e^{i t H})$ is monotone under any TI operation $\mathcal{E}_\text{TI}$:
\begin{align}\label{App:mono}
B(\rho)=\overline{Q}_2(\rho\| e^{-i t H}\rho e^{i t H}) &\ge \overline{Q}_2(\mathcal{E_\text{TI}}(\rho)\|\mathcal{E_\text{TI}}( e^{-i t H}\rho e^{i t H}))\\ &= \overline{Q}_2(\mathcal{E_\text{TI}}(\rho)\|e^{-i  t H}\mathcal{E_\text{TI}}(\rho) e^{i t H})\\ &= B(\mathcal{E_\text{TI}}(\rho))  \ ,
\end{align}
where the inequality follows from the information processing inequality for $\overline{Q}_2$, and the second line follows from the fact that $\mathcal{E}_\text{TI}$ is a TI operation.

The connection between this function and the purity of coherence follows from the fact that for small $\Delta t$,
\begin{align}\label{dhhh}
\overline{Q}_2(\rho\|e^{-i \Delta t H}\rho e^{i \Delta t H})&= \Tr(\rho^2(e^{-i \Delta t H}\rho e^{i \Delta t H})^{-1})\\ &= \Tr(\rho^2 e^{-i \Delta t H}\rho^{-1} e^{i \Delta t H} )\\ &={{\bigg[}}1+ \Delta t^2\Tr(\rho^2 H\rho^{-1} H)\nonumber-  \Delta t^2\Tr(\rho^2 H^2\rho^{-1})/2- \Delta t^2 \Tr(\rho^2 \rho^{-1} H^2)/2+\mathcal{O}(\Delta t^4) {{\bigg]}}\\ &= 1+ \Delta t^2 P_H(\rho)+\mathcal{O}(\Delta t^4) \ .
\end{align}
In other words, $P_H(\rho)$ is 2 time the second derivative of function $\overline{Q}_2(\rho\| e^{-i t H}\rho e^{i t H})$ with respect to the parameter $t$, at $t=0$. 

Then, it follows from the joint convexity of $\overline{Q}_2$ in Supplementary Eq.(\ref{app:joint}), and information processing inequality in Supplementary Eq.(\ref{app:info}), that function $P_H$ is
\begin{itemize}
\item Convex: For any $0\le p \le 1$, and any pair of states $\rho_1, \rho_2$ :
\be
p P_H(\rho_1)+ (1-p)P_H(\rho_2) \ge P_H([p \rho_1+(1-p)\rho_2])\ . 
\ee
\item Monotone: For any TI operation $\mathcal{E_\text{TI}}$
\be
P_H(\mathcal{E_\text{TI}}(\rho)) \le  P_H(\rho) . 
\ee

\end{itemize}

Furthermore, it turns out that the purity of coherence has the following useful properties:
\begin{itemize}

\item Additive: For a composite non-interacting system with the total Hamiltonian $H_\text{tot}=H_1\otimes  I_2+I_1\otimes  H_2$, the purity of coherence is additive for uncorrelated states, i.e. $F_{H_\text{tot}}(\rho_1\otimes \rho_2)=P_{H_1}(\rho_1)+P_{H_2}(\rho_2)$. This follows from the fact that $\overline{Q}_2(\rho_1\otimes \rho_2\|\sigma_1\otimes \sigma_2)=\overline{Q}_2(\rho_1\|\sigma_1)+\overline{Q}_2(\rho_2\|\sigma_2)$.   
\item Faithful:  It is non-negative, and is zero if, and only if, state is incoherent, i.e. diagonal  in the
 energy eigenbasis. 
 
 To see this consider the spectral decomposition of state $\rho$ as $\rho=\sum_j p_j |\psi_j\rangle\langle\psi_j|$. Then, we obtain the formula
\bes\label{hghg145}
\begin{align}
P_H(\rho) &=  \sum_{j,k} \frac{p^2_k-p^2_j} {p_j}\ |\langle\psi_k| H |\psi_j\rangle|^2\ \\ &=  \sum_{j,k} \frac{p_k-p_j} {p_j}\ (p_k+p_j) |\langle\psi_k| H |\psi_j\rangle|^2 ,
\\ &=  \sum_{j,k} \frac{(p_j-p_k)^2}{2p_j p_k} (p_j+p_k)  |\langle\psi_k| H |\psi_j\rangle|^2\ ,
\end{align}
\ees
where to get the third line we have used the fact that $(p_j+p_k)  |\langle\psi_k| H |\psi_j\rangle|^2$ is symmetric with respect to $k$
and $j$. The immediately implies that $P_H(\rho)\ge 0$. Furthermore, since all the terms in the summation are non-negative, the sum will be zero iff all the individual terms are zero. That is $(p_j-p_k)^2|\langle\psi_k| H |\psi_j\rangle|^2=0$ for all $j,k$, or equivalently  $(p_j-p_k)\langle\psi_k| H |\psi_j\rangle=0$. Multiplying both sides in $|\psi_k\rangle\langle\psi_j|$, and summing over $j$ and $k$ this implies $[\rho,H]=0$ and completes the proof.
\end{itemize}

\subsection*{Connection with Petz-R\'enyi  relative entropy}\label{Sec:connn}

Similar to QFI, function $ P_H(\rho)$ also determines how fast state $\rho$ becomes distinguishable from its time evolved version $e^{-i H t}\rho e^{i H t}$ and is closely related to   the Petz-R\'enyi  relative entropies. 
For $\alpha\in (0,1) \cup (1,2]$ The Petz-R\'enyi  relative entropy for  is defined as
\begin{align}
D_\alpha(\rho\|\sigma)&=\frac{1}{\alpha-1}\log \Tr(\rho^\alpha\sigma^{1-\alpha}), \ \ \ \  &&\text{supp}(\rho) \subseteq \text{supp}(\sigma), \text{or} \ \alpha\in(0,1)  \\ 
D_\alpha(\rho\|\sigma)&=\infty, \ \ \ \  &&\text{otherwise} \  \ 
\end{align}
Note that the in the special case of $\alpha=2$, we have $D_2(\rho\|\sigma)=\log \overline{Q}_2(\rho\|\sigma)$.

The relative R\'enyi  entropy can be interpreted as a measure of distinguishability of states. In particular, it is non-negative and $D_\alpha(\rho\|\sigma)$ is zero if and only if $\rho=\sigma$. Furthermore, it satisfies information processing inequality for $\alpha\in [0,2]/\{1\}$ \cite{petz1986quasi, tomamichel2015quantum}, that is for any CPTP map $\mathcal{E}$, it holds that  $D_\alpha(\rho\|\sigma) \ge  D_\alpha(\mathcal{E}(\rho)\|\mathcal{E}(\sigma)) \ .$

It can be easily seen that for small time $\Delta t$,
\begin{align}
D_\alpha(\rho\|e^{-i \Delta t H}\rho e^{i \Delta t H})&= \frac{1}{\alpha-1}\Delta t^2 [\Tr(\rho^\alpha H\rho^{1-\alpha} H)- \Tr(\rho H^2)]+\mathcal{O}(\Delta t^4)\ .
\end{align}
Then, using the arguments we used  in the case of the purity of coherence, we can see that all functions in the family 
\be
\Tr(\rho^\alpha H\rho^{1-\alpha} H)- \Tr(\rho H^2)\ ,\ \ \ \ \ \ \ \ \  1< \alpha \le 2
\ee
 satisfy all the essential properties of the purity of coherence, such as monotonicity under TI operations, additivity, and convexity. 
 
The reason that in this paper we focus on the case of $\alpha=2$, is that for higher values of $\alpha$, monotonicity under TI operations, which follows from the monotonicity of Petz-R\'enyi  relative entropy under data processing, does not hold. On the other hand, for lower values of $\alpha$, as a mixed state $\rho$ converges to a pure state,   function $\Tr(\rho^\alpha H\rho^{1-\alpha} H)- \Tr(\rho H^2)$, has slower divergence. Therefore, to see the unreachability of pure coherent states, $\alpha=2$ is the optimal choice.

\subsection*{Stochastic state conversions under TI operations (Proof of  Eq.5 in the paper)}

Suppose there exists a TI operation which transforms  $n$ copies of a system with state $\rho_1$ and Hamiltonian $H_1$ to a single copy of a system with state $\rho_2$ and Hamiltonian $H_2$,  with probability of success $p>0$. 

First, assume the transformation is deterministic, i.e. the probability of success $p$ is equal to one. Then, the monotonicity of the purity of coherence under TI operations implies  the  purity of coherence at the output is less than or equal to the purity of coherence at the input. Using the additivity of purity of coherence, we find the purity of coherence at the input is $n\times P_{H_1}(\rho_1)$. Therefore, we conclude
\begin{equation}\label{lastbound1}
n \ge  \frac{P_{H_2}(\rho_2)}{P_{H_1}(\rho_1)}\ .
\end{equation}
Next, consider the case where the transformation is stochastic, i.e $p\le 1$. This means that there exists a completely positive TI map $\tilde{\mathcal{E}}$, which is not necessarily trace-preserving, such that 
\be
\tilde{\mathcal{E}}(\rho_1^{\otimes n})=p  \rho_2\ .
\ee
Then,  define
\be
\mathcal{F}(\cdot)\equiv |0\rangle\langle 0|\otimes \tilde{\mathcal{E}}(\cdot)+[1-\Tr(\tilde{\mathcal{E}}(\cdot))] (|1\rangle\langle 1|\otimes  \sigma_\text{incoh}) \ ,
\ee
where $\sigma_\text{incoh}$ is an incoherent state (e.g. the totally mixed state), $\{|0\rangle,|1\rangle\}$ are orthonormal states of a \emph{register} qubit with Hamiltonian zero. In other words, the Hamiltonian of the output is $I\otimes H_2$, where $I$ is the identity operator on the register.

  It can be easily shown that map $\mathcal{F}$   is (i)  trace-preserving and completely positive, (ii) it is a TI operation, and (iii) implements the transformation
\be
\mathcal{F}(\rho_1^{\otimes n})=p |0\rangle\langle 0|\otimes \rho_2 +(1-p) (|1\rangle\langle 1|\otimes  \sigma_\text{incoh})\ .
\ee
Using definition $P_H(\rho)\equiv\Tr(H \rho^2 H \rho^{-1})- \Tr(\rho H^2)$, it can be easily shown that the purity of coherence of this output state is
\be
p P_{H_2}(\rho_2)+ (1-p)  P_{H_2}(\sigma_\text{incoh})= p\times  P_{H_2}(\rho_2)\ ,
\ee
Since $\mathcal{F}$ is a TI operation, the purity of coherence of its output is less than or equal to the purity of coherence of its input $\rho^{\otimes n}_1$, which is equal to $n\times  P_{H_1}(\rho_1) $. We conclude that 
\begin{equation}
n \ge p \times \frac{P_{H_2}(\rho_2)}{P_{H_1}(\rho_1)}\ .
\end{equation}
Thus, to generate a single copy of a pure state $\rho_2$ which contains coherence, we need $n=\infty$ or  $P_{H_1}(\rho_1)=\infty$.

\subsection*{States with infinite purity of coherence}

The definition of the purity of coherence immediately implies that for any pure state which is not an eigenstate of Hamiltonian the purity of coherence is infinite. As we saw above, this unboundedness, reflects the fact that given any finite copies of a generic mixed state (with full-rank density operator) it is impossible to create a single copy such pure states using TI operations. More generally, \\

\refstepcounter{Lem}\label{lemmafull}
\noindent\emph{Lemma 1}
For a bounded Hamiltonian  $H$ (i.e. $\|H\|_\infty <\infty$) the purity of coherence $P_H(\rho) < \infty$ if and only if $[\Pi_\rho,H]=0$, where 
$\Pi_\rho$ is the projector to the support of $\rho$.  In particular, the purity of coherence is bounded for states with full rank.\\
\begin{proof}
For a bounded Hamiltonian $H$,  $P_H(\rho) =\infty$ if, and only if, the operator $H\rho^2 H$ has support outside the support of $\rho$.  The support of operator $H\rho^2 H$, is equal to the support of operator $H\Pi_\rho H$. Therefore, $P_H(\rho) <\infty$ if, and only if $Q_\rho H\Pi_\rho H Q_\rho= 0$, where  $Q_\rho=I-\Pi_\rho$ is the projector to the kernel of $\rho$. The last equality holds only iff $Q_\rho H\Pi_\rho= 0$, which means $[H,\Pi_\rho]=0$. We conclude that $P_H(\rho)=\infty$, if and only if $[H,\Pi_\rho]\neq 0$.

\end{proof}
The following proposition follows immediately from this lemma together with the monotonicity of the purity of coherence under TI operations.\\

\noindent\textbf{Remark.} Let $H_\text{in}$ and $H_\text{out}$ be the Hamiltonians of the input and output systems, respectively.  
Suppose under a TI operation the input state $\rho$ is transformed to the output state $\sigma$. Let $\Pi_\rho$ and $\Pi_\sigma$ be the projectors to the supports of $\rho$ and $\sigma$, respectively. If $[\Pi_\rho, H_\text{in}]=0$ then $[\Pi_\sigma, H_\text{out}]=0$. \\

In the following, we present an interpretation and a different proof of this result in terms of the notion of unambiguous state discrimination \cite{rudolph2003unambiguous}, which clarifies the physical relevance of  condition $[H,\Pi_\rho]=0$.

Recall that two density operators can be unambiguously discriminated with a non-zero probability iff their supports are not identical \cite{rudolph2003unambiguous}. The support of state $e^{-i H t} \rho e^{i H t}$ is 
$e^{-i H t} \Pi_\rho e^{i H t}$, which is equal to $\Pi_\rho$ for all $t\in\mathbb{R}$, if and only if $[H,\Pi_\rho]=0$. 
Therefore, we conclude that there exists  $t\in\mathbb{R}$ such that two states  $e^{-i H t} \rho e^{i H t}$ and $\rho$ can be unambiguously discriminated  with a non-zero probability,  iff  $[H,\Pi_\rho]\neq 0$. 

Next, we note that if the probability of unambiguous discrimination of two states $\rho_1$ and $\rho_2$ is zero, then this probability remains zero under any  completely positive trace preserving map $\mathcal{E}$, i.e.  two states $\sigma_1=\mathcal{E}(\rho_1)$ and $\sigma_2=\mathcal{E}(\rho_2)$ will also have the same support.
This immediately implies that if the probability of unambiguous  discrimination of $\rho$ and $e^{-i H t} \rho e^{i H t}$  is zero, then for any TI operation $\mathcal{E}_\text{TI}$, the probability of unambiguous discrimination of  the two states $\sigma=\mathcal{E}_\text{TI}(\rho)$ and 
 \be
 \mathcal{E}_\text{TI}(e^{-i H t} \rho e^{i H t})=e^{-i H t} \mathcal{E}_\text{TI}(\rho) e^{i H t}=e^{-i H t} \sigma e^{i H t}
 \ee
  should also be zero. We conclude that if $[\Pi_\rho,H]=0$, and $\sigma=\mathcal{E}_\text{TI}(\rho)$ for a TI operation $\mathcal{E}_\text{TI}$, then $[\Pi_\sigma,H]=0$.

\subsection*{Purity of coherence is lower-bounded by Quantum Fisher Information}
In this section we show that the purity of coherence is lower-bounded by the Quantum Fisher Information (QFI). That is for any state $\rho$, and Hamiltonian $H$, $P_H(\rho)\ge F_H(\rho)$ .

For state $\rho$ with the spectral decomposition $\rho=\sum_j p_j |\psi_j\rangle\langle\psi_j|$, the QFI is given by
\be\label{QFI183}
F_H(\rho)= 2 \sum_{k,l} \frac{(p_k-p_l)^2 } {p_k+p_l} |\langle\psi_k| H |\psi_l\rangle|^2\ ,
\ee
Recall form Supplementary Eq.\ref{hghg145} that the purity of coherence can be rewritten as 
\begin{align}
P_H(\rho)&= \sum_{j,k} \frac{p^2_k-p^2_j} {p_j}\ |\langle\psi_k| H |\psi_j\rangle|^2=\sum_{k,l}  (p_k-p_l)^2\times \frac{p_k+p_l } {2p_l p_k} |\langle\psi_k| H |\psi_l\rangle|^2\ .
\end{align}
Using the arithmetic-geometric mean inequality applied to $p_k$ and $p_l$, we have
\be
\frac{p_k+p_l}{2p_l p_k} \ge \frac{2}{p_k+p_l}\ .
\ee
This implies that 
\be
P_H(\rho)=\sum_{k,l}  (p_k-p_l)^2 \frac{(p_k+p_l) } {2p_l p_k} |\langle\psi_k| H |\psi_l\rangle|^2\ge \sum_{k,l}  (p_k-p_l)^2 \frac{2}{p_k+p_l} |\langle\psi_k| H |\psi_l\rangle|^2  = F_H(\rho)\ , 
\ee
where in the last step we have used Supplementary Eq.\ref{QFI183} for QFI. Therefore, the inequality $P_H(\rho)\ge F_H(\rho)$ basically follows from arithmetic-geometric mean inequality \footnote{\color{black} I thank an anonymous referee for pointing this out, which simplified the proof. \color{black}}. 

\color{black}
Note that the arithmetic-geometric mean inequality holds as equality only if $p_k=p_l$. This means that, if 
$(p_k-p_l)\langle\psi_k| H |\psi_l\rangle$ is non-zero for some $l, k$, then $P_H(\rho)> F_H(\rho)$. Equivalently, this means that $P_H(\rho)=F_H(\rho)$, only if
\be
(p_k-p_l)\langle\psi_k| H |\psi_l\rangle=0\ : \forall k,l\ .
\ee
Multiplying both sides in $|\psi_k\rangle\langle \psi_l|$ and summing over $l, k $, this equation is equivalent to 
\be
\sum_k p_k |\psi_k\rangle\langle\psi_k| H \sum_l |\psi_l\rangle\langle\psi_l|=\sum_k  |\psi_k\rangle\langle\psi_k| H \sum_l p_l |\psi_l\rangle\langle\psi_l| \ ,
\ee
which is equivalent to $\rho H=H \rho$. I.e. the equality holds iff  $\rho$ is incoherent, in which case both quantities $F_H(\rho)$ and $P_H(\rho)$ are zero.  

\color{black}

\subsection*{Purity of coherence for Qubits}

Consider a general qubit state $\rho$ with the spectral decomposition 
\be
\rho=p |\psi\rangle\langle \psi| + (1-p)  |\psi^\perp\rangle\langle \psi^\perp|
\ee 
where $\langle \psi^\perp|\psi\rangle=0$. 
Then,
\be
\rho^{-1}=\frac{1}{p} |\psi\rangle\langle \psi| + \frac{1}{1-p}  |\psi^\perp\rangle\langle \psi^\perp|\ ,
\ee 
and
\be
\rho^2=p^2 |\psi\rangle\langle \psi| + (1-p)^2  |\psi^\perp\rangle\langle \psi^\perp| .
\ee 
This implies 
\begin{align}
\Tr(\rho^2 H \rho^{-1} H)&=p   \Tr(H |\psi\rangle\langle \psi| H |\psi\rangle\langle \psi|)+  \frac{p^2}{1-p}  \Tr(H |\psi\rangle\langle \psi| H|\psi^\perp\rangle\langle\psi^\perp| )\\ &+ \frac{(1-p)^2}{p} \Tr(H |\psi^\perp\rangle\langle \psi^\perp| H  |\psi\rangle\langle \psi| )+  (1-p) \Tr(H |\psi^\perp\rangle\langle \psi^\perp| H  |\psi^\perp\rangle\langle \psi^\perp| )\ .
\end{align}
Using the fact that 
\be
\Tr(H |\psi\rangle\langle \psi| H|\psi^\perp\rangle\langle\psi^\perp| )=V_H(\psi)=V_H(\psi^\perp)\ ,
\ee
we find
\begin{align}
\Tr(\rho^2 H \rho^{-1} H)=   \Big(\frac{p^2}{1-p}+  \frac{(1-p)^2}{p}\Big) \times V(\psi) +p   |\langle \psi| H  |\psi\rangle|^2 +  (1-p) |\langle \psi^\perp| H  |\psi^\perp\rangle|^2 \ .
\end{align}
Then, we find
\begin{align}
P_H(\rho)=\Tr(\rho^2 H \rho^{-1} H)-\Tr(\rho H^2)&=  \Big(\frac{p^2}{1-p}+  \frac{(1-p)^2}{p}-1\Big) \times V(\psi)\\ &= \frac{(1-2p)^2}{p(1-p)} \times V(\psi) \label{app:puity-qubit}\ .
\end{align}
Next, using the formula  for Quantum Fisher information for the family of states $e^{-i Ht} \rho e^{i Ht}$ with parameter $t$,
\be
F_H(\rho)=2\sum_{i,j}   \frac{(p_i-p_j)^2}{p_i+p_j} |\langle\psi_i|H_S|\psi_j\rangle|^2\ .
\ee
where $\sum_i p_i  |\psi_i\rangle\langle\psi_i|$ is the spectral decomposition of $\rho$. Applying this to $\rho=p |\psi\rangle\langle \psi| + (1-p)  |\psi^\perp\rangle\langle \psi^\perp|$, we find
\begin{align}
F_H(\rho)=4 (1-2p)^2\times V_H(\psi) .
\end{align}
Therefore,
\be
P_H(\rho)=\frac{F_H(\rho)}{4p(1-p)}\ .
\ee
Finally, note that
\be
1-\Tr(\rho^2)=1-[p^2 + (1-p)^2]=2p-2 p^2=2p (1-p) .
\ee 
Therefore,
\be
P_H(\rho)=\frac{F_H(\rho)}{2 [1-\Tr(\rho^2)]}\ .
\ee

\subsection*{Purity of coherence for states close to the maximally mixed state}

Any general state $\rho$ whose trace distance from the totally mixed state is $\|\rho-I/d\|_1=\epsilon\ge 0$ can be written as  
\be
\rho=\frac{I}{d}+ \epsilon A\ ,
\ee
where $A$ is a Hermitian operator with $\Tr(A)=0$ and $\|A\|_1=1$. 

 In the following we calculate  $P_H(\rho)$ and $F_H(\rho)$ in the limit of $\epsilon\ll 1$.
 
First, recall that 
\begin{align}
P_H(\rho)= Tr(H \rho^2 H \rho^{-1})- \Tr(\rho H^2) \ .
\end{align}
For $\rho=\frac{I}{d}+ \epsilon A$ we find
\begin{align}
\Tr(\rho H^2)=\frac{1}{d} \Tr(H^2)+\epsilon \Tr(H^2 A)
\end{align}
Then,
\begin{align}
\Tr(H \rho^2 H \rho^{-1})&=  \Tr(H [\frac{I}{d^2}+\frac{2}{d}\epsilon A+\epsilon^2 A^2 ] H [\frac{I}{d}+ \epsilon A]^{-1} ) \\ &=d \Tr(H [\frac{I}{d^2}+\frac{2}{d}\epsilon A+\epsilon^2 A^2 ] H \frac{1}{I+ \epsilon d A} )
\\ &=d \Tr(H [\frac{I}{d^2}+\frac{2}{d}\epsilon A+\epsilon^2 A^2 ] H [I- \epsilon d A+(\epsilon d A)^2+\mathcal{O}(\epsilon^3)] )\\ &=\frac{1}{d}\Tr(H^2)+d\Tr(H [\frac{2}{d}\epsilon A] H I)+d\Tr(H [\frac{I}{d^2}] H [- \epsilon d A] )\\ 
&\ \ \ + d \Tr(H [\frac{2}{d}\epsilon A ] H [- \epsilon d A] ) +d  \Tr(H [\epsilon^2 A^2 ] H I )+ d \Tr(H [\frac{I}{d^2}] H [(\epsilon d A)^2])+\mathcal{O}(\epsilon^3)\\ &=\frac{1}{d}\Tr(H^2)+\epsilon \Tr(H^2 A)+ d 2 \epsilon^2 [ \Tr(H^2 A^2)-\Tr(H A H A)] +\mathcal{O}(\epsilon^3) \ .
 \end{align}
Therefore
\begin{align}
P_H(\rho)= Tr(H \rho^2 H \rho^{-1})- \Tr(\rho H^2) = \epsilon^2 2 d [\Tr(H^2 A^2)-2\Tr(H A H A)] +\mathcal{O}(\epsilon^3) .
\end{align}
Next, we calculate Quantum Fisher Information for this state. Recall the formula
\be
F_H(\rho)=2\sum_{i,j}   \frac{(p_i-p_j)^2}{p_i+p_j} |\langle\psi_i|H_S|\psi_j\rangle|^2\ .
\ee
Let $A=\sum_i a_i |i\rangle\langle i|$ be the spectral decomposition of $A$. Then, 
\be
\rho=\frac{I}{d}+ \epsilon A=\sum_i (\epsilon a_i+\frac{1}{d}) |i\rangle\langle i|  \ .
\ee
This implies
\be
F_H(\rho)=2\sum_{i,j}   \frac{\epsilon^2 (a_i-a_j)^2}{2/d+\epsilon(a_i+a_j)} |\langle i|H|j\rangle|^2\ .
\ee
Expanding this we find 
\begin{align}
F_H(\rho)&=d\epsilon^2 \sum_{i,j}   \frac{ (a_i-a_j)^2}{1+\epsilon d (a_i+a_j)/2} |\langle i|H|j\rangle|^2\ \\ &=
d\epsilon^2 \sum_{i,j}    (a_i-a_j)^2 |\langle i|H|j\rangle|^2+\mathcal{O}(\epsilon^3)\\ &=
d\epsilon^2 \sum_{i,j}    (a^2_i+a^2_j-2a_i a_j) |\langle i|H|j\rangle|^2+\mathcal{O}(\epsilon^3)\\ &=
2 d \epsilon^2 \Tr(A^2 H^2) -2 d\epsilon^2 \Tr(H A H A)+\mathcal{O}(\epsilon^3)\\ &= 2 d \epsilon^2[\Tr(A^2 H^2) -\Tr(H A H A)]+\mathcal{O}(\epsilon^3)\ .
\end{align}
Comparing this with $P_H(\rho)=  \epsilon^2 2 d [\Tr(H^2 A^2)-2\Tr(H A H A)] +\mathcal{O}(\epsilon^3)$ we find  
\begin{align}\label{dcla}
\frac{P_H(\rho)}{F_H(\rho)}&=1+\mathcal{O}(\epsilon)\ .
\end{align}

\color{black}
Finally, recall that for any pair of states $\sigma_1$ and $\sigma_2$, it holds that 
\be
\|\sigma_1-\sigma_2\|_1 \le 2\sqrt{1- \text{Fid}(\sigma_1,\sigma_2)} \ ,
\ee
where $\text{Fid}(\sigma_1,\sigma_2)=\|\sqrt{\sigma_1} \sqrt{\sigma_2}\|^2_1$, and $1- \text{Fid}(\sigma_1,\sigma_2)$ is called the infidelity of $\sigma_1$ and $\sigma_2$.

Combining this with Supplementary Eq.\ref{dcla}, we conclude that if the infidelity of state $\rho$ and the maximally mixed state $I/d$ is  $\delta$,  then  
\begin{align}
\frac{P_H(\rho)}{F_H(\rho)}&=1+\mathcal{O}(\sqrt{\delta})\ .
\end{align}

\color{black}

\newpage

\section*{Supplementary Note 3: Purity of coherence for a mixed state close to a pure state}

In this section we find a useful bound on the purity of coherence for mixed states which are close to a pure state.  This bound will be used later to study coherence distillation. \\

\refstepcounter{Lem}\label{lem:app}
\noindent\emph{Lemma 2} Let $p_\text{max}=\|\sigma\|_\infty$ be the largest eigenvalue of $\sigma$ and $|\Phi\rangle$ be the corresponding eigenvector. Then,
\begin{align}
P_H(\sigma)\ge    \text{V}_H(\Phi)\times (\frac{p_\text{max}^2}{1-p_\text{max}}-1)   \label{ee2}  \ .
\end{align}\\

\begin{proof}

Recall that for any state $\rho$ with the spectral decomposition $\rho=\sum_j q_j |\phi_j\rangle\langle\phi_j|$, purity of coherence is given by 
\begin{align}\label{purity12}
P_H(\rho) &=  \sum_{j,k} \frac{q^2_k-q^2_j} {q_j}\ |\langle\phi_k| H |\phi_j\rangle|^2\ .
\end{align}
Consider the spectral decomposition of  $\sigma$, i.e.
\be
\sigma=p_\text{max} |\Phi\rangle\langle\Phi|+\sum_j p_j |\psi^\perp_j\rangle\langle \psi^\perp_j|\ ,
\ee
where $p_\text{max}=\|\sigma\|_\infty$ is the largest eigenvalue of $\sigma$, $|\Phi\rangle$ is the corresponding eigenvector, and $\{p_j\}_j$ are the rest of the eigenvalues, and  $\{|\psi^\perp_j\rangle\}$ are the corresponding eigenvectors. Then, using the general formula for purity of coherence, in Supplementary Eq.(\ref{purity12}), and using the fact that for any pair of $j$ and $k$ the sum of two terms 
\be\label{rte2}
(\frac{p^2_k-p^2_j} {p_j}+\frac{p^2_j-p^2_k} {p_k}) |\langle\psi^\perp_k| H |\psi^\perp_j\rangle|^2= \frac{(p_j-p_k)^2}{2p_j p_k}(p_j+p_k)|\langle\psi^\perp_k| H |\psi^\perp_j\rangle|^2\ge 0\ ,
\ee
is non-negative, we find 
\begin{align}
P_H(\sigma) &\ge  \sum_{j} ( \frac{p^2_\text{max}-p^2_j} {p_j}+\frac{p^2_j-p^2_\text{max}} {p_\text{max}})    \ |\langle\Phi| H |\psi^\perp_j\rangle|^2\ ,
\end{align}
where in the summation we have dropped all the terms which do not involve $|\Phi\rangle$.
Then, we find
\begin{align}
P_H(\sigma) &\ge  \sum_{j} ( \frac{p^2_\text{max}-p^2_j} {p_j}+\frac{p^2_j-p^2_\text{max}} {p_\text{max}})    \ |\langle\Phi| H |\psi^\perp_j\rangle|^2\ \\&\ge  \sum_{j} ( \frac{p^2_\text{max}} {p_j} -  [p_\text{max}+p_j])  \ |\langle\Phi| H |\psi^\perp_j\rangle|^2 \\&\ge  \sum_{j} ( \frac{p^2_\text{max}} {p_j} -1)  \ |\langle\Phi| H |\psi^\perp_j\rangle|^2  \\&\ge   ( \frac{p^2_\text{max}} {1-p_\text{max}} -1)  \ \sum_{j} |\langle\Phi| H |\psi^\perp_j\rangle|^2  \\&=   ( \frac{p^2_\text{max}} {1-p_\text{max}} -1) V_H(\Phi) \   ,
\end{align}
where  to get the third inequality we have used the fact that $p_\text{max}+p_j\le 1 $, to get the fourth inequality we have used the fact that $p_j \le 1-p_\text{max}$, and to get the last equality we have used  
\begin{align}
\sum_{j} |\langle\Phi| H |\psi^\perp_j\rangle|^2&=\langle\Phi| H  (\sum_j |\psi^\perp_j\rangle\langle\psi^\perp_j|+|\Phi\rangle\langle\Phi|) H|\Phi\rangle-\langle\Phi| H  (|\Phi\rangle\langle\Phi|) H|\Phi\rangle\\ &=\langle\Phi| H^2|\Phi\rangle-\langle\Phi| H|\Phi\rangle^2=  V_H(\Phi)\ .
 \end{align}
This completes the proof of lemma.
\end{proof}
This lemma has the following corollary. \\
\color{black}

\refstepcounter{Cor}\label{cor-var}
\noindent\emph{Corollary 1} 
Let  $\delta\equiv 1-\langle\Psi|\sigma|\Psi\rangle $ be the infidelity of pure state $\Psi$ and state $\sigma$. Let $p_\text{max}$ be the largest eigenvalue of $\sigma$, and $\Phi$ be the corresponding eigenvector. Then, the fidelity of $\Psi$ and $\Phi$ is lower bounded by $|\langle\Psi|\Phi\rangle|^2\ge 1-2\delta$, and $p_\text{max}$  
satisfies  $p_\text{max}\ge 1-\delta$. Furthermore, 
\begin{align}
P_H(\sigma)&\ge V_H(\Phi)\times (\frac{(1-\delta)^2}{\delta}-1) \label{1ododo} , \\ 
F_H(\sigma)&\ge V_H(\Phi)\times  4(1-2\delta)^2\ .\label{2ododo}
\end{align}\\

Assuming the infideliy $\delta\le 1/2$, this means that
\begin{align}
P_H(\sigma)&\ge V_H(\Phi)\times (\frac{1}{4 \delta}-1) \ , \\
\end{align}

\begin{proof}
Let
\be
\sigma=p_\text{max} |\Phi\rangle\langle\Phi|+\sum_j p_j |\psi^\perp_j\rangle\langle \psi^\perp_j|\ ,
\ee
be the eigen-decomposition of $\sigma$, where $p_\text{max}$ is the largest eigenvalue and $|\Phi\rangle$ is the corresponding eigenvector. The fact that 
$\langle\Psi|\sigma|\Psi\rangle = 1-\delta\ $ implies that
\be
1-\delta =  \langle\Psi|\sigma|\Psi\rangle=p_\text{max} |\langle\Psi|\Phi\rangle|^2+\sum_j p_j |\langle\Psi|\psi^\perp_j\rangle|^2\ \ .
\ee
Since $ |\langle\Psi|\Phi\rangle|^2+\sum_j  |\langle\Psi|\psi^\perp_j\rangle|^2=1$, we can interpret this sum as the average of eigenvalues of $\sigma$, weighted by the probability distribution defined by $\{ |\langle\Psi|\Phi\rangle|^2,  |\langle\Psi|\psi^\perp_j\rangle|^2\}$. This average is less than or equal to the maximum eigenvalue, $p_\text{max}$, i.e.
\begin{align}
1-\delta &= p_\text{max} |\langle\Psi|\Phi\rangle|^2+\sum_j p_j |\langle\Psi|\psi^\perp_j\rangle|^2\le p_\text{max} \label{eqggq23}\ ,
\end{align}
as claimed in the statement of the corollary.

Next, note that
\be\label{dgergg}
\sum_j p_j |\langle\Psi|\psi^\perp_j\rangle|^2\le  \sum_j p_j \times \sum_k |\langle\Psi|\psi^\perp_k\rangle|^2=(1-p_\text{max})(1-|\langle\Psi|\Phi\rangle|^2)\ ,
\ee
where we have used the facts that $p_\text{max}+\sum_j p_j =\Tr(\sigma)=1$,  and the fact that 
$|\langle\Psi|\Phi\rangle|^2+\sum_j |\langle\Psi|\psi^\perp_j\rangle|^2=1$, because $\{|\Phi\rangle, |\psi^\perp_j\rangle\}$ is an orthonormal basis.

Therefore,
\begin{align}
1-\delta &= \langle\Psi|\sigma|\Psi\rangle\\ &=p_\text{max} |\langle\Psi|\Phi\rangle|^2+\sum_j p_j |\langle\Psi|\psi^\perp_j\rangle|^2\ \\ &\le  p_\text{max} |\langle\Psi|\Phi\rangle|^2 + (1-p_\text{max})(1-|\langle\Psi|\Phi\rangle|^2)\\  &\le  p_\text{max} |\langle\Psi|\Phi\rangle|^2 + \delta \times (1-|\langle\Psi|\Phi\rangle|^2) \ ,
\end{align}
where to get the third line we have used Supplementary Eq.(\ref{dgergg}) and to get the fourth line we have used E.(\ref{eqggq23}), which implies $\delta \ge 1-p_\text{max}$.

This implies 
\begin{align}
|\langle\Psi|\Phi\rangle|^2 \ge \frac{1-2\delta}{p_\text{max}-\delta} \ge 1-2\delta   \ ,
\end{align}
as claimed in the statement of the corollary.

Next, to prove the lower bound on the purity of coherence in Supplementary Eq.\ref{1ododo}, we use the lower bound in lemma \ref{lem:app}  together with Supplementary Eq. \ref{eqggq23}. This implies 
\bes\label{wfeee}
\begin{align}
P_H(\sigma)&\ge V_H(\Phi)\times [\frac{p^2_\text{max}}{1-p_\text{max}}-1]\\
&\ge V_H(\Phi)\times  [\frac{(1-\delta)^2}{\delta}-1]  \ .
\end{align}
\ees
This proves Supplementary Eq.\ref{1ododo} in the corollary \ref{cor-var}.

Next, we prove Supplementary Eq.\ref{2ododo}, i.e. the lower bound on QFI. Using the spectral decomposition  $\sigma=p_\text{max} |\Phi\rangle\langle\Phi|+\sum_j p_j |\psi^\perp_j\rangle\langle \psi^\perp_j|\ $, the  formula for QFI in Supplementary Eq.\ref{QFI183} can be rewritten as
\bes\label{marma}
\begin{align}
F_H(\sigma)&=4 \sum_{j} \frac{(p_\text{max}-p_j)^2 } {p_\text{max}+p_j}\ |\langle\Phi| H |\psi^\perp_j\rangle|^2+2 \sum_{j, k} \frac{(p_k-p_j)^2 } {p_k+p_j}\ |\langle\psi^\perp_k| H |\psi^\perp_j\rangle|^2\\ &\ge 4 \sum_{j} \frac{(p_\text{max}-p_j)^2 } {p_\text{max}+p_j}\ |\langle\Phi| H |\psi^\perp_j\rangle|^2\\ &\ge 4 \sum_{j}(p_\text{max}-p_j)^2\ |\langle\Phi| H |\psi^\perp_j\rangle|^2 \ ,
\end{align}
\ees
where to get the second line  we ignore all the terms which do not contain  $p_\text{max}$, and to get the third line we have used $p_j+p_\text{max}\le 1$. Also, using $p_j+p_\text{max}\le 1$, or equivalently, $p_j\le 1-p_\text{max}$, 
we have
\be
p_\text{max}-p_j\ge 2p_\text{max}-1\ge 2(1-\delta)-1=1-2\delta\ , 
\ee
where the second inequality follows from Supplementary Eq.\ref{eqggq23}. 

Combining this with Supplementary Eq.\ref{marma}, we find 
\begin{align}
F_H(\sigma)&\ge 4  \sum_{j}(p_\text{max}-p_j)^2\ |\langle\Phi| H |\psi^\perp_j\rangle|^2\\ &\ge 4(1-2\delta)^2 \sum_{j} |\langle\Phi| H |\psi^\perp_j\rangle|^2\\ &=4(1-2\delta)^2 \times V_H(\Phi)\ .
\end{align}

\end{proof}

\color{black}

\newpage

\section*{Supplementary Note 4: QFI and purity of coherence in the iid regime\\ (Proof of Eq.10 and  Eq.11 in the paper) }\label{AppD}

\refstepcounter{Thm}\label{kkeod}
\noindent\emph{Theorem 2} 
Consider $m$ non-interacting systems, each with Hamiltonian $H$, and with the total Hamiltonian $H_\text{tot}=\sum_{i=1}^m H^{(i)}$. Let  $\sigma_m$ be their joint state. Suppose the fidelity of this state with state $|\phi\rangle^{\otimes m}$ is $\langle\phi|^{\otimes m}\sigma_m|\phi\rangle^{\otimes m}=1-\epsilon_m$, where $|\phi\rangle$ is a pure state with positive energy variance, i.e. $V_H(\phi)> 0$. Then, for sufficiently large $m$, e.g.  $m \ge  70 \frac{|\langle\phi|H^3|\phi\rangle|^2}{V^3_H(\phi)}$  and  sufficiently small $\epsilon_m$, e.g. $\epsilon_m\le 10^{-3}$, the QFI of state $\sigma_m$ and its purity of coherence, relative to the total Hamiltonian $H_\text{tot}$ are lower bounded by
\bes
\begin{align}
F_{H_\text{tot}}(\sigma_m)&\ge 4\times c \times m\times F_H(\phi)\ , \\ 
P_{H_\text{tot}}(\sigma_m)&\ge  c \times m\times F_H(\phi) \times \frac{1}{\epsilon_m}  \ ,
\end{align}
\ees
where $c$ is a positive constant, e.g. $c= 10^{-2}$.\\

\begin{proof}
We use corollary \ref{cor-var}. According to this result, the fact that $\langle\phi|^{\otimes m}\sigma_m|\phi\rangle^{\otimes m}= 1-\epsilon_m$ implies that 
\be\label{dgkkkk1}
|\langle\phi^{\otimes m}|\Theta_m\rangle|^2\ge 1-2\epsilon_m\ ,
\ee
 where  $|\Theta_m\rangle$ is the  eigenvector of $\sigma_m$ with the largest eigenvalue. Furthermore, the corollary implies that
\begin{align}
P_{H_\text{tot}}(\sigma_m)&\ge V_{H_\text{tot}}(|\Theta_m\rangle)\times (\frac{(1-\epsilon_m)^2}{\epsilon_m}-1)   \ ,
\end{align}
Assuming $\epsilon_m\le 10^{-3}$,  this implies 
\begin{align}\label{sffk}
P_{H_\text{tot}}(\sigma_m)&\ge V_{H_\text{tot}}(|\Theta_m\rangle)\times \frac{1}{\epsilon_m}   \times (0.997)\ .
\end{align}
Next, we use the following lemma which is a lower bound on the energy variance of pure states which are close to an iid pure state (This lemma is proven in Sec.\ref{Sec:124}). \\

\refstepcounter{Lem}\label{lem91919}
\noindent\emph{Lemma 3} 
Consider $m$ copies of a system with Hamiltonian $H$ and pure state $\phi$ with positive energy variance, i.e. $V_H(\phi)>0$.  Assume  $m$ is sufficiently large, e.g. $m \ge  70 \frac{|\langle\phi|H^3|\phi\rangle|^2}{V^3_H(\phi)}$.  Consider state $|\Theta_m\rangle$ whose fidelity with state $|\phi\rangle^{\otimes m}$ satisfies $|\langle\Theta_m|\phi\rangle^{\otimes m}|^2\ge 1-2.5\times 10^{-3}$.  Then, the energy variance of  $|\Theta_m\rangle$ is lower bounded by
\begin{align}\label{lkjd}
V_{H_\text{tot}}(|\Theta_m\rangle)\ge C\times m \times V_H(\phi) \ ,
\end{align}
where $C$ is a positive constant, e.g. $C=0.05$. \\

To apply this lemma, we first note that for $\epsilon_m\le 10^{-3}$,  
\be
|\langle\phi^{\otimes m}|\Theta_m\rangle|^2\ge 1-2\epsilon_m\ge 1-2\times 10^{-3}\ge 1-2.5\times 10^{-3} \ ,
\ee
where we have used Supplementary Eq.\ref{dgkkkk1}. Therefore, if $m \ge  70 \frac{|\langle\phi|H^3|\phi\rangle|^2}{V^3_H(\phi)}$, then we can apply the lemma, which implies
\begin{align}
V_{H_\text{tot}}(|\Theta_m\rangle)\ge C\times m \times V_H(\phi) \ ,
\end{align}
where $C$ is a positive constant, e.g. $C=0.05$. Putting this into Supplementary Eq.\ref{sffk}, we find
\begin{align}
P_{H_\text{tot}}(\sigma_m)&\ge V_{H_\text{tot}}(|\Theta_m\rangle)\times \frac{1}{\epsilon_m}   \times 0.997\\ &\ge C\times m \times V_H(\phi)\times \frac{1}{\epsilon_m}   \times 0.997\\ &\ge \frac{C \times 0.997}{4}\times  \frac{m \  F_H(\phi)}{\epsilon_m} \\ &\ge c \times \frac{m \  F_H(\phi)}{\epsilon_m}  \ ,
\end{align}
where $c\ge 10^{-2}$. Here, to get the third line we have used $F_H(\phi)=4 V_H(\phi)$. 
This proves the lower bound  $P_{H_\text{tot}}(\sigma_m)\ge 10^{-2} \times \frac{m \  F_H(\phi)}{\epsilon_m}$, stated in  theorem \ref{kkeod}. The lower bound on QFI in this theorem can also be proven in a similar way by combining lemma \ref{lem91919} above and the lower bound on fidelity in  corollary \ref{cor-var}. In particular, using this corollary we have
\begin{align}
F_{H_\text{tot}}(\sigma_m)&\ge V_{H_\text{tot}}(|\Theta_m\rangle)\times 4(1-2\epsilon_m)^2   \ ,
\end{align}
which together with Supplementary Eq.\ref{lkjd} implies
\begin{align}
F_{H_\text{tot}}(\sigma_m)&\ge C\times m \times V_H(\phi)\times 4(1-2\epsilon_m)^2\\ &= {C}\times m \times F_H(\phi)\times (1-2\epsilon_m)^2 \\ &\ge 4 c \times m \times F_H(\phi) \ ,
\end{align}
where $c\ge 10^{-2}$  (Note that $\epsilon_m\le 10^{-3}$, and therefore $(1-2\epsilon_m)^2\ge 0.996$, and $C\ge  0.05$) . 
Therefore, to complete the proof of theorem \ref{kkeod}, we only need to prove lemma \ref{lem91919}.

\subsection*{Proof of lemma \ref{lem91919} (A lower bound on the energy variance of pure states which are close to an iid pure state)}\label{Sec:124}
Let
\be
H_\text{tot}=\sum_{i=1}^m H^{(i)}= \sum_{E\in \text{spec}(H_\text{tot})}\  E\ \Pi_E
\ee
 be the spectral decomposition of Hamiltonian $H_\text{tot}$, where $\text{spec}(H_\text{tot})$ is the set of eigenvalues, and $\Pi_E$ is the projector to the subspace with eigenvalue $E$. 
 Let 
 \begin{align}
p_{m}(E)&=\langle\phi|^{\otimes m}\Pi_E|\phi\rangle^{\otimes m}\ , \\
q_{m}(E)&=\langle\Theta_m|\Pi_E|\Theta_m\rangle\ , \
\end{align}
be, respectively, the energy distributions of states  $|\phi\rangle^{\otimes m}$ and $|\Theta_m\rangle$, relative to Hamiltonian $H_\text{tot}$. 

Note that $p_{m}$, i.e. the energy distribution for state $|\phi\rangle^{\otimes m}$, corresponds to the distribution of sum of $m$ independent and identically distributed random variables, each with non-zero variance $V_H(\phi)$. Therefore, the variance of the distribution $p_{m}$ is $m\times V_H(\phi)$. On the other hand, the variance of distribution $q_{m}$ is  equal to $V_{H_\text{tot}}(|\Theta_m\rangle)$. Next, we argue that the assumption 
$|\langle\Theta_m|\phi\rangle^{\otimes m}|^2\ge 1-2.5\times 10^{-3}$ in the statement of lemma implies an upper bound on the total variation distance of  $p_{m}$ and $q_{m}$, and then use this to find a lower bound on the variance of distribution $q_m$, or equivalently, a lower bound on $V_{H_\text{tot}}(|\Theta_m\rangle)$. 

 The total variation distance between $p_{m}$ and $q_m$ is upper bounded by
\begin{align}\label{fdfdfew}
d_\text{TV}(p_{m}, q_m)&\equiv \frac{1}{2}\ \sum_{E\in \text{spec}(H_\text{tot})}  |p_m(E)-q_m(E)|\\ &\le  \frac{1}{2}\ \Big\||\phi\rangle\langle\phi|^{\otimes m}-  |\Theta_m\rangle\langle \Theta_m| \Big\|_1\\ &=\sqrt{1-|\langle\Theta_m|\phi\rangle^{\otimes m}|^2}\label{wddwd}\  ,
\end{align}
where to get the second line we have used the monotonicity of $l_1-$norm under CPTP maps, and the fact that measurement in the energy basis is a CPTP map. The equality in the third line holds for any general pair of normalized pure states. This implies that if  $|\langle\Theta_m|\phi\rangle^{\otimes m}|^2\ge 1-(0.05)^2=1-2.5\times 10^{-3}$, then 
\be
d_\text{TV}(p_{m}, q_m)\le 0.05\ .
\ee
Next, we use the following lemma proven at the end of this section, using Berry-Ess\'een's theorem.\\

\refstepcounter{Lem}\label{lem:classic}
\noindent\emph{Lemma 4} 
Let $X_1,\cdots , X_M$ be $M$ independent  identically distributed random variables, each with variance $\sigma_X^2>0$ and   bounded third moment $\xi=\mathbb{E}(X^3)<\infty$. Let $Z_M=\frac{1}{\sqrt{M}}\sum_{i=1}^M X_i$. Suppose $M$ is sufficiently large, e.g. $M\ge  \frac{ 70\times  |\xi|^2}{ \sigma^6_X}$. Then, any random variable  $Y_M$ whose total variation distance from $Z_M$ is sufficiently small,  e.g. $d_\text{TV}(Z_M, Y_M)\le 0.05$, has variance $\sigma^2_{Y_M}$ which is lower bounded by
\be
\sigma^2_{Y_M} \ge C \times \sigma_X^2  ,
\ee
where $C$ is a positive constant, e.g. $C=0.05$. \\

To apply the lemma, we assume $X$ is the random variable which takes values in  the set of eigenvalues of $H$, with the distribution defined by the weight of state  $|\phi\rangle$ in 
the energy eigen-subspaces of $H$. This means that the distribution of the random variable $\tilde{Z}_m=\sum_{i=1}^m X_i= \sqrt{m} {Z}_m$ is given by
$p_m$ and its variance is $m\times V_H(\phi)$. 

Now suppose the energy distribution of state $|\Theta_m\rangle$ is described by the random variable $\tilde{Y}_m$, which has distribution $q_m$. This means that its variance  is equal to 
\be
\sigma_{\tilde{Y}_m}^2= V_{H_\text{tot}}(|\Theta_m\rangle) \ .
\ee
Define 
\be
Y_m\equiv \frac{\tilde{Y}_m}{\sqrt{m}}\ .
\ee
Lemma \ref{lem:classic} implies that if $d_\text{TV}(p_{m}, q_m)\le 0.05$, then 
\be
\sigma_{{Y}_m}^2 \ge C \sigma_X^2= C V_H(\phi)\ ,
\ee
which in turn implies
\be
 V_{H_\text{tot}}(|\Theta_m\rangle)= \sigma_{\tilde{Y}_m}^2=m\times \sigma_{{Y}_m}^2\ge C m \sigma_X^2= C m V_H(\phi)  \ .
\ee
This proves lemma  \ref{lem91919}. Therefore, to complete the proof of lemma \ref{lem91919} and theorem \ref{kkeod}, 
we only need to prove lemma \ref{lem:classic}.

\end{proof}

\subsection*{Proof of lemma \ref{lem:classic}}

Without loss of generality we assume the expectation of the random variable $X$ is zero, i.e. $\mathbb{E}(X)=0$ (Otherwise, we can always add a constant to the random variable and make its expectation zero). 

Let $p_{M}$ and $q_M$ be, respectively, the probability distributions associated to the random variables $Z_M=\frac{1}{\sqrt{M}}\sum_{i=1}^M X_i$ and  $Y_M$. To simplify the notation, we assume they have discrete supports. Suppose their total variation distance is $\delta$, i.e.
\be
\frac{1}{2}\sum_y |p_{M}(y)-q_M(y)|=\delta\ .
\ee
It follows that for any set $S$, 
\begin{align}
\sum_{y\in S }  q_M(y) &\ge \sum_{y\in S}  p_M(y) -\delta\ .
\end{align}
In particular, for both sets $S_+\equiv \{y: y> \sigma_X\} $ and   $S_-\equiv \{y: y< -\sigma_X\} $, we have
\bes\label{wgggg}
\begin{align}
\sum_{y\in S_{+} }  q_M(y) &\ge \sum_{y\in S_{+}}  p_M(y) -\delta\ , \\ \sum_{y\in S_{-} }  q_M(y) &\ge \sum_{y\in S_{-}}  p_M(y) -\delta\ .
\end{align}
\ees

Let 
\be
\mu_{Y_M}=\mathbb{E}(Y_M)= \sum_y q_M(y) y  \ ,
\ee
be the expectation of the random variable $Y_M$. Then, the variance of this random variable is 
\begin{align}
\sigma^2_{Y_M}&=\sum_{y}  q_M(y) (y- \mu_{Y_M})^2\\ &\ge \sum_{y\in S_+}  q_M(y) (y- \mu_{Y_M})^2 +  \sum_{y\in S_-}  q_M(y) (y- \mu_{Y_M})^2 \  ,
\end{align}
where the bound follows from the fact that all the terms in the summation are  non-negative. Given that the sets $S_+=\{y: y>  \sigma_X\}$ and $S_-=\{y: y<  -\sigma_X\}$ are separated by, at least, $2\times \sigma_X$,  we can easily see that for any value of $\mu_{Y_M}$, at least, one of the followings holds:
$$ \forall y\in S_+ :\  |y- \mu_{Y_M}|> \sigma_X , $$
or
$$ \forall y\in S_-:\   |y- \mu_{Y_M}|> \sigma_X  \ . $$
Therefore, 
\begin{align}
\sigma^2_{Y_M}&\ge \sum_{y\in S_+}  q_M(y) (y- \mu_{Y_M})^2 +  \sum_{y\in S_-}  q_M(y) (y- \mu_{Y_M})^2  \\ &\ge \sigma_X^2  \times \min\Big\{ \sum_{y\in S_+}  q_M(y)  \  ,  \sum_{y\in S_-}  q_M(y) \Big\}\ .
\end{align}
Combining this with Eqs.\ref{wgggg} we find
\begin{align}\label{lkjd2}
\sigma^2_{Y_M}\ge  \sigma_X^2   \times \min\Big\{ \sum_{y\in S_+}  p_M(y)  \  ,  \sum_{y\in S_-}  p_M(y) \Big\}-\sigma_X^2\times \delta\ .
\end{align}
Next, we find a lower bound on $\sum_{y\in S_+}  p_M(y)$, and $\sum_{y\in S_-}  p_M(y)$. Recall that $p_M$ is the distribution of the random variable $Z_M=\frac{1}{\sqrt{M}}\sum_{i=1}^M X_i$. For sufficiently large $M$, this distribution converges to the Gaussian distribution, and we can use Berry-Ess\'een's theorem to find lower bounds on the tails of this distribution. \\

\noindent \emph{Berry-Ess\'een's theorem} \cite{berry1941accuracy, durrett2019probability}: Let $X_1, \cdots X_M$ be independent and identically distributed random variables, with means $0$, i.e. $\mathbb{E}(X_i)=0$ and variance $\sigma_X^2=\mathbb{E}(X^2_i)>0$,  and finite third moment $\xi=\mathbb{E}(X^3_i)<\infty$.  Let ${W}_M=\frac{1}{\sigma_X \sqrt{M}} \sum_{i=1}^M X_i$. Then, for any $a\in \mathbb{R}$, the probability that $W_M< a$, denoted by $\textbf{P}(W_M< a)$, satisfies 
\be
\Big|\textbf{P}(W_M< a)-\textbf{P}(N< a)\Big| < c \frac{|\xi|}{\sigma_X^3 \sqrt{M}}\ ,
\ee
where $N$ is the random variable with the standard Normal distribution, i.e. with mean zero and variance one, $\textbf{P}(N< a)$ is the corresponding cumulative distribution,  and $c$ is an order one positive constant (E.g. we can choose $c=1/2$). \\

Recall that $p_M$ is the distribution associated to the random variable 
\be
Z_M= \frac{1}{ \sqrt{M}} \sum_{i=1}^M X_i= 
\sigma_X\times W_M\ ,
\ee
which means
\be
 \sum_{y\in S_-}  p_M(y)=\sum_{y< - \sigma_X}  p_M(y)= \textbf{P}({Z}_M< -\sigma_X)=\textbf{P}(W_M< -1)
\ee
and
\be
 \sum_{y\in S_+}  p_M(y)= \sum_{y>  \sigma_X}  p_M(y)= \textbf{P}({Z}_M>\sigma_X)=\textbf{P}(W_M>1)\ .
\ee
Then, applying  Berry-Ess\'een theorem, we find
\be
\Big| \sum_{y\in S_-}  p_M(y)-\textbf{P}(N<-1)\Big|=\Big|\Big(\sum_{y< - \sigma_X}  p_M(y)\Big) -\textbf{P}(N<-1)\Big| < c \frac{|\xi|}{\sigma_X^3 \sqrt{M}}\ ,
\ee
and
\be
\Big| \sum_{y\in S_+}  p_M(y)-\textbf{P}(N>1)\Big|=\Big|\Big(\sum_{y> \sigma_X}  p_M(y)\Big) -\textbf{P}(N>1)\Big| < c \frac{|\xi|}{\sigma^3 \sqrt{M}}\ .
\ee
Given that for the normal distribution $N$ with mean zero and variance one, $P(N>1)=P(N<-1)> 0.16$, we find
\bes\label{sggggg}
\begin{align}
\sum_{y\in S_-}  p_M(y)=\sum_{y< - \sigma_X}  p_M(y) &\ge 0.16 - c  \frac{|\xi|}{\sigma_X^3 \sqrt{M}}\ , \\ 
\sum_{y\in S_+}  p_M(y)=\sum_{y>  \sigma_X}  p_M(y) &\ge 0.16 - c  \frac{|\xi|}{\sigma_X^3 \sqrt{M}}\ .
\end{align}
\ees
Now  assume $M$ is  sufficiently large, e.g.
\be
M\ge \frac{ 70\times  |\xi|^2}{ \sigma^6_X}\ .
\ee
Since $c=1/2$, this guarantees that $c\times \frac{|\xi|}{\sigma_X^3 \sqrt{M}}\le 0.06$. Putting this in Supplementary Eq.(\ref{sggggg}), we find 
\begin{align}
\sum_{y\in S_-}  p_M(y)=\sum_{y< - \sigma_X}  p_M(y)\ge 0.1 \ \ \ \ \ , \ \text{and}\ \ \ \  \sum_{y\in S_+}  p_M(y)=\sum_{y>  \sigma_X}  p_M(y)  &\ge 0.1\ .
\end{align}
Putting this back into Supplementary Eq.(\ref{lkjd2}), we find
\begin{align}
\sigma^2_{Y_M}&\ge  \sigma_X^2  \times  \min\Big\{ \sum_{y\in S_+}  p_M(y)  \  ,  \sum_{y\in S_-}  p_M(y) \Big\} -\sigma^2_X \delta\\ &\ge \sigma^2_X \times 0.1\ -\sigma_X^2 \delta\ .
\end{align}
Assuming the total variation distance $\delta= \frac{1}{2}\sum_y |p_{M}(y)-q_M(y)|\le 0.05$, we find
\begin{align}
\sigma^2_{Y_M}&\ge 0.05 \times \sigma^2_X  .
\end{align}
This proves lemma \ref{lem:classic}, and completes the proof of  lemma \ref{lem91919} and theorem \ref{kkeod}.

\color{black}

 \newpage

\section*{Supplementary Note 5: Extension of the main theorem: Finite helper systems do not help}\label{AppW}

It turns out that our no-go theorem on coherence distillation can be extended  to the case where one is allowed to use a finite \emph{helper} system at the input to implement the transformation   
\be
\rho^{\otimes n}\otimes \chi 
\xrightarrow{TI}\stackrel{}{ \approx}  \psi^{\otimes  \lceil R n\rceil} ,
\ee
where $\chi $ is the state of the helper system, in a finite-dimensional Hilbert space with a bounded Hamiltonian. The  helper system can be in a pure state, in which case the purity of coherence of the input can be $\infty$, even for finite $n$. Therefore, in this case it is not clear that how we can put a restriction on the output based on the purity of coherence of the input. Nevertheless, we can overcome this issue, and prove an extension of our no-go theorem, which implies distillable coherence remains zero for states with bounded purity of coherence, even if one allows  a finite helper system at the input.  This result follows from the following lemma together with an argument similar to the one which proved our no-go theorem.\\

\refstepcounter{Lem}\label{lem163}
\noindent\emph{Lemma 5}
Suppose there exists a TI operation $\mathcal{E}_n$ which transforms $n$ copies of system with state $\rho$ and Hamiltonian $H$ and a helper system in state $\chi$ and Hamiltonian $H_\text{help}$, to $m$ copies of a system with Hamiltonian $H$ and state $\psi$ with error $\epsilon_n$ in trace distance, such that
\be
\Big\|\mathcal{E}_n(\rho^{\otimes n}\otimes \chi)-  \psi^{\otimes m} \Big\|_1\le \epsilon_n \ . 
\ee 
Then, there exists a pure state $|\Theta_n\rangle$ (namely the eigenstate of $\mathcal{E}_n(\rho^{\otimes n}\otimes \chi)$ with the largest eigenvalue) whose overlap with the desired state $\psi^{\otimes m}$ is 
\be
|\langle\Theta_n|\psi\rangle^{\otimes m}|^2\ge 1-4\epsilon_n ,
 \ee
 and satisfies
\begin{align}
\epsilon_n P_H(\rho)+ 2(d_\chi-1)\ \frac{1}{n}V_{H_\text{help}}(\chi)  \ge \frac{1}{n} V_{H_\text{tot}}(|\Theta_n\rangle) \times ({(1-\epsilon_n)^2}-\epsilon_n)\  ,
\end{align}
where $d_\chi$ is the dimension of the Hilbert space of the helper system, and $H_\text{tot}=\sum_{i=1}^{m} H^{(i)}$ is the sum of the Hamiltonians of the output systems.\\

Suppose in the limit $n$ goes to $\infty$, error $\epsilon_n$ goes to zero. If $ P_H(\rho)$ is bounded, then the left-hand side of the above bound vanishes, which implies 
\be\label{sffffk}
\lim_{n\rightarrow \infty} \frac{1}{n} V_{H_\text{tot}}(|\Theta_n\rangle)=0\ . 
\ee
Recall that $|\Theta_n\rangle$ is the eigenvector of the output state $\mathcal{E}_n(\rho^{\otimes n}\otimes \chi)$. But, applying lemma \ref{lem91919} we know that for sufficiently large $m$ and sufficiently small $\epsilon_m$,  
\be
 V_{H_\text{tot}}(|\Theta_n\rangle)\ge   C\times m \times V_H(\psi) \ ,
\ee
where $C$ is a positive constant, e.g. $C=0.05$.  Combining this with Supplementary Eq.\ref{sffffk}, we conclude that 
\be
V_H(\psi) \times \lim_{n\rightarrow \infty} \frac{m}{n}  =0\ . 
\ee
Therefore, assuming $P_H(\rho)$ is bounded and $V_H(\psi)>0$, then to have a vanishing error $\epsilon_n\rightarrow 0$, we also need to have a vanishing rate. In conclusion, the distillable coherence of states with finite purity of coherence remains zero, even at the presence of finite-dimensional helper systems.

\subsection*{Proof of lemma \ref{lem163}}
In general, at the presence of the helper state, the purity of coherence of the input can be   $\infty$  for a finite $n$, in which case we cannot put any constraint on the output based on its purity of coherence.  To rectify this issue we use the following trick, which can be used more generally when one deals with the purity of coherence for pure states:  assume instead of using the helper state in the pure state $\chi$, we use $\tau_\chi$, a noisy version of $\chi$  obtained by mixing 
$\chi$ with the totally mixed state, with a ratio such that the trace distance between $\chi$ and $\tau_\chi$ is exactly $\epsilon_n$. Now suppose in the process $\rho^{\otimes n} \otimes \chi\xrightarrow{\text{TI}} \psi^{\otimes m} $, we use  $\tau_\chi$ instead of $\chi$. Then, we introduce an additional error in the process. Using the fact that the trace distance satisfies the triangle inequality, and is non-increasing under CPTP maps,   this  additional error can be bounded by $\epsilon_n$. Therefore, the total error at the output will be bounded by $2\epsilon_n$. To summarize,  if 
\be
\Big\|\mathcal{E}_n(\rho^{\otimes n}\otimes\chi)-  \psi^{\otimes m} \Big\|_1\le \epsilon_n \ , 
\ee
then, 
\be
\Big\|\mathcal{E}_n(\rho^{\otimes n}\otimes\tau_\chi)-\psi^{\otimes m}\Big\|_1\le 2\epsilon_n \ . 
\ee
In this transformation, the purity of coherence for the input is $n P_H(\rho)+P_{H_\text{help}}(\tau_\chi)$. Later, we show that $P_{H_\text{help}}(\tau_\chi)$ is upper bounded by  
\be\label{bounf1}
P_{H_\text{help}}(\tau_\chi)\le \frac{2(d_\chi-1)}{\epsilon_n}\ V_{H_\text{help}}(\chi)\ ,
\ee
where $d_\chi$ is the dimension of the Hilbert space of $\chi$. Therefore, the total purity of coherence for the input $\rho^{\otimes n}\otimes\tau_\chi$ is upper bounded by  
\be
n P_H(\rho)+ \frac{2(d_\chi-1)}{\epsilon_n}\ V_{H_\text{help}}(\chi)\ .
\ee

Next, we focus on the purity of coherence of the output, and use corollary \ref{cor-var}, which provides a lower bound on the purity of coherence for mixed states close to pure states. Let $\sigma=\mathcal{E}_n(\rho^{\otimes n}\otimes \tau_\chi)$ be the actual output state. By assumption, $\|\psi^{\otimes m}-\sigma \|_1\le 2\epsilon_n$. 
Using the standard relation between the trace distance and fidelity \cite{wilde2013quantum}, $\|\psi^{\otimes m}-\sigma \|_1\le 2\epsilon_n$ implies that the infidelity $1-\langle\psi|^{\otimes m}\sigma|\psi\rangle^{\otimes m} \le \epsilon_n$.
Then, according to the corollary \ref{cor-var}, there exists a pure state $|\Theta_n\rangle$ (namely the eigenvector of $\sigma$ with the largest eigenvalue) which satisfies  both inequalities 
\be
|\langle\Theta_n|\psi\rangle^{\otimes m}|^2\ge 1-2\epsilon_n\ ,
\ee
and
\begin{align}
P_{H_\text{tot}}(\sigma)&\ge V_{H_\text{tot}}(|\Theta_n\rangle)\times (\frac{(1-\epsilon_n)^2}{\epsilon_n}-1)   \ ,
\end{align}
where $H_\text{tot}=\sum_{i=1}^{m} H^{(i)}$  is the sum of the Hamiltonians of the output systems. 

Therefore, using the monotonicity of the purity of coherence, we conclude
\begin{align}
V_{H_\text{tot}}(|\Theta_n\rangle)\times (\frac{(1-\epsilon_n)^2}{\epsilon_n}-1)  \le P_{H_\text{tot}}(\sigma)\le  n P_H(\rho)+ \frac{2(d_\chi-1)}{\epsilon_n}\ V_{H_\text{help}}(\chi) \ ,
\end{align}
or, equivalently, 
\begin{align}
\epsilon_n P_H(\rho)+ 2(d_\chi-1)\ \frac{1}{n}V_{H_\text{help}}(\chi)  \ge \frac{1}{n} V_{H_\text{tot}}(|\Theta_n\rangle) \times ((1-\epsilon_n)^2-\epsilon_n)\ .
\end{align}
To complete the proof, in the following we prove Supplementary Eq.(\ref{bounf1}): Let  $\tau_\chi$ be the state obtained by mixing the pure state $\chi$ and the totally mixed state $I/d_\chi$, such that the trace distance between $\tau_\chi$ and $\chi$ is $\epsilon$. Then, 
\be\label{mixedrr}
\tau_\chi=(1-\frac{\epsilon}{2})|\chi\rangle\langle \chi |+ \frac{\epsilon}{2(d_\chi-1)} (I-|\chi\rangle\langle \chi |) \ .
\ee
Recall that  for any Hamiltonian $H$ and state $\rho$ with spectral decomposition $\rho=\sum_j p_j |\psi_j\rangle\langle\psi_j|$, we have
$P_H(\rho)=  \sum_{k,l} \frac{p^2_k-p^2_l} {p_l} |\langle\psi_k| H |\psi_l\rangle|^2\ $. Therefore, for any $j$ and $k$ whose corresponding eigenvalues are equal, the corresponding term $\frac{p^2_k-p^2_l} {p_l} |\langle\psi_k| H |\psi_l\rangle|^2$ does not contribute in the summation. Using this  for state $\tau_\chi$ in Supplementary Eq.(\ref{mixedrr}), we find 
\begin{align}
P_{H_\text{help}}(\tau_\chi)&\le  \frac{ (1-\frac{\epsilon}{2})^2-(\frac{\epsilon}{2(d_\chi-1)})^2 }{\frac{\epsilon}{2(d_\chi-1)}} \sum_{l: \psi_l\neq \chi} |\langle\chi| H_{\text{help}} |\psi_l\rangle|^2  \\ &\le \frac{2(d_\chi-1)}{\epsilon} V_{H_\text{help}}(\chi)\ .
\end{align} 
This proves Supplementary Eq.(\ref{bounf1}) and completes the proof of the lemma.

\newpage

\newpage
\color{black}

\section*{Supplementary Note 6: Mixed states with distillable coherence}\label{AppF}

In this section we study examples of mixed states for which  the distillable coherence is non-zero.

First, we consider states in the form
\be
\rho=p|\psi\rangle\langle \psi|+(1-p) \sigma\ ,
\ee
where $0<p<1$. We assume the support of $\sigma$ is limited to a proper subspace of the Hilbert space, and the pure state $\psi$ does not belong to this subspace. 

Consider the subspace spanned by all the energy levels $\{|E_i\rangle:\ \langle E_i|\sigma |E_i\rangle>0\}$, i.e. energy levels in which $\sigma$ has a non-zero probability. Let $P_\sigma$ be the projector to this subspace, and  $P_\sigma^\perp=I-P_\sigma$ be the projector to the orthogonal subspace. This means that 
\be
[P_\sigma^\perp, H]=0,\ \ \ \text{and},\ \ \ \ P_\sigma^\perp \sigma= \sigma P_\sigma^\perp=0\ .
\ee

Then, the two-outcome projective measurement $\{P_\sigma^\perp, P_\sigma \}$ is a TI operation. By performing this TI operation on the input state $\rho$,  we project the system to the pure state 
\be
|\tilde{\psi}\rangle=\frac{P_\sigma^{\perp}|\psi\rangle}{\sqrt{\langle\psi|P_\sigma^{\perp}|\psi\rangle}} \ ,
\ee 
with probability 
\be
\Tr(\rho P^\perp_\sigma)= p\times \langle\psi|P_\sigma^{\perp}|\psi\rangle\ .
\ee
From results of \cite{schuch2004nonlocal, schuch2004quantum, gour2008resource, marvian2018coherence} we know that copies of state $|\tilde{\psi}\rangle$ can be transformed to copies of any other pure state $\phi_\text{coh}$, which has the same period, with the optimal rate $V_H(\tilde{\psi})/V_H(\phi_\text{coh})$, where $V_H(\tilde{\psi})= \langle\tilde{\psi}|H^2|\tilde{\psi}\rangle-\langle\tilde{\psi}|H|\tilde{\psi}\rangle^2  \ $, is the energy variance.

It follows that using this strategy, we obtain copies of state $\phi_\text{coh}$ from states $\rho$ with the rate
\begin{align}
R =p\times \langle\psi|P_\sigma^{\perp}|\psi\rangle \times \frac{V_H(\tilde{\psi})}{V_H(\phi_\text{coh})}\ .
\end{align}
In the special case where  $\sigma$ is incoherent, i.e. $[\sigma,H]=0$, it turns out that this rate can be written nicely in terms of the function
\be
Q_H(\rho)\equiv \Tr(H\rho H\Pi^\perp_\rho)\ ,
\ee
where $\Pi_\rho^\perp=I-\Pi_\rho$  is the projector to the kernel of $\rho$. As we show later, $Q_H(\rho)$ is closely related to Petz-R\'enyi  relative entropies.  Also, note that for any pure state $\phi$, this function is equal to energy variance, i.e. $Q_H(\phi)=V_H(\phi)$. In the following, we show that  if $\sigma$ is incoherent,  then for state $\rho=p|\psi\rangle\langle \psi|+(1-p) \sigma$,
\be\label{dvdv}
Q_H(\rho)=p\times \langle\psi|P_\sigma^{\perp}|\psi\rangle \times V_H(\tilde{\psi})\ ,
\ee
and therefore, the above rate $R$ can be rewritten as
\begin{align}
R &=p\times \langle\psi|P_\sigma^{\perp}|\psi\rangle \times \frac{V_H(\tilde{\psi})}{V_H(\phi_\text{coh})}  \\  &=\frac{Q_H(\rho)}{V_H(\phi_\text{coh})} \\ &=\frac{Q_H(\rho) }{Q_H(\phi_\text{coh})}  \ .
\end{align}
To see Supplementary Eq.\ref{dvdv},   note that if $\sigma$ is incoherent, then it can be diagonalized in the energy eigenbasis, as
\be
\sigma=\sum_{i\in S} q_i |E_i\rangle\langle E_i|\ ,
\ee
where the summation is over all energy eigenstates  with non-zero probability, i.e. $q_i>0$. Then, the support of $\rho=p|\psi\rangle\langle \psi|+(1-p) \sigma$ is the subspace spanned by $\{ |E_i\rangle: i\in S\} \cup \{|{\psi}\rangle\}$.  Since, 
\be
P_\sigma=\sum_{i\in S} |E_i\rangle\langle E_i|\ .
\ee
and $P^\perp_\sigma=I-P_\sigma$,  the subspace spanned by $\{ |E_i\rangle: i\in S\} \cup \{|{\psi}\rangle\}$ is equal to the subspace spanned by $\{ |E_i\rangle: i\in S\} \cup \{P^\perp_\sigma |{\psi}\rangle\}$. Therefore, the projector to the support of $\rho$ is
 \be
 \Pi_\rho=P_\sigma+|\tilde{\psi}\rangle\langle\tilde{\psi}|\ ,
 \ee
 and the projector to its kernel is
\be\label{wggg20}
 \Pi^\perp_\rho=I- \Pi_\rho= P^\perp_\sigma-|\tilde{\psi}\rangle\langle\tilde{\psi}| .
 \ee
Note that
 \be
 \Pi^\perp_\rho=P^\perp_\sigma \Pi^\perp_\rho P^\perp_\sigma\ .
 \ee
Using this formula we have
\begin{align}
Q_H(\rho)\equiv \Tr(H\rho H \Pi^\perp_\rho)&=\Tr\Big(H[p|\psi\rangle\langle \psi|+(1-p) \sigma] H [P^\perp_\sigma \Pi^\perp_\rho P^\perp_\sigma]\Big) \\ &=p\Tr\Big(H|\psi\rangle\langle \psi| H [P^\perp_\sigma \Pi^\perp_\rho P^\perp_\sigma]\Big) +(1-p)\Tr\Big(H^2\sigma [P^\perp_\sigma \Pi^\perp_\rho P^\perp_\sigma]\Big) \\ &=p\Tr\Big(H P^\perp_\sigma|\psi\rangle\langle \psi|P^\perp_\sigma H \Pi^\perp_\rho\Big) \\ &=p\times \langle\psi|P^\perp_\sigma|\psi\rangle  \times \Tr\Big(H |\tilde{\psi}\rangle\langle \tilde{\psi}| H \Pi^\perp_\rho\Big)   \\ &= p\times \langle\psi|P^\perp_\sigma|\psi\rangle  \times \Tr\Big(H |\tilde{\psi}\rangle\langle \tilde{\psi}| H [P^\perp_\sigma-|\tilde{\psi}\rangle\langle\tilde{\psi}|]\Big)   \\ &= p\times \langle\psi|P^\perp_\sigma|\psi\rangle  \times V_H(\tilde{\psi})\ ,  \end{align}
where to get the second line we have used the fact that $[H,\sigma]=0$, to get the third line we have used the fact that $P_\sigma^\perp\sigma=\sigma P_\sigma^\perp=0$, to get the fourth line we have used the definition  $|\tilde{\psi}\rangle= P^\perp_\sigma|{\psi}\rangle/\sqrt{\langle\psi|P^\perp_\sigma|{\psi}\rangle}$, to get the fifth line we have used Supplementary Eq.\ref{wggg20}, and to get the last line we have the facts that   $P^\perp_\sigma$ commutes with $H$, and $P^\perp_\sigma|\tilde{\psi}\rangle=|\tilde{\psi}\rangle$.  This proves Supplementary Eq.\ref{dvdv}.

It is worth noting that function $Q_H(\rho)$ can  be obtained from Petz-R\'enyi  relative entropies, in the same way we derived the purity of coherence: For $\alpha=0$, the  Petz-R\'enyi  relative entropy can be defined by taking the limit $\alpha\rightarrow 0$ of  
\begin{align}
D_\alpha(\rho\|\sigma)&=\frac{1}{\alpha-1}\log \Tr(\rho^\alpha\sigma^{1-\alpha})\ ,
\end{align}
which yields 
\begin{align}
D_0(\rho\|\sigma)&=-\log \Tr(\Pi_\rho\sigma)\ .
\end{align}
Choosing $\sigma=e^{-i H \Delta t} \rho e^{i H \Delta t} $, for small $\Delta t$, we find
\begin{align}
D_0(\rho\|\sigma)&=(\Delta t)^2\times  [\Tr(\rho H^2)-\Tr( H \rho H \Pi_\rho)]+\mathcal{O}(\Delta t^4)=(\Delta t)^2\times  Q_H(\rho)+\mathcal{O}(\Delta t^4) \ .
\end{align}
In other words,
\begin{align}
Q_H(\rho)= \frac{1}{2} \frac{d^2}{dt^2} D_0(\rho\| e^{-i H \Delta t} \rho e^{i H \Delta t})\Big|_{t=0}\ .
\end{align}
Then, following the same arguments we used in the case of the  purity of coherence, one can easily show that   $Q_H(\rho)$ is additive and monotone under TI operations. Furthermore, $Q_H(\rho)$ is zero, iff  $[\Pi_\rho, H]=0$, i.e. iff the purity of coherence of $\rho$ is bounded. To see this note that $Q_H(\rho)>0$, iff $H\rho H$ has support in the kernel of $\rho$. On the other hand, $P_H(\rho)=\Tr(H\rho^2 H\rho^{-1})-\Tr(H^2 \rho)=\infty$, iff $H\rho^2 H$ has support in the kernel of $\rho$. Because the support of $H\rho H$ is equal to the support of $H\rho^2 H$, then these two conditions are equivalent.

\subsection*{Generalization of the above example}

The previous example can be generalized extensively. Specifically, suppose the Hilbert space $\mathcal{H}$ of the system with Hamiltonian $H$ and state $\rho$ can be decomposed to
\be
\mathcal{H}=\bigoplus_\mu \mathcal{H}^{\text{pure}}_\mu\otimes \mathcal{H}^{\text{mixed}}_\mu \ ,
\ee
 such that \\
\indent (i) Hamiltonian $H$ is block-diagonal with respect to the subspaces $\{ \mathcal{H}^{\text{pure}}_\mu\otimes \mathcal{H}^{\text{mixed}}_\mu\}_\mu$\ , i.e. 
\be
\forall \mu: \ [H,\Pi_\mu]=0\ ,
\ee
where $\Pi_\mu$ is the projector to the subspace   $ \mathcal{H}^{\text{pure}}_\mu\otimes \mathcal{H}^{\text{mixed}}_\mu$.\\
\indent (ii) Furthermore, the operator $\Pi_\mu H \Pi_\mu$  does not introduce  interactions between subsystems $\mathcal{H}^{\text{pure}}_\mu$ and $\mathcal{H}^{\text{mixed}}_\mu$ (i.e. it can be written as the sum of two terms, each acting non-trivially on, at most, one of $\mathcal{H}^{\text{mixed}}_\mu$ and $\mathcal{H}^{\text{pure}}_\mu$).\\
\indent (iii) The reduced (unnormalized) state of $\mathcal{H}^{\text{pure}}_\mu$, defined by
\be
|\psi_\mu\rangle\langle\psi_\mu|=\Tr_{\mathcal{H}^{\text{mixed}}_\mu}(\Pi_\mu \rho \Pi_\mu)\ ,
\ee
 is a pure state, and the reduced (unnormalized) state of $\mathcal{H}^{\text{mixed}}_\mu$, defined by
\be
\tau_\mu=\Tr_{\mathcal{H}^{\text{pure}}_\mu}(\Pi_\mu \rho \Pi_\mu)\ ,
\ee
is a mixed state. Note that we can always choose $\mathcal{H}^{\text{pure}}_\mu$ or $\mathcal{H}^{\text{mixed}}_\mu$ to be one-dimensional subsystems. For one-dimensional subsystems, the unnormalized reduced state is both pure and mixed.
 
Suppose for a subspace corresponding to label $\mu^\ast$, the reduced state of the subsystems $\mathcal{H}^{\text{pure}}_{\mu^\ast}$, i.e. state  $|\psi_{\mu^\ast}\rangle\langle\psi_{\mu^\ast}|$, contains coherence, i.e. does not commute with the Hamiltonian induced on $\mathcal{H}^{\text{pure}}_{\mu^\ast}$. Then, by performing the projective measurement corresponding to projectors $\{\Pi_\mu\}_\mu$, which is a TI operation, and discarding the subsystem $\mathcal{H}^{\text{mixed}}_\mu$, we obtain the (unnormalized) pure state $|\psi_{\mu^\ast}\rangle\langle\psi_{\mu^\ast}|$. Then, because this pure state contains coherence, we can use it to distill coherence. Therefore, in this case the distillable coherence for state $\rho$ is non-zero.

The question of classifying all states with non-zero distillable coherence remains open. In particular, it is not clear if the above family of states includes all states with non-zero distillable coherence.

\newpage

\section*{Supplementary Note 7: Sub-linear distillation with a measure-and-prepare TI process\\ (Proof of Eq.14 in the paper)}\label{AppG}

In this section we study a coherence distillation process which works based on a measure-and-prepare TI process. 

\subsection*{Covariant estimators}

Consider the following parameter estimation problem:  Suppose we are given $n$ copies of state $e^{-i H t}\rho e^{i H t}$, where $t\in [0,\tau)$ is unknown, and $\tau$ is the period. By performing a measurement on these systems we can find
 an estimate $t_\text{est}\in[0,\tau)$ of $t$, with probability density $p(t_\text{est}|t)$.  This estimator can be described by the POVM 
 $$\{ M_{t_\text{est}} dt_\text{est} :\ \ t_\text{est}\in[0,\tau) \} \ , $$  
such that
\be
p(t_\text{est}|t) =\Tr(M_{t_\text{est}} (e^{-i H t}\rho e^{i H t})^{\otimes n})\ .
\ee

Given any such POVM, we can construct a new POVM defined by 
\be
\tilde{M}_{t_\text{est}} = \frac{1}{\tau}\int_0^\tau ds\  (e^{i H s })^{\otimes n} M_{t_\text{est}+s} (e^{-i H s })^{\otimes n} \ ,
\ee
where $t_\text{est}+s$ is mod $\tau$.  This POVM describes the  estimator which first shifts the received state state by $(e^{-i H s })^{\otimes n}$, where $s$ is chosen uniformly at random, then apply the original estimator,  and at the end, cancels the shift $s$ at the output. 

Such estimators  are guaranteed to be invariant under time-translations in the following sense 
\be\label{covpovm}
(e^{i H r })^{\otimes n} \tilde{M}_{t_\text{est}+r} (e^{-i H r })^{\otimes n}= \tilde{M}_{t_\text{est}}  \ : \forall r\in [0,
\tau)\ ,
\ee
where $t_\text{est}+r$ is mod $\tau$. The probability  density of outcome $t_\text{est}$ for this POVM, i.e. 
\be
\tilde{p}(t_\text{est}|t) \equiv\Tr(\tilde{M}_{t_\text{est}} (e^{-i H t}\rho e^{i H t})^{\otimes n})\ ,
\ee
satisfies
\be
\tilde{p}(t_\text{est}|t)=\frac{1}{\tau}\int_0^\tau ds\  {p}(t_\text{est}+s|t+s)
\ee
and is invariant under time translations, i.e.
\be\label{cov109}
\tilde{p}(t_\text{est}|t)= \tilde{p}(t_\text{est}+r|t+r)  \ : \forall r\in [0,
\tau)\ .
\ee
Therefore, the Mean Squared Error (MSE) of the new estimator is independent of $t$, and is equal to
\begin{align}
 \langle\delta t^2\rangle=\int_0^\tau dt_\text{est}\ \tilde{p}(t_\text{est}|t) (t-t_\text{est})^2&= \int_0^\tau dt_\text{est}\ \frac{1}{\tau}\int_0^\tau ds\ p(t_\text{est}+s|t+s) (t-t_\text{est})^2 \\ &=\frac{1}{\tau}\int_0^\tau ds \int_0^\tau dr\ \ p(r|t+s) (t-r+s)^2 \\ &=  \frac{1}{\tau}\int_0^\tau ds\ \Big[ \int_0^\tau dr\ p(r|s) (s-r)^2\Big] \ ,
 \end{align}
 which is the average of MSE for the original estimator. 
 
In the following we always assume that the estimator satisfies the covariance condition in Supplementary Eq.\ref{covpovm}, or equivalently, \ref{cov109}; otherwise, if an estimator does not satisfy this condition we can always construct a new estimator which satisfies this condition. Then,  the MSE for the new estimator is independent of the parameter $t$,  and is equal to the average MSE for the original estimator.
  
 \subsection*{A TI measure-and-prepare channel}
 
Next, suppose after estimating $t$ we prepare $m$ copies of state $e^{-i H t_\text{est}}|\phi\rangle$. Overall this process implements the state transformation
\be
(e^{-i H t}\rho e^{i H t})^{\otimes n}\longrightarrow  \int_0^\tau dt_\text{est}\ p(t_\text{est}|t)\ \Big(e^{-i H t_\text{est}}|\phi\rangle\langle\phi|e^{i H t_\text{est}}\Big)^{\otimes m}\ ,
\ee
and is described by the TI quantum channel $\mathcal{E}_\text{TI}$, defined by
\be
\mathcal{E}_\text{TI}(\sigma)=\int_0^\tau dt_\text{est}\ \Tr(\sigma  M_{t_\text{est}}) \Big(e^{-i H t_\text{est}}|\phi\rangle\langle\phi|e^{i H t_\text{est}}\Big)^{\otimes m}\ .
\ee
The fact that this channel is TI follows immediately from the covariance condition for POVM in Supplementary Eq.\ref{covpovm}.

Applying this TI channel  to input $\rho^{\otimes n}$, we obtain
\be
\mathcal{E}_\text{TI}(\rho^{\otimes n})=\int_0^\tau dt_\text{est}\ \Tr(\rho^{\otimes n}  M_{t_\text{est}}) \Big(e^{-i H t_\text{est}}|\phi\rangle\langle\phi|e^{i H t_\text{est}}\Big)^{\otimes m}=
\int_0^\tau dt_\text{est}\ p(t_\text{est}|t=0)\ \Big(e^{-i H t_\text{est}}|\phi\rangle\langle\phi|e^{i H t_\text{est}}\Big)^{\otimes m}\ ,
\ee
where
\be
p(t_\text{est}|t=0)=\Tr(M_{t_\text{est}}\rho^{\otimes n})\ .
\ee
The fidelity of this state with  $|\phi\rangle^{\otimes m}$ is
\begin{align}
\langle\phi|^{\otimes m}\mathcal{E}_\text{TI}(\rho^{\otimes n})|\phi\rangle^{\otimes m}&=
\int_0^\tau dt_\text{est}\ p(t_\text{est}|t=0)\ |\langle\phi|e^{i H t_\text{est}}|\phi\rangle|^{2 m}\ . 
\end{align}
To bound this fidelity, we use the fact that 
\be\label{wfefee}
 |\langle\phi|e^{i H r}|\phi\rangle|^{2m}\ge 1-r^2 m\times V_H(\phi)\ ,
\ee 
which is proven at the end of this section. Using this bound, we find 
\begin{align}
\langle\phi|^{\otimes m}\mathcal{E}_\text{TI}(\rho^{\otimes n})|\phi\rangle^{\otimes m}&=\int_0^\tau dt_\text{est}\ p(t_\text{est}|t=0)\ |\langle\phi|e^{i H t_\text{est}}|\phi\rangle|^{2 m} \\ &\ge \int_0^\tau dt_\text{est}\ p(t_\text{est}|t=0) \times [1-t_\text{est}^2 m\times V_H(\phi)]\\ &= 1- m\times V_H(\phi)\times \langle\delta t^2\rangle \label{gegr}
\end{align}
where $\langle\delta t^2\rangle=\int_0^\tau dt_\text{est}\ p(t_\text{est}|t) (t-t_\text{est})^2$ 
  is the MSE of the estimator and is independent of $t$ (Recall that we have assumed the estimator is invariant under time translations).   
    
  Let
 \be
 \epsilon_n=1-\langle\phi|^{\otimes m}\mathcal{E}_\text{TI}(\rho^{\otimes n})|\phi\rangle^{\otimes m}\ ,
 \ee
  be the infidelity between state $|\phi\rangle^{\otimes m}$ and $\mathcal{E}_\text{TI}(\rho^{\otimes n})$. Then, the above result means that
 \begin{align}
  \frac{m}{\epsilon_n}\ge \frac{1}{V_H(\phi)\times \langle\delta t^2\rangle}\ .
\end{align}
 Diving both sides by $n$, we find
 \begin{align}
  \frac{r(n)}{\epsilon_n}\ge \frac{1}{V_H(\phi)\times n\times \langle\delta t^2\rangle}\ ,
\end{align}
 where $r(n)=m(n)/n$ is the yield. Finally, using $F_H(\phi)=4\times V_H(\phi)$, we find
 \begin{align}
  \frac{r(n)}{\epsilon_n}\ge \frac{4}{F_H(\phi)\times n\times \langle\delta t^2\rangle}\ .
\end{align}
  The MSE $\langle \delta t^2\rangle$ for any reasonable estimator  scales as $1/n$. Therefore, as $n$ goes to infinity, the above lower bound remains a positive non-zero constant. In particular, as shown in \cite{Braunstein:94, barndorff2000fisher}, there exists  an estimator, based on the Maximum Likelihood (ML) estimator, which achieves MSE equal to 
  \be\label{sgg020}
  \langle\delta t^2\rangle=\frac{1}{n F_H(\rho)}+o(\frac{1}{n})\ ,
  \ee
   i.e. saturates the Quantum Cram\'er-Rao bound  \cite{helstrom1969quantum, Holevo:book, Braunstein:94}. Using this estimator we obtain 
  \begin{align}
  \frac{r(n)}{\epsilon_n}\ge 4\frac{F_H(\rho)}{F_H(\phi)}\times [1-o(1)]\ ,
\end{align}
which is the lower bound on maximum achievable yield in Eq. 13 of the main paper. 
 To complete the proof, we need to prove the bound in Supplementary Eq.\ref{wfefee}, which is presented in the following section.  
 
It is worth noting that if the Hamiltonian $H$ is  bounded and $m$ is fixed, then the above bound is tight, up to corrections of $o(1/n)$. In other words, the measure-and-prepare TI process working based on the ML estimator transforms input $\rho^{\otimes n}$ to an output state whose infidelity with  $|\phi\rangle^{\otimes m}$ is equal to 
 \be\label{keke1}
 \epsilon_n=m V_H(\phi)\times  \langle\delta t^2\rangle+o(\frac{1}{n})=\frac{1}{n}\times \frac{m F_H(\phi)}{4 F_H(\rho)}  +o(\frac{1}{n})\ .
 \ee
To see this note that for ML estimator (as well as any other reasonable estimator) the second movement $\langle\delta t^2\rangle$ scales as $1/n$, and the higher moments scales as $o(1/n)$. Hence, assuming the Hamiltonian $H$ is  bounded and $m$ is fixed,  the Supplementary Eq.\ref{gegr}  holds as equality, up to correction of order $o(1/n)$, i.e. $ \epsilon_n=m V_H(\phi)\times  \langle\delta t^2\rangle+o(\frac{1}{n})$ (This can be seen by Taylor expanding $|\langle\phi|e^{i H t_\text{est}}|\phi\rangle|^{2 m}$ in powers of $t_{\text{est}}$). The second equality  in Eq.(\ref{keke1}) follows from the fact that ML estimator asymptotically achieves the Quantum Cram\'er-Rao bound, up to corrections of order  $o(1/n)$.

 \subsection*{Proof of Supplementary Eq.\ref{wfefee}}

In the following, we show that for any Hamiltonian $G$, state $|\eta\rangle$, and $r\in\mathbb{R}$,
\be\label{wfefee2}
 |\langle\eta| e^{i G r}|\eta\rangle|^{2}\ge 1-r^2 V_G(\eta)\ ,
\ee
where $V_G(\eta)=\langle\eta|G^2|\eta\rangle-\langle\eta|G|\eta\rangle^2$. To see this note that
\begin{align}
 \Big| \langle\eta|e^{i G r} |\eta\rangle\Big|^2
&=1+\int_0^r ds_1 \int_0^{s_1} ds_2\ \frac{d^2}{ds^2_2}\  \Big| \langle\eta|e^{i G s_2} |\eta\rangle\Big|^2\\ &\ge 1+\frac{r^2}{2}\times \min_{s_2\in[0,s]}\frac{d^2}{ds_2^2}   \Big| \langle\eta|e^{i G s_2} |\eta\rangle\Big|^2\label{wg139} .
\end{align}
Then, we note that
\begin{align}
\frac{d}{ds}|\langle\eta|e^{i G s} |\eta\rangle|^2&=\frac{d}{ds} \Tr(|\eta\rangle\langle\eta| e^{-i G s}  |\eta\rangle\langle \eta|e^{i G s})\\ &=-i\Tr(|\eta\rangle\langle\eta|  \Big[G,  e^{-i G s}  |\eta\rangle\langle \eta|e^{i G s}\Big])\\ &=-i\Tr(\Big[|\eta\rangle\langle\eta|, G\Big]  e^{-i G s}  |\eta\rangle\langle \eta|e^{i G s})\ .
\end{align}
Therefore,
\begin{align}
\frac{d^2}{ds^2}|\langle\eta|e^{i G s} |\eta\rangle|^2&=-i\frac{d}{ds} \Tr(\Big[|\eta\rangle\langle\eta|, G\Big]  e^{-i G s}  |\eta\rangle\langle \eta|e^{i G s}) \\ &= \Tr(\Big[|\eta\rangle\langle\eta|, G\Big]   e^{-i G s} \Big[|\eta\rangle\langle \eta|, G\Big]  e^{i G s})
\end{align}
Therefore,
\begin{align}
\Big|\frac{d^2}{ds^2}|\langle\eta|e^{i G s} |\eta\rangle|^2\Big|&=\Big|  \Tr(\Big[|\eta\rangle\langle\eta|, G\Big]   e^{-i G s} \Big[|\eta\rangle\langle \eta|, G\Big]  e^{i G s})\Big| \\ &\le \Big| \Tr(\Big[|\eta\rangle\langle\eta|, G\Big]^2) \Big|\\ &=2V_G(\eta)\ ,
\end{align}
where the  bound follows from Cauchy-Schwartz inequality.  This means that
\be
 \min_{s_2\in[0,s]}\frac{d^2}{ds_2^2}   \Big| \langle\eta|e^{i G s_2} |\eta\rangle\Big|\ge -2V_G(\eta)\ .
\ee
Putting this back into Supplementary Eq.\ref{wg139} we find Supplementary Eq.\ref{wfefee2}\ .

Now consider $m$ copies of a system with state $|\phi\rangle$ and Hamiltonian $H$, i.e. a composite system with the joint state  $|\eta\rangle=|\phi\rangle^{\otimes m}$, and  the total Hamiltonian $H_\text{tot}=\sum_{i=1}^m H^{(i)}$. Then, the total energy variance  with respect to the total Hamiltonian $H_\text{tot}$ is $m\times V_H(\phi)$. Therefore, we conclude that 
\be
 |\langle\phi|^{\otimes m} e^{i H_\text{tot} r}|\phi\rangle^{\otimes m}|^{2}= |\langle\phi|e^{i H r}|\phi\rangle|^{2m}\ge 1-r^2 m\times V_H(\phi)\ ,
\ee 
which proves Supplementary Eq.\ref{wfefee}. 
 
\color{black}

\newpage

\color{black}
\section*{Supplementary Note 8: Purity of coherence of the output of Measure-and-Prepare TI channels is upper bounded by QFI of the input}\label{AppC:Eng}

Recall that a quantum channel is called a Measure-and-Prepare channel if it can be written as
\be
\mathcal{E}_{\text{MP}}(\rho)=\sum_{x\in\mathcal{X}} \Tr(M_x \rho) \sigma_x\ ,
\ee
where $\{M_x: x\in \mathcal{X}\}$ is a POVM and  $\{\sigma_x: x\in \mathcal{X}\}$ is a set of density operators \cite{wilde2013quantum}. Furthermore, assuming the input Hilbert space is finite-dimensional, 
the set of outcomes $\mathcal{X}$ can be chosen to have finite elements  \cite{wilde2013quantum}. It turns out that a quantum channel is Measure-and-Prepare  if, and only if, it is entanglement-breaking.

In this section we prove that for any TI Measure-and-Prepare process $\mathcal{E}_\text{TI-MP}$, and any input $\rho$ it holds that 
\be
P_{H_{\text{out}}}(\mathcal{E}_\text{TI-MP}(\rho))\le F_{H_\text{in}}(\rho)\le P_{H_\text{in}}(\rho)\ ,
\ee
where $H_\text{in}$ and $H_\text{out}$ are, respectively, the input and output Hamiltonians. We have shown the inequality $F_{H_\text{in}}(\rho)\le P_{H_\text{in}}(\rho)$ before. In the following we prove $P_{H_{\text{out}}}(\mathcal{E}_\text{TI-MP}(\rho))\le F_{H_\text{in}}(\rho)$.


 Define the channels
\be
\mathcal{E}_{\text{meas}}(\cdot)=\sum_{x\in\mathcal{X}} \Tr(M_x \cdot) |x\rangle\langle x|
\ee
and
\be
\mathcal{E}_{\text{prep}}(\cdot)=\sum_{x\in\mathcal{X}} \Tr(|x\rangle\langle x| \cdot)\sigma_x \ ,
\ee
where $\{|x\rangle\}_x$ is a set of orthonormal states. Then, for any input state $\rho$, we have
\begin{align}
\sigma=\mathcal{E}_{\text{MP}}(\rho)=\mathcal{E}_{\text{prep}} \circ \mathcal{E}_{\text{meas}}(\rho) = \sum_{x\in\mathcal{X}} \Tr(M_x \rho) \sigma_x\ .
\end{align}
For arbitrary time $t$, define 
\begin{align}
\rho(t)&\equiv e^{-i H_\text{in} t} \rho e^{i H_\text{in} t}\ ,\\ \omega_\text{cl}(t)& \equiv  \mathcal{E}_{\text{meas}}(\rho(t))=\sum_{x\in\mathcal{X}} p_t(x) |x\rangle\langle x|\ ,
\\ \sigma(t)&\equiv e^{-i H_\text{out} t} \sigma e^{i H_\text{out} t}= \mathcal{E}_{\text{MP-TI}}(\rho(t))=   \mathcal{E}_\text{prep}(\omega_\text{cl}(t))\ ,
\end{align}
where 
\be
p_t(x)=\Tr(M_x \rho(t))\ .
\ee
To summarize, we have
\be
\forall t:\ \ \  \rho(t)\ \xrightarrow{\text{Measurement}}\  \omega_{\text{cl}}(t)\  \xrightarrow{\text{Preperation}} \ \sigma(t)\ .
\ee
In the following, we show that
\be
F_{H_\text{in}}(\rho) \ge I_{{t=0}} \ge P_{H_\text{out}}(\sigma)  \ ,
\ee
where $I_{{t=0}} $ is the classical Fisher information for the family of distribution $p_t$,  corresponding to parameter $t$, at $t=0$. We note that the bound $ I_{{t=0}} \ge P_{H_\text{out}}(\sigma)$ also follows from the general result of  \cite{matsumoto2005reverse}. The bound $F_{H_\text{in}}(\rho) \ge I_\text{t=0}$ is a consequence of the monotonicity of QFI under data processing. Here, for completeness, we prove this directly using the relation between QFI and the fidelity. 

Before presenting the proof, we note that if the input Hilbert space is finite-dimensional and the Hamiltonian is bounded, then the output probabilities $p_t(x)$ are analytic  functions of $t$. Furthermore, without loss of generality we can assume  for any outcome $x\in \mathcal{X}$, outcome probability $p_t(x)$ is non-zero for some $t$; otherwise, we can combine all the POVM elements with zero probability with a POVM element with non-zero probability, without changing the action of the channel $\mathcal{E}_\text{TI-MP}$ on the input state $\rho$.  Moreover, we assume for all  outcomes $x\in \mathcal{X}$, the corresponding probabilities $p_t(x)$ are non-zero at $t=0$; if this is not the case, 
then we can shift the point $t=0$ by a properly chosen $s\in \mathbb{R}$, or equivalently, we can replace the input $\rho$ with its translated version, namely  state  $\rho(s)=e^{-i H_{\text{in}} s} \rho e^{i H_{\text{in}} s}$, where $s\in \mathbb{R}$ is chosen  such that the probabilities $p_{s}(x)=\Tr(M_x \rho(s))>0$  for all outcomes $x\in \mathcal{X}$ (Since all the probabilities $\{p_{s}(x): x\in\mathcal{X}\}$ are non-zero analytic  functions of $s$, and   $\mathcal{X}$ is a finite set, there always exists $s\in \mathbb{R}$, which satisfies this property). Then, the following argument proves that  $P_{H_\text{out}}(\mathcal{E}_\text{TI-MP}(e^{-i H_{\text{out}} s} \rho e^{i H_{\text{out}} s}))\le F_{H_\text{in}}(e^{-i H_{\text{in}} s} \rho e^{i H_{\text{in}} s})$, which immediately implies  $P_{H_\text{out}}(\mathcal{E}_\text{TI-MP}(\rho))\le F_{H_\text{in}}(\rho)$.  Therefore,  without loss of generality, in the following we assume 
\be
p(x)\equiv p_{t=0}(x)\equiv \Tr(M_x \rho) >0, \ \forall x\in \mathcal{X}\ .
\ee

Recall that the fidelity between two states $\rho_1$ and $\rho_2$ is defined as\footnote{Note that sometimes fidelity is defined as the square root of this formula.}, $\text{Fid}(\rho_1,\rho_2)=\|\sqrt{\rho_1} \sqrt{\rho_2} \|^2_1\ $. Then, QFI is equal to
\be
F_{H_\text{in}}(\rho)=-{4}\frac{d^2}{dt^2}\sqrt{\text{Fid}(\rho, \rho(t))}\ \Big|_{t=0}\ .
\ee
Note that the first derivative vanishes at $t=0$.
 
Recall that fidelity is monotone under CPTP maps, i.e. for any quantum channel $\mathcal{E}$,
\be
\text{Fid}(\mathcal{E}(\rho_1),\mathcal{E}(\rho_2))\ge \text{Fid}(\rho_1,\rho_2)\ .
\ee
It follows that for any quantum channel $\mathcal{E}$, 
\be
F_{H_\text{in}}(\rho)=-{4}\frac{d^2}{dt^2}\sqrt{\text{Fid}(\rho,\rho(t))}\ \Big|_{t=0}\ \ge -{4}\frac{d^2}{dt^2}\sqrt{\text{Fid}(\mathcal{E}(\rho),\mathcal{E}(\rho(t)))}\ \Big|_{t=0}\ .
\ee
We apply  this to channel $\mathcal{E}_{\text{meas}}$.  For this channel we have,
\be
\sqrt{\text{Fid}(\mathcal{E}_{\text{meas}}(\rho),\mathcal{E}_{\text{meas}}(\rho(t)))} =\sum_{x\in\mathcal{X}} \sqrt{p(x) p_t(x)} \ .
\ee
Let $\dot{p}_t(x)=\frac{d}{dt}p_t(x)$ and $\ddot{p}_t(x)=\frac{d^2}{dt^2}p_t(x)$ be, respectively, the first and second derivatives of ${p}_t(x)$ with respect to time $t$. Then,
\begin{align}
\frac{d^2}{dt^2}\sqrt{\text{Fid}(\mathcal{E}_{\text{meas}}(\rho),\mathcal{E}_{\text{meas}}(e^{-i H t}\rho e^{i H t}))}\ &= \frac{d^2}{dt^2} \sum_{x\in\mathcal{X}} \sqrt{p(x) p_t(x)}\\ &= \frac{d}{dt} \Big( \frac{d}{dt} \sum_{x\in\mathcal{X}} \sqrt{p(x) p_t(x)} \Big)\\ &= \frac{1}{2}\frac{d}{dt} \Big(  \sum_{x\in\mathcal{X}} \sqrt{p(x)} \dot{p}_t(x) [p_t(x)]^{-1/2} \Big)\\ &= -\frac{1}{4}   \sum_{x\in\mathcal{X}} \sqrt{p(x)} \dot{p}^2_t(x) [p_t(x)]^{-3/2} +\frac{1}{2}\sum_{x\in\mathcal{X}} \sqrt{p(x)} \ddot{p}_t(x) [p_t(x)]^{-1/2} \ .
\end{align}
At $t=0$, we find
\begin{align}
\frac{d^2}{dt^2}\sqrt{\text{Fid}(\mathcal{E}_{\text{meas}}(\rho),\mathcal{E}_{\text{meas}}(e^{-i H t}\rho e^{i H t}))}\ \Big|_{t=0}&= -\frac{1}{4}   \sum_{x\in\mathcal{X}} \sqrt{p(x)} \dot{p}^2_t(x) [p_t(x)]^{-3/2} \Big|_{t=0} +\frac{1}{2}\sum_{x\in\mathcal{X}} \sqrt{p(x)} \ddot{p}_t(x) [p_t(x)]^{-1/2}\Big|_{t=0} \ .
\\ &= -\frac{1}{4}   \sum_{x\in\mathcal{X}}  \frac{\dot{p}_t^2(x)}{p(x)}\Big|_{t=0}  +\frac{1}{2}\sum_{x\in\mathcal{X}} \ddot{p}_t(x) \Big|_{t=0}\\ &= -\frac{1}{4}   \sum_{x\in\mathcal{X}}  \frac{\dot{p}_t^2(x)}{p(x)}\Big|_{t=0}  \ ,
\end{align}
where in the last step we have used the fact that $\sum_{x\in\mathcal{X}} p_t(x)=1$ for all $t$, and therefore $\sum_{x\in\mathcal{X}} \ddot{p}_t(x)=0$ . We conclude that
\begin{align}
F_{H_\text{in}}(\rho)&=-{4}\frac{d^2}{dt^2}\sqrt{\text{Fid}(\rho,e^{-i H t}\rho e^{i H t})}\ \Big|_{t=0}\ \\ &\ge -{4}\frac{d^2}{dt^2}\sqrt{\text{Fid}(\mathcal{E}_{\text{meas}}(\rho),\mathcal{E}_{\text{meas}}(e^{-i H t}\rho e^{i H t}))}\ \Big|_{t=0}\\  &=\sum_{x\in\mathcal{X}}  \frac{\dot{p}_t^2(x)}{p(x)}\Big|_{t=0}  \ .
\end{align} 
The quantity in the last line is in fact $I_{{t=0}} $, the (classical) Fisher information associated to parameter $t$, for the family of probability distributions $p_t$, at $t=0$. The above calculation basically shows that the classical Fisher information at the output of $\mathcal{E}_\text{meas}$  is upper bound by the Quantum Fisher information at the input. To summarize, we explicitly checked that
\be\label{whw0}
F_{H_\text{in}}(\rho) \ge I_{t=0}\equiv \sum_{x\in\mathcal{X}}  \frac{\dot{p}_t^2(x)}{p(x)}\Big|_{t=0} \ .
\ee

Next, we prove the bound $ I_{{t=0}} \ge P_{H_\text{out}}(\sigma)$ (See also \cite{matsumoto2005reverse}).  First, recall the connection between the purity of coherence and  Petz-R\'enyi  relative entropy. In particular, using Supplementary Eq.\ref{dhhh}, the purity of coherence of $\sigma=\mathcal{E}_{\text{MP-TI}}(\rho)$ is given by
\begin{align}
P_{H_\text{out}}(\sigma)&=\frac{1}{2}\frac{d^2}{dt^2}\Big[ \overline{Q}\Big(\sigma\|\sigma(t)\Big)\Big]_{t=0}=\frac{1}{2}\frac{d^2}{dt^2} \Big[\overline{Q}\Big(\sigma(t)\|\sigma\Big)\Big]_{t=0}\ ,
\end{align}
where 
\be
 \overline{Q}_2(\rho_1\|\rho_2)=\Tr(\rho_1^2\rho_2^{-1})\ ,
\ee
is monotone under CPTP maps, i.e. for any CPTP map $\mathcal{E}$, we have
\be\label{jsdbf}
 \overline{Q}_2(\mathcal{E}(\rho_1)\|\mathcal{E}(\rho_2))\le  \overline{Q}_2(\rho_1\|\rho_2)\ .
\ee
Recall that $\sigma(t)=\mathcal{E}_\text{prep}(\omega_\text{cl}(t))$.  Applying Supplementary Eq.\ref{jsdbf}  to channel  $\mathcal{E}_\text{prep}$, and the input states $\omega_\text{cl}(t)$ and $\omega_\text{cl}(0)$, we find
\be
 \overline{Q}_2\Big(\sigma(t)\|\sigma\Big)= \overline{Q}_2\Big(\sigma(t)\|\sigma(0)\Big)=
 \overline{Q}_2\Big(\mathcal{E}_\text{prep}(\omega_\text{cl}(t))\|\mathcal{E}_\text{prep}(\omega_\text{cl}(0))\Big)\le  \overline{Q}_2(\omega_\text{cl}(t)\|\omega_\text{cl}(0))\ ,
\ee

Taking the second derivative of both sides, and using the fact that the first derivatives vanish, we find
\be
P_{H_\text{out}}(\sigma)\le \frac{1}{2} \frac{d^2}{dt^2}  \overline{Q}_2(\omega_\text{cl}(t)\|\omega_\text{cl}(0))\Big|_{t=0}\ .
\ee

Next, we note that for state $\omega_{\text{cl}}(t)$, 
\be
  \overline{Q}_2(\omega_\text{cl}(t)\|\omega_\text{cl}(0))=\sum_{x\in\mathcal{X}} \frac{p^2_t(x)}{p(x)}   \ .
\ee
Taking the derivative of both sides with respect to $t$, we find
\begin{align}
\frac{d^2}{dt^2}\overline{Q}_2(\omega_\text{cl}(t)\|\omega_\text{cl}(0))\Big|_{t=0} &= \frac{d^2}{dt^2} \sum_{x\in\mathcal{X}} \frac{p^2_t(x)}{p(x)}\Big|_{t=0}\\  &= 2\sum_{x\in\mathcal{X}} \frac{\dot{p}^2_t(x)+p_t(x)\ddot{p}_t(x)}{p(x)})\Big|_{t=0} \\ &= 2\sum_{x\in\mathcal{X}} \frac{\dot{p}^2_t(x)}{p(x)}\Big|_{t=0}+ 2\sum_{x\in\mathcal{X}} \ddot{p}_t(x) \Big|_{t=0}\\ &=2 I_{t=0}\ ,
\end{align}
where to get the last line we again used the fact that the fact that $\sum_{x\in\mathcal{X}} \ddot{p}(x)=1$. Therefore, we conclude that
\be
P_{H_\text{out}}(\sigma)\le \frac{1}{2} \frac{d^2}{dt^2}  \overline{Q}_2(\omega_\text{cl}(t)\|\omega_\text{cl}(0))\Big|_{t=0}=  \sum_{x\in\mathcal{X}} \frac{\dot{p}^2_t(x)}{p(x)}\Big|_{t=0}= I_{t=0}\ .
\ee
Combining this with Supplementary Eq.\ref{whw0}, we find
\be
P_{H_\text{out}}(\sigma)\le \sum_{x\in\mathcal{X}} \frac{\dot{p}^2_t(x)}{p(x)}\Big|_{t=0}= I_{t=0}\le F_{H_\text{in}}(\rho)\ ,
\ee
which completes the proof.

\color{black}
\newpage

\section*{Supplementary Note 9: Distillation in the single-shot regime}\label{Sec:sing}

\subsection*{Maximum achievable fidelity with a pure state (Proof of  Eq.16 in the paper)}\label{app:sec:min}
In this section, we use the approach of \cite{gour2017quantum}, to find a simple formula for the maximum achievable fidelity $\max_{\mathcal{E}_\text{TI}}\  \langle\psi| \mathcal{E}_\text{TI}(\rho^{\otimes n})|\psi\rangle$, in terms of the \emph{conditional min-entropy}. 

Recall the definition of the conditional min-entropy \cite{renner2008security, Konig, tomamichel2015quantum}, $H_{\min}(\text{B}|\text{A})_\Omega$, of a bipartite state $\Omega_{\text{AB}}$, 
\begin{align}
2^{-H_{\text{min}} (\text{B}|\text{A})_\Omega} = \inf_{\tau^{\text{A}} \ge 0} \{ \Tr({\tau^{\text{A}}}) : \tau^{\text{A}}\otimes I^{\text{B}} \ge \Omega_{\text{AB}}\}\ .
\end{align}\\

\refstepcounter{Thm}
\noindent\emph{Theorem 3}
Let $H_{\text{A}}$ and $H_{\text{B}}$ be, respectively, the Hamiltonians of the input and output systems $A$ and $B$. Let  $\sigma_{\text{A}}$ and  be an arbitrary state of $A$ and  $|\psi\rangle_{\text{B}}$ be a pure state of system $B$.  Then,
\begin{align}
\max_{\mathcal{E}_\text{TI}}\  \langle\psi| \mathcal{E}_\text{TI}(\sigma_{\text{A}})|\psi\rangle_{\text{B}} &=2^{- H_\text{min}(\text{B}|\text{A})_\Omega}= 2^{- H_\text{min}(\text{B}|\text{A})_\Gamma} ,
\end{align}
where the maximization is over the set of all TI operations, and state $\Omega_{\text{AB}}$ and $\Gamma_{\text{AB}}$ are defined as 
\begin{align}
\Omega_{\text{AB}}&= 
\lim_{T\rightarrow \infty} \frac{1}{T}\int_0^T dt \ e^{-i H_{\text{A}} t}\otimes e^{i H_{\text{B}} t}[\sigma_{\text{A}}\otimes |{\psi}\rangle\langle {\psi}|_{\text{B}}]  e^{i H_{\text{A}} t}\otimes e^{-i H_{\text{B}} t}=\sum_E \Pi_E [\sigma_{\text{A}}\otimes |{\psi}\rangle\langle{\psi}|_{\text{B}}] \Pi_E\ , \\ \Gamma_{\text{AB}}&= 
\lim_{T\rightarrow \infty} \frac{1}{T}\int_0^T dt \ e^{-i H_{\text{A}} t}\otimes e^{i H_{\text{B}} t}[\sigma_{\text{A}}\otimes |\overline{\psi}\rangle\langle \overline{\psi}|_{\text{B}}]  e^{i H_{\text{A}} t}\otimes e^{-i H_{\text{B}} t}=\sum_E \Pi_E [\sigma_{\text{A}}\otimes |\overline{\psi}\rangle\langle\overline{\psi}|_{\text{B}}] \Pi_E\ ,
\end{align}
 where  $\Pi_E$ is the projector to the eigen-subspace of $H_{\text{A}}\otimes I_{\text{B}}-I_{\text{A}}\otimes H_{\text{B}}$ with energy $E$,
and $|\overline{\psi}\rangle=\sum_i   \overline{\langle E_i|\psi\rangle} |E_i\rangle=\sum_i \langle\psi|E_i\rangle |E_i\rangle\ $ is the complex conjugate of $|\psi\rangle$ in  the eigenbasis of Hamiltonian $H_{\text{B}}$, denoted by $\{|E_i\rangle: i=1, \cdots , d_{\text{B}} \}$, and $d_{\text{B}}$ is the dimension of Hilbert space of B.\\

In other words, state $\Omega_{\text{AB}}$ is the state obtained by dephasing  $\sigma_{\text{A}}\otimes |{\psi}\rangle\langle{\psi}|_{\text{B}}$ in the eingenbasis of the Hamiltonian $H_{\text{A}}\otimes I_{\text{B}}-I_{\text{A}}\otimes H_{\text{B}}$. 
Note that if the input system $A$ is $n$ copies of a system with Hamiltonian $H$ and state $\rho$, then state 
$\Omega_{\text{AB}}$ will be given by
\be
\sum_E \Pi_E [(\rho^{\otimes n})_{\text{A}}\otimes |{\psi}\rangle\langle{\psi}|_{\text{B}}] \Pi_E\ ,
\ee
where $\Pi_E$ is the projector to the eigen-subpaces of Hamiltonian $H_\text{tot} \otimes I_{\text{B}}-I_\text{tot}\otimes H_{\text{B}}$. Here, $H_\text{tot}=\sum_i H^{(i)}$ is the total Hamiltonian of the input systems, $H^{(i)}=I^{\otimes (i-1)}\otimes H\otimes I^{\otimes (n-i-1)}$, and $I_\text{tot}=I^{\otimes n}$ is the identity operator on the input systems.

\begin{proof}

Let $B'$ be an auxiliary system with dimension equal to $d_{\text{B}}$, the dimension $B$.Define
\be
|\gamma_{\text{BB}'}\rangle=\frac{1}{\sqrt{d_{\text{B}}}} \sum_{i=1}^{d_{\text{B}}} |E_i E_i\rangle_{\text{BB}'}\ 
\ee
 be a maximally entangled state of  $B$ and the auxiliary system $B'$. Then, for any pair of operators $X$ and $Y$ defined on $B$, we have 
$ \Tr(XY)=d_{\text{B}}\times  \langle\gamma_{\text{BB}'}  |[ X\otimes Y^T] |\gamma_{\text{BB}'}\rangle\ ,$ where $T$ denotes transpose in the energy eigenbasis, $\{|E_i\rangle_{\text{B}}: i=1, \cdots , d_{\text{B}} \}$. 
 
 This implies that for any 
 quantum channel $\mathcal{E}_\text{TI}$ we have
\be
\langle\psi|\mathcal{E}_\text{TI}(\sigma)|\psi\rangle=d_{\text{B}}\times  \langle\gamma_{\text{BB}'}  | \left[\mathcal{E}_\text{TI}(\sigma)\otimes |\overline{\psi}\rangle\langle \overline{\psi}|\right] |\gamma_{\text{BB}'}\rangle\ ,
\ee
 where $ |\overline{\psi}\rangle$ is the complex conjugate of $ |{\psi}\rangle$ in the energy eigenbasis.

Next, we note that 
 \begin{align}
\langle\psi|\mathcal{E}_\text{TI}(\sigma)|\psi\rangle&=d_{\text{B}}\times  \langle\gamma_{\text{BB}'}  | \left[\mathcal{E}_\text{TI}(\sigma)\otimes |\overline{\psi}\rangle\langle \overline{\psi}|\right] |\gamma_{\text{BB}'}\rangle\\ &= d_{\text{B}}\times  \langle\gamma_{\text{BB}'}  |(e^{i H_{\text{B}} t}\otimes e^{-i H_{\text{B}} t}) \left[\mathcal{E}_\text{TI}(\sigma)\otimes |\overline{\psi}\rangle\langle \overline{\psi}|\right] (e^{-i H_{\text{B}} t}\otimes e^{i H_{\text{B}} t}) |\gamma_{\text{BB}'}\rangle \\ &= d_{\text{B}}\times  \langle\gamma_{\text{BB}'}  | (\mathcal{E}_\text{TI}\otimes \mathcal{I}_{B'}) \left( [e^{i H_{\text{A}} t}\otimes e^{-i H_{\text{B}} t}][\sigma\otimes |\overline{\psi}\rangle\langle \overline{\psi}|] [e^{-i H_{\text{A}} t}\otimes e^{i H_{\text{B}} t}]  \right) |\gamma_{\text{BB}'}\rangle\ ,
\end{align}
where $\mathcal{I}_{B'}$ is the identity super-operator on $B'$. Here, to get the second line we have used the fact that
\be\label{app:sym1746}
(e^{-i H_{\text{B}} t}\otimes e^{i H_{\text{B}} t})|\gamma_{\text{BB}'}\rangle=|\gamma_{\text{BB}'}\rangle \  ,
\ee
and to get the last line we have used the fact that $\mathcal{E}_\text{TI}$ satisfies the covariance condition 
\be\label{app:cov165}
\mathcal{E}_\text{TI}(e^{-i H_{\text{A}} t} (\cdot) e^{i H_{\text{A}} t} )=e^{-i H_{\text{B}} t}\mathcal{E}_\text{TI}(\cdot)e^{i H_{\text{B}} t}, \ \forall t\in \mathbb{R} .
\ee  
Then, taking the average over $t$, we find that
 \begin{align}
\langle\psi|\mathcal{E}_\text{TI}(\sigma)|\psi\rangle= d_{\text{B}}\times  \langle\gamma_{\text{BB}'}  | \mathcal{E}_\text{TI}\otimes \mathcal{I}_{B'}(\Gamma_{\text{AB}'}) |\gamma_{\text{BB}'}\rangle\ ,
\end{align}
where
\begin{align}\label{deflh}
\Gamma_{\text{AB}'}&\equiv\lim_{T\rightarrow \infty} \frac{1}{T}\int_0^T dt\  \left( [e^{i H_{\text{A}} t}\otimes e^{-i H_{\text{B}} t}][\sigma\otimes |{\overline{\psi}}\rangle\langle{\overline{\psi}}|] [e^{-i H_{\text{A}} t}\otimes e^{i H_{\text{B}} t}]  \right)\ \\ &=\sum_E \Pi_E [\sigma_{\text{A}}\otimes |{\overline{\psi}}\rangle\langle{\overline{\psi}}|_{\text{B}}] \Pi_E\ ,
 \end{align}
 where  $\Pi_E$ is the projector to the eigensubspace of $H_{\text{A}}\otimes I_{\text{B}}-I_{\text{A}}\otimes H_{\text{B}}$ with energy $E$.

Therefore, 
 \begin{align}\label{eq9182}
\max_{\mathcal{E}_\text{TI}}\langle\psi|\mathcal{E}_\text{TI}(\sigma)|\psi\rangle= d_{\text{B}}\times \max_{\mathcal{E}_\text{TI}}   \langle\gamma_{\text{BB}'}  | \mathcal{E}_\text{TI}\otimes \mathcal{I}_{B'}(\Gamma_{\text{AB}'}) |\gamma_{\text{BB}'}\rangle\ .
\end{align}
Next, we argue that in the right-hand side, instead of maximizing over the set of all TI operations from $A$ to $B$, we can maximize over the larger set of \emph{all} quantum operations from $A$ to $B$ (i.e. all CPTP maps) and still the equality remains valid. Let $\mathcal{E}$ be an arbitrary CPTP map from $A$ to $B$. Using Supplementary Eq.(\ref{app:sym1746}) we have
\begin{align}\label{ergie}
 \langle\gamma_{\text{BB}'}  | \mathcal{E}\otimes \mathcal{I}_{B'}(\Gamma_{\text{AB}'}) |\gamma_{\text{BB}'}\rangle= \langle\gamma_{\text{BB}'}  |(e^{i H_{\text{B}} t}\otimes e^{-i H_{\text{B}} t}) \mathcal{E}\otimes \mathcal{I}_{B'}(\Gamma_{\text{AB}'}) (e^{-i H_{\text{B}} t}\otimes e^{i H_{\text{B}} t})|\gamma_{\text{BB}'}\rangle\ .
\end{align}
Next, we note that $\Gamma_{\text{AB}'}$ has the symmetry,
\be
\forall t\in\mathbb{R}: \ \ \ \  [e^{-i H_{\text{A}} t}\otimes e^{i H_{\text{B}} t}]\Gamma_{\text{AB}'} [e^{i H_{\text{A}} t}\otimes e^{-i H_{\text{B}} t}] =\Gamma_{\text{AB}'}  .
\ee
Combining this with Supplementary Eq.(\ref{ergie}), we find
\begin{align}
 \langle\gamma_{\text{BB}'}  | \mathcal{E}\otimes \mathcal{I}_{B'}(\Gamma_{\text{AB}'}) |\gamma_{\text{BB}'}\rangle &= \langle\gamma_{\text{BB}'}  |(e^{i H_{\text{B}} t}\otimes e^{-i H_{\text{B}} t}) \mathcal{E}\otimes \mathcal{I}_{B'}(\Gamma_{\text{AB}'}) (e^{-i H_{\text{B}} t}\otimes e^{i H_{\text{B}} t})|\gamma_{\text{BB}'}\rangle\ \\ &= \langle\gamma_{\text{BB}'}  |(e^{i H_{\text{B}} t}\otimes e^{-i H_{\text{B}} t}) \mathcal{E}\otimes \mathcal{I}_{B'}\left([e^{-i H_{\text{A}} t}\otimes e^{i H_{\text{B}} t}]\Gamma_{\text{AB}'} [e^{i H_{\text{A}} t}\otimes e^{-i H_{\text{B}} t}] \right) (e^{-i H_{\text{B}} t}\otimes e^{i H_{\text{B}} t})|\gamma_{\text{BB}'}\rangle\\ &= \langle\gamma_{\text{BB}'}  |(e^{i H_{\text{B}} t}\otimes I_{B'}) \mathcal{E}\otimes \mathcal{I}_{B'}\left([e^{-i H_{\text{A}} t}\otimes  I_{B'}]\Gamma_{\text{AB}'} [e^{i H_{\text{A}} t}\otimes  I_{B'}] \right) (e^{-i H_{\text{B}} t}\otimes  I_{B'})|\gamma_{\text{BB}'}\rangle  ,
\end{align}
where $ I_{B'}$ is the identity operator on system $B'$. Taking the average over $t$, we find
\begin{align}
 \langle\gamma_{\text{BB}'}  | \mathcal{E}\otimes \mathcal{I}_{B'}(\Gamma_{\text{AB}'}) |\gamma_{\text{BB}'}\rangle =\langle\gamma_{\text{BB}'}  |\tilde{\mathcal{E}}_\text{TI}\otimes \mathcal{I}_{B'}(\Gamma_{\text{AB}'})|\gamma_{\text{BB}'}\rangle  ,
\end{align}
where we have defined  
\be
\tilde{\mathcal{E}}_\text{TI}(X)\equiv\lim_{T\rightarrow \infty} \frac{1}{T}\int_0^T dt\ e^{i H_{\text{B}} t} \mathcal{E}(e^{-i H_{\text{A}} t}(X)e^{i H_{\text{A}} t})e^{-i H_{\text{B}} t}\ ,
\ee
which is a TI operation.

This implies that in the right-hand side of Supplementary Eq.(\ref{eq9182}), maximization over TI quantum operations, can be replaced by maximization over \emph{all} quantum operations, i.e. 
 \begin{align}
\max_{\mathcal{E}_\text{TI}}\langle\psi|\mathcal{E}_\text{TI}(\sigma)|\psi\rangle&= d_{\text{B}}\times \max_{\mathcal{E}_\text{TI}}   \langle\gamma_{\text{BB}'}  | \mathcal{E}_\text{TI}\otimes \mathcal{I}_{B'}(\Gamma_{\text{AB}'}) |\gamma_{\text{BB}'}\rangle\\ &= d_{\text{B}}\times \max_{\mathcal{E}}   \langle\gamma_{\text{BB}'}  | \mathcal{E}\otimes \mathcal{I}_{B'}(\Gamma_{\text{AB}'}) |\gamma_{\text{BB}'}\rangle  .
\end{align}
Finally, using the result of \cite{Konig} we note that 
\begin{align}\label{Eq1871}
2^{-H_{\min} (\text{B}'|\text{A})_\Gamma} &=d_{\text{B}}\times\max_{\mathcal{E}} \langle\gamma_{\text{BB}'}| \mathcal{E}\otimes \mathcal{I}_{B'}(\Gamma_{\text{AB}'})|\gamma_{\text{BB}'}\rangle\ ,
\end{align}
where  the maximization is over all CPTP maps from system $A$ to system $B'$. Therefore, we conclude that
\begin{align}\label{efgthehertg}
2^{-H_{\min} (\text{B}'|\text{A})_\Gamma} &= \max_{\mathcal{E}_\text{TI}}\langle\psi|\mathcal{E}_\text{TI}(\sigma)|\psi\rangle \ .
\end{align}
Next, suppose in definition of state $\Gamma_{\text{AB}'}$, we replace  $|\overline{\psi}\rangle\langle\overline{\psi}|$ with $|{\psi}\rangle\langle\psi|$, and define state  
\begin{align}
\Omega_{\text{AB}'}&\equiv\lim_{T\rightarrow \infty} \frac{1}{T}\int_0^T dt\  \left( [e^{i H_{\text{A}} t}\otimes e^{-i H_{\text{B}} t}][\sigma\otimes |{\psi}\rangle\langle{\psi}|] [e^{-i H_{\text{A}} t}\otimes e^{i H_{\text{B}} t}]  \right)\ \\ &=\sum_E \Pi_E [\sigma_{\text{A}}\otimes |{\psi}\rangle\langle{\psi}|_{\text{B}}] \Pi_E\ ,
 \end{align}
 where  $\Pi_E$ is the projector to the eigensubspace of $H_{\text{A}}\otimes I_{\text{B}}-I_{\text{A}}\otimes H_{\text{B}}$ with energy $E$.

Then, using Supplementary Eq.(\ref{efgthehertg}), we know that  if we consider $H_\text{min}$ for state $\Omega_{\text{AB}'}$ rather than state $\Gamma_{\text{AB}'}$, we find
\begin{align}\label{wfegwefgw}
2^{-H_{\min} (\text{B}'|\text{A})_\Omega} &= \max_{\mathcal{E}_\text{TI}}\langle\overline{\psi}|\mathcal{E}_\text{TI}(\sigma)|\overline{\psi}\rangle \ .
\end{align}
Finally, we use the following fact, which is proven later:  The maximum achievable fidelity with state $|\psi\rangle$ and state $|\overline{\psi}\rangle$ are equal, i.e.
\be\label{comps}
\max_{\mathcal{E}_\text{TI}}\  \langle\psi| \mathcal{E}_\text{TI}(\sigma_{\text{A}})|\psi\rangle_{\text{B}}= \max_{\mathcal{E}_\text{TI}}\  \langle\overline{\psi}| \mathcal{E}_\text{TI}(\sigma_{\text{A}})|\overline{\psi}\rangle_{\text{B}}\ .
\ee
This combined with Supplementary Eq.(\ref{wfegwefgw}) implies 
\be
\max_{\mathcal{E}_\text{TI}}\  \langle\psi| \mathcal{E}_\text{TI}(\sigma_{\text{A}})|\psi\rangle_{\text{B}}= \max_{\mathcal{E}_\text{TI}}\  \langle\overline{\psi}| \mathcal{E}_\text{TI}(\sigma_{\text{A}})|\overline{\psi}\rangle_{\text{B}}=2^{-H_{\min} (\text{B}'|\text{A})_\Omega}   \ ,
\ee
 which proves the theorem (Note that in the statement of theorem we have replaced label $B'$ by $B$). 
 
 To complete the proof, we need to prove Supplementary Eq.(\ref{comps}), which is presented in the following:  We use  the fact that for any state $|\psi\rangle$  there exists a unitary  $V_\psi$, which commutes with $H_{\text{B}}$ and transforms $|\psi\rangle$ to $|\overline{\psi}\rangle=V_\psi|\psi\rangle$. In particular, we can choose
\be
V_\psi= \sum_i \frac{\langle\psi|E_i\rangle}{\langle E_i|\psi\rangle} |E_i\rangle\langle E_i| ,
\ee
where we assume $\frac{\langle\psi|E_i\rangle}{\langle E_i|\psi\rangle}=1$ if $\langle\psi|E_i\rangle=0$ (Note that, in general, this unitary depends on $|\psi\rangle$. A transformation which maps $|\psi\rangle$ to $|\overline{\psi}\rangle$ for all $|\psi\rangle$ should be anti-linear).
Clearly, $[V_\psi, H_{\text{B}}]=0$ and $V_\psi|\psi\rangle\equiv |\overline{\psi}\rangle$. Let $\mathcal{V}_\psi[\cdot]=V_\psi(\cdot)V^\dag_\psi$ be the super-operator corresponding to the unitary $V_\psi$.  Clearly $\mathcal{V}_\psi$ is a TI operation.  

Next, we note that for any TI operation $\mathcal{E}_\text{TI}$,
\begin{align}
\langle\psi|\mathcal{E}_\text{TI}(\sigma_{\text{A}})|\psi\rangle &= \langle\overline{\psi}|\mathcal{V}_\psi\circ{\mathcal{E}}_\text{TI}(\sigma_{\text{A}})|\overline{\psi}\rangle\ \\
\langle\psi|\mathcal{V}^\dag_\psi\circ \mathcal{E}_\text{TI}(\sigma_{\text{A}})|\psi\rangle &= \langle\overline{\psi}|{\mathcal{E}}_\text{TI}(\sigma_{\text{A}})|\overline{\psi}\rangle\ .
\end{align}
Since both $\mathcal{V}_\psi$ and $\mathcal{V}^\dag_\psi$ are TI operations, and TI operations are closed under composition, we conclude that both $\mathcal{V}^\dag_\psi\circ \mathcal{E}_\text{TI}$ and $\mathcal{V}_\psi\circ{\mathcal{E}}_\text{TI}$ are also TI operation. Therefore, the above equations together imply 
\be
\max_{\mathcal{E}_\text{TI}}\  \langle\psi| \mathcal{E}_\text{TI}(\sigma_{\text{A}})|\psi\rangle_{\text{B}}= \max_{\mathcal{E}_\text{TI}}\  \langle\overline{\psi}| \mathcal{E}_\text{TI}(\sigma_{\text{A}})|\overline{\psi}\rangle_{\text{B}}\ .
\ee
This proves Supplementary Eq.(\ref{comps}) and completes the proof of the theorem.

\end{proof}

\noindent
\textbf{Remark.}
This result can be easily extended to the case of symmetries described by a  finite or compact Lie group $G$. Let $G\ni g\rightarrow U_{A/B}(g)$ be the unitary representations of symmetry $G$ on the input system $A$ and the output system $B$ $G$. Then,   
\be
\max_{\mathcal{E}_\text{cov}}\  \langle\psi| \mathcal{E}_\text{cov}(\sigma_{\text{A}})|\psi\rangle_{\text{B}}=2^{- H_\text{min}(\text{B}|\text{A})_\Gamma}\ ,
\ee
where the maximization is over the set of covariant operations, i.e. operations satisfying  the covariance condition
\be
\forall g\in G:\ \ \ \ \  U_{\text{B}}(g) \mathcal{E}_\text{cov}(\cdot)\ U^\dag_{\text{B}}(g)= \mathcal{E}_\text{cov}\left(U_{\text{A}}(g)(\cdot) U^\dag_{\text{A}}(g)\right)\ \ ,
\ee
and state 
\be
\Gamma_{\text{AB}}=\int dg\ [U_{\text{A}}(g)\otimes U_{\text{B}}(g)][\rho_{\text{A}}\otimes |\overline{\psi}\rangle\langle\overline{\psi}|_{\text{B}}][U^\dag_{\text{A}}(g)\otimes U^\dag_{\text{B}}(g)]\ .
\ee

\newpage

\section*{Supplementary Note 10: Qubit example (Proof of Eq.17 in the paper)}\label{app:sec:ex}

The smallest  quantum clock is a qubit with two different energy levels. Without loss of generality we assume the  Hamiltonian of this system is $H=\pi \sigma_z/\tau$. Suppose  we want to prepare this clock in a state close to the pure state $|\Phi\rangle_{\text{c-bit}}=(|0\rangle+|1\rangle)/\sqrt{2}$, but we have access to the noisy version of this state, i.e.  state 
\be
\rho=\lambda |\Phi\rangle\langle \Phi|_{\text{c-bit}}+(1-\lambda) I/2\ , 
\ee
with  $0<\lambda < 1$. The goal is to obtain a qubit state $\sigma$ which has higher fidelity with $|\Phi\rangle_{\text{c-bit}}$, by combining $n$ copies of this state via a TI operation.  How close can we get to state $|\Phi\rangle_{\text{c-bit}}$? In other words, what is the maximum achievable fidelity,
\be
\max_{\mathcal{E}_\text{TI}}\  \langle\Phi| \mathcal{E}_\text{TI}(\rho^{\otimes n})|\Phi\rangle_{\text{c-bit}} \ ,
\ee
where the maximization is over all  TI operations.

For any TI operation $\mathcal{E}_\text{TI}$, let $\sigma=\mathcal{E}_\text{TI}(\rho^{\otimes n})$ be the actual output state of the transformation. Then,  using the monotonicity and  the additivity of the purity of coherence, we find 
\be
P_H(\sigma)\le P_{H_\text{tot}}(\rho^{\otimes n})=n\times P_H(\rho)\ ,
\ee
where $H_\text{tot}=\sum_i H^{(i)}$, and $H^{(i)}=I^{\otimes (i-1)}\otimes H\otimes I^{\otimes (n-i-1)}$.

As we saw in Supplementary Eq.(\ref{app:puity-qubit}), for a general qubit state $\rho$ with the spectral decomposition 
$\rho=p |\psi\rangle\langle \psi| + (1-p)  |\psi^\perp\rangle\langle \psi^\perp|$, the purity of coherence is given by
\be
P_H(\rho)= \frac{(1-2p)^2}{p(1-p)} \times V_H(\psi)\ .
\ee
For state $\rho=\lambda |\Phi\rangle\langle \Phi|_{\text{c-bit}}+(1-\lambda) I/2$, we have $p=(1+\lambda)/2$, and $\psi=\Phi_{\text{c-bit}}$. Therefore, 
\begin{align}\label{ewvewrg}
P_H(\rho)= \frac{4\lambda^2}{1-\lambda^2} \times V_H(\Phi_{\text{c-bit}})\ .
\end{align}
We conclude that for the output state $\sigma$, it holds that
\be\label{Eq1492}
P_H(\sigma)\le \frac{4 n \lambda^2}{1-\lambda^2} \times V_H(\Phi_{\text{c-bit}}) \ . 
\ee
Next, we focus on the purity of coherence of the output state $\sigma=\mathcal{E}_\text{TI}(\rho^{\otimes n})$ and find a lower bound on $P_H(\sigma)$. Define state 
\be
\tilde{\sigma}= \frac{1}{2}(\sigma+ X\sigma X)\ ,
\ee
where $X=|0\rangle\langle 1|+|1\rangle\langle 0|$ is the Pauli-x operator. Using the fact that state $|\Phi_{\text{c-bit}}\rangle=(|0\rangle+|1\rangle)/\sqrt{2}$  is an eigenvector of $X$,  it can be easily seen that the fidelity of $|\Phi_{\text{c-bit}}\rangle$ with states $\tilde{\sigma}$  and ${\sigma}$ are equal, i.e. 
\be\label{fisbd}
\langle\Phi|\tilde{\sigma}|\Phi\rangle_{\text{c-bit}}=\langle\Phi|{\sigma}|\Phi\rangle_{\text{c-bit}}= \langle\Phi| \mathcal{E}_\text{TI}(\rho^{\otimes n})|\Phi\rangle_{\text{c-bit}}\ .
\ee
On the other hand, convexity of the purity of coherence implies
\be
P_H(\tilde{\sigma})=P_H(\frac{1}{2}[\sigma+ X\sigma X]) \le \frac{1}{2} P_H(\sigma)+ \frac{1}{2} P_H(X \sigma X) .
\ee
We can easily show that $P_H(X \sigma X)=P_H(\sigma) .$ This follows, for instance, by noting that $P_H$ is an 
even function of $H$, i.e. $P_H=P_{-H}$. Therefore $P_{H}(X \sigma X)=P_{-H}(X \sigma X)=P_{XHX}(X \sigma X)$, where we use the fact that the Pauli operator $X$, flips the sign of Hamiltonian $H=\pi \sigma_z/\tau$.  Finally, we note that for any unitary $U$, $P_{UHU^\dag}(U \sigma U^\dag)$. Therefore, we conclude that $P_{H}(X \sigma X)=P_{-H}(X \sigma X)=P_{XHX}(X \sigma X)=P_{H}(\sigma)$, which implies
\be
P_H(\tilde{\sigma})=P_H(\frac{1}{2}[\sigma+ X\sigma X]) \le \frac{1}{2} P_H(\sigma)+ \frac{1}{2} P_H(X \sigma X)=P_H(\sigma)\ .
\ee

Next, we note that $\tilde{\sigma}$ commutes with $X$, and therefore it can be written as 
\be
\tilde{\sigma}=\tilde{\lambda} |\Phi\rangle\langle \Phi|_{\text{c-bit}}+(1-\tilde{\lambda}) I/2 \ ,
\ee
for some $0\le \tilde{\lambda}\le 1 $.  This means that, given a fixed value of the purity of coherence, the state with this purity of coherence, which has the maximum fidelity with state  $|\Phi\rangle_{\text{c-bit}}$ is in the form $\tilde{\lambda} |\Phi\rangle\langle \Phi|_{\text{c-bit}}+(1-\tilde{\lambda}) I/2$.

Applying  Supplementary Eq.(\ref{ewvewrg}) for state $\tilde{\sigma}=\tilde{\lambda} |\Phi\rangle\langle \Phi|_{\text{c-bit}}+(1-\tilde{\lambda}) I/2$, we find
 \begin{align}
P_H(\tilde{\sigma})= \frac{4\tilde{\lambda}^2}{1-\tilde{\lambda}^2} \times V_H(\Phi_{\text{c-bit}})\ .
\end{align}
Therefore, we conclude that
\be
P_H(\sigma) \ge P_H(\tilde{\sigma}) = \frac{4\tilde{\lambda}^2}{1-\tilde{\lambda}^2} \times V_H(\Phi_{\text{c-bit}})\ .
\ee
Putting this into Supplementary Eq.(\ref{Eq1492}) we find
\begin{align}
\frac{\tilde{\lambda}^2}{1-\tilde{\lambda}^2}  \le n\times \frac{\lambda^2}{1-\lambda^2} \ .
\end{align}
which implies
\begin{align}
\tilde{\lambda}^2\le \frac{n\lambda^2}{1+(n-1)\lambda^2}= \frac{1}{1+\frac{1}{n}(\frac{1}{\lambda^2}-1)} \ .
\end{align}
For a fixed $\lambda>0$, in the large $n$ limit this implies
\begin{align}\label{bssb}
\tilde{\lambda}^2\le \frac{1}{1+\frac{1}{n}(\frac{1}{\lambda^2}-1)}=1-\frac{1}{n}(\frac{1-\lambda^2}{\lambda^2})+\mathcal{O}(\frac{1}{n^2}) \ .
\end{align}
This means that, among all states whose  
purity of coherence are equal to the purity of coherence of the input, such that  $P_H(\sigma)=n\times P_H(\rho)$,   state $\sigma=\tilde{\lambda} |\Phi\rangle\langle \Phi|_{\text{c-bit}}+(1-\tilde{\lambda}) I/2$, with 
\be
\tilde{\lambda}=\sqrt{1-\frac{1-\lambda^2}{n\lambda^2}+\mathcal{O}(\frac{1}{n^2})}= 1-\frac{1-\lambda^2}{2n\lambda^2}+\mathcal{O}(\frac{1}{n^2})   \ ,
\ee
has the minimum infidelity with state $ |\Phi\rangle_{\text{c-bit}}$ . This minimum infidelity is 
\be
1-\langle\Phi|\tilde{\sigma} |\Phi\rangle_{\text{c-bit}}=\frac{1-\tilde{\lambda}}{2}=\frac{1-\lambda^2}{4n\lambda^2}+\mathcal{O}(\frac{1}{n^2})\ .
\ee
 Therefore, for any TI process $\mathcal{E}_\text{TI}$,
 \begin{align}
1-\langle\Phi| \mathcal{E}_\text{TI}(\rho^{\otimes n})|\Phi\rangle_{\text{c-bit}}&\ge \frac{1-\lambda^2}{4n\lambda^2}+\mathcal{O}(\frac{1}{n^2}) \ .
\end{align}

Remarkably, this bound is tight (up to a factor of 2).   Using \cite{cirac1999optimal}, we find that   there exists a quantum operation $\mathcal{E}_{\text{Schur}}$ (related to the Schur transformation) which is   covariant  with respect to the full unitary group $\text{SU}(2)$, for which the infidelity $1-\langle \Phi|  \mathcal{E}_{\text{Schur}}(\rho^{\otimes n})  |\Phi\rangle_{\text{c-bit}} $ is equal to $2/(1+\lambda)$ times the right-hand side of this bound, i.e.
\begin{align}
1-\langle\Phi| \mathcal{E}_{\text{Schur}}(\rho^{\otimes n})|\Phi\rangle_{\text{c-bit}}= \frac{1-\lambda}{2n \lambda^2}+\mathcal{O}(\frac{1}{n^2}) \ .
\end{align}
But, since this operation is covariant with respect to the full unitary group, it is also  covariant with respect to time translations.  

\newpage

\renewcommand{\bibname}{\textbf{References}}
 \bibliography{Ref_2018}

\begin{thebibliography}{88}
\expandafter\ifx\csname natexlab\endcsname\relax\def\natexlab#1{#1}\fi
\expandafter\ifx\csname bibnamefont\endcsname\relax
  \def\bibnamefont#1{#1}\fi
\expandafter\ifx\csname bibfnamefont\endcsname\relax
  \def\bibfnamefont#1{#1}\fi
\expandafter\ifx\csname citenamefont\endcsname\relax
  \def\citenamefont#1{#1}\fi
\expandafter\ifx\csname url\endcsname\relax
  \def\url#1{\texttt{#1}}\fi
\expandafter\ifx\csname urlprefix\endcsname\relax\def\urlprefix{URL }\fi
\providecommand{\bibinfo}[2]{#2}
\providecommand{\eprint}[2][]{\url{#2}}

\bibitem[{\citenamefont{Lostaglio
  et~al.}(2015{\natexlab{a}})\citenamefont{Lostaglio, Jennings, and
  Rudolph}}]{lostaglio2015description}
\bibinfo{author}{\bibfnamefont{M.}~\bibnamefont{Lostaglio}},
  \bibinfo{author}{\bibfnamefont{D.}~\bibnamefont{Jennings}}, \bibnamefont{and}
  \bibinfo{author}{\bibfnamefont{T.}~\bibnamefont{Rudolph}},
  \bibinfo{journal}{Nature communications} \textbf{\bibinfo{volume}{6}}
  (\bibinfo{year}{2015}{\natexlab{a}}).

\bibitem[{\citenamefont{Lostaglio
  et~al.}(2015{\natexlab{b}})\citenamefont{Lostaglio, Korzekwa, Jennings, and
  Rudolph}}]{lostaglio2015quantumPRX}
\bibinfo{author}{\bibfnamefont{M.}~\bibnamefont{Lostaglio}},
  \bibinfo{author}{\bibfnamefont{K.}~\bibnamefont{Korzekwa}},
  \bibinfo{author}{\bibfnamefont{D.}~\bibnamefont{Jennings}}, \bibnamefont{and}
  \bibinfo{author}{\bibfnamefont{T.}~\bibnamefont{Rudolph}},
  \bibinfo{journal}{Physical Review X} \textbf{\bibinfo{volume}{5}},
  \bibinfo{pages}{021001} (\bibinfo{year}{2015}{\natexlab{b}}).

\bibitem[{\citenamefont{Korzekwa et~al.}(2016)\citenamefont{Korzekwa,
  Lostaglio, Oppenheim, and Jennings}}]{korzekwa2016extraction}
\bibinfo{author}{\bibfnamefont{K.}~\bibnamefont{Korzekwa}},
  \bibinfo{author}{\bibfnamefont{M.}~\bibnamefont{Lostaglio}},
  \bibinfo{author}{\bibfnamefont{J.}~\bibnamefont{Oppenheim}},
  \bibnamefont{and} \bibinfo{author}{\bibfnamefont{D.}~\bibnamefont{Jennings}},
  \bibinfo{journal}{New Journal of Physics} \textbf{\bibinfo{volume}{18}},
  \bibinfo{pages}{023045} (\bibinfo{year}{2016}).

\bibitem[{\citenamefont{Narasimhachar and Gour}(2015)}]{narasimhachar2015low}
\bibinfo{author}{\bibfnamefont{V.}~\bibnamefont{Narasimhachar}}
  \bibnamefont{and} \bibinfo{author}{\bibfnamefont{G.}~\bibnamefont{Gour}},
  \bibinfo{journal}{Nature communications} \textbf{\bibinfo{volume}{6}},
  \bibinfo{pages}{7689} (\bibinfo{year}{2015}).

\bibitem[{\citenamefont{Winter and Yang}(2016)}]{winter2016operational}
\bibinfo{author}{\bibfnamefont{A.}~\bibnamefont{Winter}} \bibnamefont{and}
  \bibinfo{author}{\bibfnamefont{D.}~\bibnamefont{Yang}},
  \bibinfo{journal}{Physical review letters} \textbf{\bibinfo{volume}{116}},
  \bibinfo{pages}{120404} (\bibinfo{year}{2016}).

\bibitem[{\citenamefont{Streltsov et~al.}(2017)\citenamefont{Streltsov, Adesso,
  and Plenio}}]{streltsov2017colloquium}
\bibinfo{author}{\bibfnamefont{A.}~\bibnamefont{Streltsov}},
  \bibinfo{author}{\bibfnamefont{G.}~\bibnamefont{Adesso}}, \bibnamefont{and}
  \bibinfo{author}{\bibfnamefont{M.~B.} \bibnamefont{Plenio}},
  \bibinfo{journal}{Reviews of Modern Physics} \textbf{\bibinfo{volume}{89}},
  \bibinfo{pages}{041003} (\bibinfo{year}{2017}).

\bibitem[{\citenamefont{Chitambar and Gour}(2019)}]{chitambar2019quantum}
\bibinfo{author}{\bibfnamefont{E.}~\bibnamefont{Chitambar}} \bibnamefont{and}
  \bibinfo{author}{\bibfnamefont{G.}~\bibnamefont{Gour}},
  \bibinfo{journal}{Reviews of Modern Physics} \textbf{\bibinfo{volume}{91}},
  \bibinfo{pages}{025001} (\bibinfo{year}{2019}).

\bibitem[{\citenamefont{Chitambar}(2018)}]{chitambar2018dephasing}
\bibinfo{author}{\bibfnamefont{E.}~\bibnamefont{Chitambar}},
  \bibinfo{journal}{Physical Review A} \textbf{\bibinfo{volume}{97}},
  \bibinfo{pages}{050301} (\bibinfo{year}{2018}).

\bibitem[{\citenamefont{Devetak and Winter}(2005)}]{devetak2005distillation}
\bibinfo{author}{\bibfnamefont{I.}~\bibnamefont{Devetak}} \bibnamefont{and}
  \bibinfo{author}{\bibfnamefont{A.}~\bibnamefont{Winter}}, in
  \emph{\bibinfo{booktitle}{Proceedings of the Royal Society of London A:
  Mathematical, Physical and Engineering Sciences}} (\bibinfo{organization}{The
  Royal Society}, \bibinfo{year}{2005}), vol. \bibinfo{volume}{461}, pp.
  \bibinfo{pages}{207--235}.

\bibitem[{\citenamefont{Devetak et~al.}(2008)\citenamefont{Devetak, Harrow, and
  Winter}}]{devetak2008resource}
\bibinfo{author}{\bibfnamefont{I.}~\bibnamefont{Devetak}},
  \bibinfo{author}{\bibfnamefont{A.~W.} \bibnamefont{Harrow}},
  \bibnamefont{and} \bibinfo{author}{\bibfnamefont{A.~J.}
  \bibnamefont{Winter}}, \bibinfo{journal}{IEEE Transactions on Information
  Theory} \textbf{\bibinfo{volume}{54}}, \bibinfo{pages}{4587}
  (\bibinfo{year}{2008}).

\bibitem[{\citenamefont{Devetak and Winter}(2004)}]{devetak2004relating}
\bibinfo{author}{\bibfnamefont{I.}~\bibnamefont{Devetak}} \bibnamefont{and}
  \bibinfo{author}{\bibfnamefont{A.}~\bibnamefont{Winter}},
  \bibinfo{journal}{Physical Review Letters} \textbf{\bibinfo{volume}{93}},
  \bibinfo{pages}{080501} (\bibinfo{year}{2004}).

\bibitem[{\citenamefont{Salecker and Wigner}(1958)}]{salecker1958quantum}
\bibinfo{author}{\bibfnamefont{H.}~\bibnamefont{Salecker}} \bibnamefont{and}
  \bibinfo{author}{\bibfnamefont{E.}~\bibnamefont{Wigner}},
  \bibinfo{journal}{Physical Review} \textbf{\bibinfo{volume}{109}},
  \bibinfo{pages}{571} (\bibinfo{year}{1958}).

\bibitem[{\citenamefont{Peres}(1980)}]{peres1980measurement}
\bibinfo{author}{\bibfnamefont{A.}~\bibnamefont{Peres}},
  \bibinfo{journal}{American Journal of Physics} \textbf{\bibinfo{volume}{48}},
  \bibinfo{pages}{552} (\bibinfo{year}{1980}).

\bibitem[{\citenamefont{Bartlett et~al.}(2007)\citenamefont{Bartlett, Rudolph,
  and Spekkens}}]{QRF_BRS_07}
\bibinfo{author}{\bibfnamefont{S.~D.} \bibnamefont{Bartlett}},
  \bibinfo{author}{\bibfnamefont{T.}~\bibnamefont{Rudolph}}, \bibnamefont{and}
  \bibinfo{author}{\bibfnamefont{R.~W.} \bibnamefont{Spekkens}},
  \bibinfo{journal}{Reviews of Modern Physics} \textbf{\bibinfo{volume}{79}},
  \bibinfo{pages}{555} (\bibinfo{year}{2007}).

\bibitem[{\citenamefont{Giovannetti et~al.}(2001)\citenamefont{Giovannetti,
  Lloyd, and Maccone}}]{Lloyd_Nature_Clock}
\bibinfo{author}{\bibfnamefont{V.}~\bibnamefont{Giovannetti}},
  \bibinfo{author}{\bibfnamefont{S.}~\bibnamefont{Lloyd}}, \bibnamefont{and}
  \bibinfo{author}{\bibfnamefont{L.}~\bibnamefont{Maccone}},
  \bibinfo{journal}{Nature} \textbf{\bibinfo{volume}{412}},
  \bibinfo{pages}{417} (\bibinfo{year}{2001}).

\bibitem[{\citenamefont{Giovannetti et~al.}(2002)\citenamefont{Giovannetti,
  Lloyd, Maccone, and Shahriar}}]{Limits_Lloyd_Clock}
\bibinfo{author}{\bibfnamefont{V.}~\bibnamefont{Giovannetti}},
  \bibinfo{author}{\bibfnamefont{S.}~\bibnamefont{Lloyd}},
  \bibinfo{author}{\bibfnamefont{L.}~\bibnamefont{Maccone}}, \bibnamefont{and}
  \bibinfo{author}{\bibfnamefont{M.}~\bibnamefont{Shahriar}},
  \bibinfo{journal}{Physical Review A} \textbf{\bibinfo{volume}{65}},
  \bibinfo{pages}{062319} (\bibinfo{year}{2002}).

\bibitem[{\citenamefont{Bu{\v z}ek et~al.}(1999)\citenamefont{Bu{\v z}ek,
  Derka, and Massar}}]{Clock_Buzek}
\bibinfo{author}{\bibfnamefont{V.}~\bibnamefont{Bu{\v z}ek}},
  \bibinfo{author}{\bibfnamefont{R.}~\bibnamefont{Derka}}, \bibnamefont{and}
  \bibinfo{author}{\bibfnamefont{S.}~\bibnamefont{Massar}},
  \bibinfo{journal}{Physical review letters} \textbf{\bibinfo{volume}{82}},
  \bibinfo{pages}{2207} (\bibinfo{year}{1999}).

\bibitem[{\citenamefont{Chiribella et~al.}(2013)\citenamefont{Chiribella, Yang,
  and Yao}}]{chiribella2013quantum}
\bibinfo{author}{\bibfnamefont{G.}~\bibnamefont{Chiribella}},
  \bibinfo{author}{\bibfnamefont{Y.}~\bibnamefont{Yang}}, \bibnamefont{and}
  \bibinfo{author}{\bibfnamefont{A.~C.-C.} \bibnamefont{Yao}},
  \bibinfo{journal}{Nature communications} \textbf{\bibinfo{volume}{4}},
  \bibinfo{pages}{2915} (\bibinfo{year}{2013}).

\bibitem[{\citenamefont{Nielsen and Chuang}(2000)}]{nielsen2000quantum}
\bibinfo{author}{\bibfnamefont{M.}~\bibnamefont{Nielsen}} \bibnamefont{and}
  \bibinfo{author}{\bibfnamefont{I.}~\bibnamefont{Chuang}},
  \emph{\bibinfo{title}{Quantum Computation and Quantum Information}},
  Cambridge Series on Information and the Natural Sciences
  (\bibinfo{publisher}{Cambridge University Press}, \bibinfo{year}{2000}), ISBN
  \bibinfo{isbn}{9780521635035}.

\bibitem[{\citenamefont{Wilde}(2013)}]{wilde2013quantum}
\bibinfo{author}{\bibfnamefont{M.~M.} \bibnamefont{Wilde}},
  \emph{\bibinfo{title}{Quantum information theory}}
  (\bibinfo{publisher}{Cambridge University Press}, \bibinfo{year}{2013}).

\bibitem[{\citenamefont{Marvian and Spekkens}(2013)}]{marvian2013theory}
\bibinfo{author}{\bibfnamefont{I.}~\bibnamefont{Marvian}} \bibnamefont{and}
  \bibinfo{author}{\bibfnamefont{R.~W.} \bibnamefont{Spekkens}},
  \bibinfo{journal}{New Journal of Physics} \textbf{\bibinfo{volume}{15}},
  \bibinfo{pages}{033001} (\bibinfo{year}{2013}).

\bibitem[{\citenamefont{Marvian}(2012)}]{Marvian_thesis}
\bibinfo{author}{\bibfnamefont{I.}~\bibnamefont{Marvian}}, Ph.D. thesis,
  \bibinfo{school}{University of Waterloo},
  \bibinfo{address}{https://uwspace.uwaterloo.ca/handle/10012/7088}
  (\bibinfo{year}{2012}).

\bibitem[{\citenamefont{Horodecki and
  Oppenheim}(2013{\natexlab{a}})}]{horodecki2013quantumness}
\bibinfo{author}{\bibfnamefont{M.}~\bibnamefont{Horodecki}} \bibnamefont{and}
  \bibinfo{author}{\bibfnamefont{J.}~\bibnamefont{Oppenheim}},
  \bibinfo{journal}{International Journal of Modern Physics B}
  \textbf{\bibinfo{volume}{27}}, \bibinfo{pages}{1345019}
  (\bibinfo{year}{2013}{\natexlab{a}}).

\bibitem[{\citenamefont{Brand{\~a}o and Gour}(2015)}]{brandao2015reversible}
\bibinfo{author}{\bibfnamefont{F.~G.} \bibnamefont{Brand{\~a}o}}
  \bibnamefont{and} \bibinfo{author}{\bibfnamefont{G.}~\bibnamefont{Gour}},
  \bibinfo{journal}{Physical review letters} \textbf{\bibinfo{volume}{115}},
  \bibinfo{pages}{070503} (\bibinfo{year}{2015}).

\bibitem[{\citenamefont{Coecke et~al.}(2016)\citenamefont{Coecke, Fritz, and
  Spekkens}}]{coecke2016mathematical}
\bibinfo{author}{\bibfnamefont{B.}~\bibnamefont{Coecke}},
  \bibinfo{author}{\bibfnamefont{T.}~\bibnamefont{Fritz}}, \bibnamefont{and}
  \bibinfo{author}{\bibfnamefont{R.~W.} \bibnamefont{Spekkens}},
  \bibinfo{journal}{Information and Computation}
  \textbf{\bibinfo{volume}{250}}, \bibinfo{pages}{59} (\bibinfo{year}{2016}).

\bibitem[{\citenamefont{Gour et~al.}(2015)\citenamefont{Gour, M{\"u}ller,
  Narasimhachar, Spekkens, and Halpern}}]{gour2015resource}
\bibinfo{author}{\bibfnamefont{G.}~\bibnamefont{Gour}},
  \bibinfo{author}{\bibfnamefont{M.~P.} \bibnamefont{M{\"u}ller}},
  \bibinfo{author}{\bibfnamefont{V.}~\bibnamefont{Narasimhachar}},
  \bibinfo{author}{\bibfnamefont{R.~W.} \bibnamefont{Spekkens}},
  \bibnamefont{and} \bibinfo{author}{\bibfnamefont{N.~Y.}
  \bibnamefont{Halpern}}, \bibinfo{journal}{Physics Reports}
  \textbf{\bibinfo{volume}{583}}, \bibinfo{pages}{1} (\bibinfo{year}{2015}).

\bibitem[{\citenamefont{{A. S. Holevo}}(1982)}]{Holevo:book}
\bibinfo{author}{\bibnamefont{{A. S. Holevo}}},
  \emph{\bibinfo{title}{{Probabilistic and Statistical Aspects of Quantum
  Theory}}} (\bibinfo{publisher}{{North-Holland}},
  \bibinfo{address}{{Amsterdam}}, \bibinfo{year}{1982}).

\bibitem[{\citenamefont{{C. W. Helstrom}}(1976)}]{Helstrom:book}
\bibinfo{author}{\bibnamefont{{C. W. Helstrom}}},
  \emph{\bibinfo{title}{{Quantum Detection and Estimation Theory}}}
  (\bibinfo{publisher}{{Academic Press}}, \bibinfo{address}{{New York}},
  \bibinfo{year}{1976}).

\bibitem[{\citenamefont{Paris}(2009)}]{paris2009quantum}
\bibinfo{author}{\bibfnamefont{M.~G.} \bibnamefont{Paris}},
  \bibinfo{journal}{International Journal of Quantum Information}
  \textbf{\bibinfo{volume}{7}}, \bibinfo{pages}{125} (\bibinfo{year}{2009}).

\bibitem[{\citenamefont{Braunstein and
  Caves}(1994)}]{braunstein1994statistical}
\bibinfo{author}{\bibfnamefont{S.~L.} \bibnamefont{Braunstein}}
  \bibnamefont{and} \bibinfo{author}{\bibfnamefont{C.~M.} \bibnamefont{Caves}},
  \bibinfo{journal}{Physical Review Letters} \textbf{\bibinfo{volume}{72}},
  \bibinfo{pages}{3439} (\bibinfo{year}{1994}).

\bibitem[{\citenamefont{Chiribella et~al.}(2005)\citenamefont{Chiribella,
  D'ariano, and Sacchi}}]{chiribella2005optimal}
\bibinfo{author}{\bibfnamefont{G.}~\bibnamefont{Chiribella}},
  \bibinfo{author}{\bibfnamefont{G.}~\bibnamefont{D'ariano}}, \bibnamefont{and}
  \bibinfo{author}{\bibfnamefont{M.}~\bibnamefont{Sacchi}},
  \bibinfo{journal}{Physical Review A} \textbf{\bibinfo{volume}{72}},
  \bibinfo{pages}{042338} (\bibinfo{year}{2005}).

\bibitem[{\citenamefont{Marvian and Spekkens}(2016)}]{marvian2016quantify}
\bibinfo{author}{\bibfnamefont{I.}~\bibnamefont{Marvian}} \bibnamefont{and}
  \bibinfo{author}{\bibfnamefont{R.~W.} \bibnamefont{Spekkens}},
  \bibinfo{journal}{Physical Review A} \textbf{\bibinfo{volume}{94}},
  \bibinfo{pages}{052324} (\bibinfo{year}{2016}).

\bibitem[{\citenamefont{Gour and Spekkens}(2008)}]{gour2008resource}
\bibinfo{author}{\bibfnamefont{G.}~\bibnamefont{Gour}} \bibnamefont{and}
  \bibinfo{author}{\bibfnamefont{R.~W.} \bibnamefont{Spekkens}},
  \bibinfo{journal}{New Journal of Physics} \textbf{\bibinfo{volume}{10}},
  \bibinfo{pages}{033023} (\bibinfo{year}{2008}).

\bibitem[{\citenamefont{Marvian and
  Spekkens}(2014{\natexlab{a}})}]{marvian2014asymmetry}
\bibinfo{author}{\bibfnamefont{I.}~\bibnamefont{Marvian}} \bibnamefont{and}
  \bibinfo{author}{\bibfnamefont{R.~W.} \bibnamefont{Spekkens}},
  \bibinfo{journal}{Physical Review A} \textbf{\bibinfo{volume}{90}},
  \bibinfo{pages}{014102} (\bibinfo{year}{2014}{\natexlab{a}}).

\bibitem[{\citenamefont{Janzing et~al.}(2000)\citenamefont{Janzing, Wocjan,
  Zeier, Geiss, and Beth}}]{janzing2000thermodynamic}
\bibinfo{author}{\bibfnamefont{D.}~\bibnamefont{Janzing}},
  \bibinfo{author}{\bibfnamefont{P.}~\bibnamefont{Wocjan}},
  \bibinfo{author}{\bibfnamefont{R.}~\bibnamefont{Zeier}},
  \bibinfo{author}{\bibfnamefont{R.}~\bibnamefont{Geiss}}, \bibnamefont{and}
  \bibinfo{author}{\bibfnamefont{T.}~\bibnamefont{Beth}},
  \bibinfo{journal}{Int. J. Theor. Phys.} \textbf{\bibinfo{volume}{39}},
  \bibinfo{pages}{2717} (\bibinfo{year}{2000}).

\bibitem[{\citenamefont{Horodecki and
  Oppenheim}(2013{\natexlab{b}})}]{FundLimitsNature}
\bibinfo{author}{\bibfnamefont{M.}~\bibnamefont{Horodecki}} \bibnamefont{and}
  \bibinfo{author}{\bibfnamefont{J.}~\bibnamefont{Oppenheim}},
  \bibinfo{journal}{Nat. Commun.} \textbf{\bibinfo{volume}{4}},
  \bibinfo{pages}{1} (\bibinfo{year}{2013}{\natexlab{b}}).

\bibitem[{\citenamefont{Brandao et~al.}(2013)\citenamefont{Brandao, Horodecki,
  Oppenheim, Renes, and Spekkens}}]{brandao2013resource}
\bibinfo{author}{\bibfnamefont{F.~G.} \bibnamefont{Brandao}},
  \bibinfo{author}{\bibfnamefont{M.}~\bibnamefont{Horodecki}},
  \bibinfo{author}{\bibfnamefont{J.}~\bibnamefont{Oppenheim}},
  \bibinfo{author}{\bibfnamefont{J.~M.} \bibnamefont{Renes}}, \bibnamefont{and}
  \bibinfo{author}{\bibfnamefont{R.~W.} \bibnamefont{Spekkens}},
  \bibinfo{journal}{Physical review letters} \textbf{\bibinfo{volume}{111}},
  \bibinfo{pages}{250404} (\bibinfo{year}{2013}).

\bibitem[{\citenamefont{{\AA}berg}(2013)}]{aaberg2013truly}
\bibinfo{author}{\bibfnamefont{J.}~\bibnamefont{{\AA}berg}},
  \bibinfo{journal}{Nature communications} \textbf{\bibinfo{volume}{4}},
  \bibinfo{pages}{1925} (\bibinfo{year}{2013}).

\bibitem[{\citenamefont{Goold et~al.}(2016)\citenamefont{Goold, Huber, Riera,
  del Rio, and Skrzypczyk}}]{goold2016role}
\bibinfo{author}{\bibfnamefont{J.}~\bibnamefont{Goold}},
  \bibinfo{author}{\bibfnamefont{M.}~\bibnamefont{Huber}},
  \bibinfo{author}{\bibfnamefont{A.}~\bibnamefont{Riera}},
  \bibinfo{author}{\bibfnamefont{L.}~\bibnamefont{del Rio}}, \bibnamefont{and}
  \bibinfo{author}{\bibfnamefont{P.}~\bibnamefont{Skrzypczyk}},
  \bibinfo{journal}{Journal of Physics A: Mathematical and Theoretical}
  \textbf{\bibinfo{volume}{49}}, \bibinfo{pages}{143001}
  (\bibinfo{year}{2016}).

\bibitem[{\citenamefont{Keyl and Werner}(1999)}]{keyl1999optimal}
\bibinfo{author}{\bibfnamefont{M.}~\bibnamefont{Keyl}} \bibnamefont{and}
  \bibinfo{author}{\bibfnamefont{R.~F.} \bibnamefont{Werner}},
  \bibinfo{journal}{Journal of Mathematical Physics}
  \textbf{\bibinfo{volume}{40}}, \bibinfo{pages}{3283} (\bibinfo{year}{1999}).

\bibitem[{\citenamefont{Bennett
  et~al.}(1996{\natexlab{a}})\citenamefont{Bennett, DiVincenzo, Smolin, and
  Wootters}}]{bennett1996mixed}
\bibinfo{author}{\bibfnamefont{C.~H.} \bibnamefont{Bennett}},
  \bibinfo{author}{\bibfnamefont{D.~P.} \bibnamefont{DiVincenzo}},
  \bibinfo{author}{\bibfnamefont{J.~A.} \bibnamefont{Smolin}},
  \bibnamefont{and} \bibinfo{author}{\bibfnamefont{W.~K.}
  \bibnamefont{Wootters}}, \bibinfo{journal}{Physical Review A}
  \textbf{\bibinfo{volume}{54}}, \bibinfo{pages}{3824}
  (\bibinfo{year}{1996}{\natexlab{a}}).

\bibitem[{\citenamefont{Bennett
  et~al.}(1996{\natexlab{b}})\citenamefont{Bennett, Brassard, Popescu,
  Schumacher, Smolin, and Wootters}}]{BBP+96}
\bibinfo{author}{\bibfnamefont{C.~H.} \bibnamefont{Bennett}},
  \bibinfo{author}{\bibfnamefont{G.}~\bibnamefont{Brassard}},
  \bibinfo{author}{\bibfnamefont{S.}~\bibnamefont{Popescu}},
  \bibinfo{author}{\bibfnamefont{B.}~\bibnamefont{Schumacher}},
  \bibinfo{author}{\bibfnamefont{J.~A.} \bibnamefont{Smolin}},
  \bibnamefont{and} \bibinfo{author}{\bibfnamefont{W.~K.}
  \bibnamefont{Wootters}}, \bibinfo{journal}{Phys. Rev. Lett.}
  \textbf{\bibinfo{volume}{78}}, \bibinfo{pages}{2031}
  (\bibinfo{year}{1996}{\natexlab{b}}).

\bibitem[{\citenamefont{Schuch et~al.}(2004{\natexlab{a}})\citenamefont{Schuch,
  Verstraete, and Cirac}}]{schuch2004nonlocal}
\bibinfo{author}{\bibfnamefont{N.}~\bibnamefont{Schuch}},
  \bibinfo{author}{\bibfnamefont{F.}~\bibnamefont{Verstraete}},
  \bibnamefont{and} \bibinfo{author}{\bibfnamefont{J.~I.} \bibnamefont{Cirac}},
  \bibinfo{journal}{Physical review letters} \textbf{\bibinfo{volume}{92}},
  \bibinfo{pages}{087904} (\bibinfo{year}{2004}{\natexlab{a}}).

\bibitem[{\citenamefont{Schuch et~al.}(2004{\natexlab{b}})\citenamefont{Schuch,
  Verstraete, and Cirac}}]{schuch2004quantum}
\bibinfo{author}{\bibfnamefont{N.}~\bibnamefont{Schuch}},
  \bibinfo{author}{\bibfnamefont{F.}~\bibnamefont{Verstraete}},
  \bibnamefont{and} \bibinfo{author}{\bibfnamefont{J.~I.} \bibnamefont{Cirac}},
  \bibinfo{journal}{Physical Review A} \textbf{\bibinfo{volume}{70}},
  \bibinfo{pages}{042310} (\bibinfo{year}{2004}{\natexlab{b}}).

\bibitem[{\citenamefont{Marvian}(2018)}]{marvian2018coherence}
\bibinfo{author}{\bibfnamefont{I.}~\bibnamefont{Marvian}},
  \bibinfo{journal}{arXiv preprint arXiv:1805.01989}  (\bibinfo{year}{2018}).

\bibitem[{\citenamefont{Marvian and
  Spekkens}(2014{\natexlab{b}})}]{marvian2014extending}
\bibinfo{author}{\bibfnamefont{I.}~\bibnamefont{Marvian}} \bibnamefont{and}
  \bibinfo{author}{\bibfnamefont{R.~W.} \bibnamefont{Spekkens}},
  \bibinfo{journal}{Nature communications} \textbf{\bibinfo{volume}{5}},
  \bibinfo{pages}{3821} (\bibinfo{year}{2014}{\natexlab{b}}).

\bibitem[{\citenamefont{Girolami}(2014)}]{girolami2014observable}
\bibinfo{author}{\bibfnamefont{D.}~\bibnamefont{Girolami}},
  \bibinfo{journal}{Physical review letters} \textbf{\bibinfo{volume}{113}},
  \bibinfo{pages}{170401} (\bibinfo{year}{2014}).

\bibitem[{\citenamefont{Yadin and Vedral}(2016)}]{yadin2016general}
\bibinfo{author}{\bibfnamefont{B.}~\bibnamefont{Yadin}} \bibnamefont{and}
  \bibinfo{author}{\bibfnamefont{V.}~\bibnamefont{Vedral}},
  \bibinfo{journal}{Physical Review A} \textbf{\bibinfo{volume}{93}},
  \bibinfo{pages}{022122} (\bibinfo{year}{2016}).

\bibitem[{\citenamefont{Marvian and
  Spekkens}(2014{\natexlab{c}})}]{marvian2014modes}
\bibinfo{author}{\bibfnamefont{I.}~\bibnamefont{Marvian}} \bibnamefont{and}
  \bibinfo{author}{\bibfnamefont{R.~W.} \bibnamefont{Spekkens}},
  \bibinfo{journal}{Physical Review A} \textbf{\bibinfo{volume}{90}},
  \bibinfo{pages}{062110} (\bibinfo{year}{2014}{\natexlab{c}}).

\bibitem[{\citenamefont{Piani et~al.}(2016)\citenamefont{Piani, Cianciaruso,
  Bromley, Napoli, Johnston, and Adesso}}]{piani2016robustness}
\bibinfo{author}{\bibfnamefont{M.}~\bibnamefont{Piani}},
  \bibinfo{author}{\bibfnamefont{M.}~\bibnamefont{Cianciaruso}},
  \bibinfo{author}{\bibfnamefont{T.~R.} \bibnamefont{Bromley}},
  \bibinfo{author}{\bibfnamefont{C.}~\bibnamefont{Napoli}},
  \bibinfo{author}{\bibfnamefont{N.}~\bibnamefont{Johnston}}, \bibnamefont{and}
  \bibinfo{author}{\bibfnamefont{G.}~\bibnamefont{Adesso}},
  \bibinfo{journal}{Physical Review A} \textbf{\bibinfo{volume}{93}},
  \bibinfo{pages}{042107} (\bibinfo{year}{2016}).

\bibitem[{\citenamefont{Gour et~al.}(2009)\citenamefont{Gour, Marvian, and
  Spekkens}}]{gour2009measuring}
\bibinfo{author}{\bibfnamefont{G.}~\bibnamefont{Gour}},
  \bibinfo{author}{\bibfnamefont{I.}~\bibnamefont{Marvian}}, \bibnamefont{and}
  \bibinfo{author}{\bibfnamefont{R.~W.} \bibnamefont{Spekkens}},
  \bibinfo{journal}{Physical Review A} \textbf{\bibinfo{volume}{80}},
  \bibinfo{pages}{012307} (\bibinfo{year}{2009}).

\bibitem[{\citenamefont{Vaccaro et~al.}(2008)\citenamefont{Vaccaro, Anselmi,
  Wiseman, and Jacobs}}]{vaccaro2008tradeoff}
\bibinfo{author}{\bibfnamefont{J.~A.} \bibnamefont{Vaccaro}},
  \bibinfo{author}{\bibfnamefont{F.}~\bibnamefont{Anselmi}},
  \bibinfo{author}{\bibfnamefont{H.~M.} \bibnamefont{Wiseman}},
  \bibnamefont{and} \bibinfo{author}{\bibfnamefont{K.}~\bibnamefont{Jacobs}},
  \bibinfo{journal}{Physical Review A} \textbf{\bibinfo{volume}{77}},
  \bibinfo{pages}{032114} (\bibinfo{year}{2008}).

\bibitem[{\citenamefont{Petz}(1996)}]{petz1996monotone}
\bibinfo{author}{\bibfnamefont{D.}~\bibnamefont{Petz}},
  \bibinfo{journal}{Linear algebra and its applications}
  \textbf{\bibinfo{volume}{244}}, \bibinfo{pages}{81} (\bibinfo{year}{1996}).

\bibitem[{\citenamefont{Petz and Ghinea}(2011)}]{petz2011introduction}
\bibinfo{author}{\bibfnamefont{D.}~\bibnamefont{Petz}} \bibnamefont{and}
  \bibinfo{author}{\bibfnamefont{C.}~\bibnamefont{Ghinea}}, in
  \emph{\bibinfo{booktitle}{Quantum probability and related topics}}
  (\bibinfo{publisher}{World Scientific}, \bibinfo{year}{2011}), pp.
  \bibinfo{pages}{261--281}.

\bibitem[{\citenamefont{Petz}(1986)}]{petz1986quasi}
\bibinfo{author}{\bibfnamefont{D.}~\bibnamefont{Petz}},
  \bibinfo{journal}{Reports on mathematical physics}
  \textbf{\bibinfo{volume}{23}}, \bibinfo{pages}{57} (\bibinfo{year}{1986}).

\bibitem[{\citenamefont{Tomamichel}(2015)}]{tomamichel2015quantum}
\bibinfo{author}{\bibfnamefont{M.}~\bibnamefont{Tomamichel}},
  \emph{\bibinfo{title}{Quantum Information Processing with Finite Resources:
  Mathematical Foundations}}, vol.~\bibinfo{volume}{5}
  (\bibinfo{publisher}{Springer}, \bibinfo{year}{2015}).

\bibitem[{\citenamefont{Zanardi
  et~al.}(2007{\natexlab{a}})\citenamefont{Zanardi, Giorda, and
  Cozzini}}]{zanardi2007information}
\bibinfo{author}{\bibfnamefont{P.}~\bibnamefont{Zanardi}},
  \bibinfo{author}{\bibfnamefont{P.}~\bibnamefont{Giorda}}, \bibnamefont{and}
  \bibinfo{author}{\bibfnamefont{M.}~\bibnamefont{Cozzini}},
  \bibinfo{journal}{Physical review letters} \textbf{\bibinfo{volume}{99}},
  \bibinfo{pages}{100603} (\bibinfo{year}{2007}{\natexlab{a}}).

\bibitem[{\citenamefont{Zanardi et~al.}(2008)\citenamefont{Zanardi, Paris, and
  Venuti}}]{zanardi2008quantum}
\bibinfo{author}{\bibfnamefont{P.}~\bibnamefont{Zanardi}},
  \bibinfo{author}{\bibfnamefont{M.~G.} \bibnamefont{Paris}}, \bibnamefont{and}
  \bibinfo{author}{\bibfnamefont{L.~C.} \bibnamefont{Venuti}},
  \bibinfo{journal}{Physical Review A} \textbf{\bibinfo{volume}{78}},
  \bibinfo{pages}{042105} (\bibinfo{year}{2008}).

\bibitem[{\citenamefont{Zanardi
  et~al.}(2007{\natexlab{b}})\citenamefont{Zanardi, Campos~Venuti, and
  Giorda}}]{zanardi2007bures}
\bibinfo{author}{\bibfnamefont{P.}~\bibnamefont{Zanardi}},
  \bibinfo{author}{\bibfnamefont{L.}~\bibnamefont{Campos~Venuti}},
  \bibnamefont{and} \bibinfo{author}{\bibfnamefont{P.}~\bibnamefont{Giorda}},
  \bibinfo{journal}{Physical Review A} \textbf{\bibinfo{volume}{76}},
  \bibinfo{pages}{062318} (\bibinfo{year}{2007}{\natexlab{b}}).

\bibitem[{\citenamefont{Campos~Venuti and Zanardi}(2007)}]{campos2007quantum}
\bibinfo{author}{\bibfnamefont{L.}~\bibnamefont{Campos~Venuti}}
  \bibnamefont{and} \bibinfo{author}{\bibfnamefont{P.}~\bibnamefont{Zanardi}},
  \bibinfo{journal}{Physical Review Letters} \textbf{\bibinfo{volume}{99}},
  \bibinfo{pages}{095701} (\bibinfo{year}{2007}).

\bibitem[{\citenamefont{Pires et~al.}(2016)\citenamefont{Pires, Cianciaruso,
  C{\'e}leri, Adesso, and Soares-Pinto}}]{pires2016generalized}
\bibinfo{author}{\bibfnamefont{D.~P.} \bibnamefont{Pires}},
  \bibinfo{author}{\bibfnamefont{M.}~\bibnamefont{Cianciaruso}},
  \bibinfo{author}{\bibfnamefont{L.~C.} \bibnamefont{C{\'e}leri}},
  \bibinfo{author}{\bibfnamefont{G.}~\bibnamefont{Adesso}}, \bibnamefont{and}
  \bibinfo{author}{\bibfnamefont{D.~O.} \bibnamefont{Soares-Pinto}},
  \bibinfo{journal}{Physical Review X} \textbf{\bibinfo{volume}{6}},
  \bibinfo{pages}{021031} (\bibinfo{year}{2016}).

\bibitem[{\citenamefont{Kwon et~al.}(2018)\citenamefont{Kwon, Jeong, Jennings,
  Yadin, and Kim}}]{kwon2018clock}
\bibinfo{author}{\bibfnamefont{H.}~\bibnamefont{Kwon}},
  \bibinfo{author}{\bibfnamefont{H.}~\bibnamefont{Jeong}},
  \bibinfo{author}{\bibfnamefont{D.}~\bibnamefont{Jennings}},
  \bibinfo{author}{\bibfnamefont{B.}~\bibnamefont{Yadin}}, \bibnamefont{and}
  \bibinfo{author}{\bibfnamefont{M.}~\bibnamefont{Kim}},
  \bibinfo{journal}{Physical review letters} \textbf{\bibinfo{volume}{120}},
  \bibinfo{pages}{150602} (\bibinfo{year}{2018}).

\bibitem[{\citenamefont{Morozova and Chentsov}(1991)}]{morozova1991markov}
\bibinfo{author}{\bibfnamefont{E.~A.} \bibnamefont{Morozova}} \bibnamefont{and}
  \bibinfo{author}{\bibfnamefont{N.~N.} \bibnamefont{Chentsov}},
  \bibinfo{journal}{Journal of Soviet Mathematics}
  \textbf{\bibinfo{volume}{56}}, \bibinfo{pages}{2648} (\bibinfo{year}{1991}).

\bibitem[{\citenamefont{Synak-Radtke and
  Horodecki}(2006)}]{synak2006asymptotic}
\bibinfo{author}{\bibfnamefont{B.}~\bibnamefont{Synak-Radtke}}
  \bibnamefont{and}
  \bibinfo{author}{\bibfnamefont{M.}~\bibnamefont{Horodecki}},
  \bibinfo{journal}{Journal of Physics A: Mathematical and General}
  \textbf{\bibinfo{volume}{39}}, \bibinfo{pages}{L423} (\bibinfo{year}{2006}).

\bibitem[{\citenamefont{Regula et~al.}(2018)\citenamefont{Regula, Fang, Wang,
  and Adesso}}]{regula2018one}
\bibinfo{author}{\bibfnamefont{B.}~\bibnamefont{Regula}},
  \bibinfo{author}{\bibfnamefont{K.}~\bibnamefont{Fang}},
  \bibinfo{author}{\bibfnamefont{X.}~\bibnamefont{Wang}}, \bibnamefont{and}
  \bibinfo{author}{\bibfnamefont{G.}~\bibnamefont{Adesso}},
  \bibinfo{journal}{Physical review letters} \textbf{\bibinfo{volume}{121}},
  \bibinfo{pages}{010401} (\bibinfo{year}{2018}).

\bibitem[{\citenamefont{Zhao et~al.}(2018)\citenamefont{Zhao, Liu, Yuan,
  Chitambar, and Winter}}]{zhao2018one}
\bibinfo{author}{\bibfnamefont{Q.}~\bibnamefont{Zhao}},
  \bibinfo{author}{\bibfnamefont{Y.}~\bibnamefont{Liu}},
  \bibinfo{author}{\bibfnamefont{X.}~\bibnamefont{Yuan}},
  \bibinfo{author}{\bibfnamefont{E.}~\bibnamefont{Chitambar}},
  \bibnamefont{and} \bibinfo{author}{\bibfnamefont{A.}~\bibnamefont{Winter}},
  \bibinfo{journal}{arXiv preprint arXiv:1808.01885}  (\bibinfo{year}{2018}).

\bibitem[{\citenamefont{Lami et~al.}(2019)\citenamefont{Lami, Regula, and
  Adesso}}]{lami2019generic}
\bibinfo{author}{\bibfnamefont{L.}~\bibnamefont{Lami}},
  \bibinfo{author}{\bibfnamefont{B.}~\bibnamefont{Regula}}, \bibnamefont{and}
  \bibinfo{author}{\bibfnamefont{G.}~\bibnamefont{Adesso}},
  \bibinfo{journal}{Physical review letters} \textbf{\bibinfo{volume}{122}},
  \bibinfo{pages}{150402} (\bibinfo{year}{2019}).

\bibitem[{\citenamefont{Lami}(2019)}]{lami2019completing}
\bibinfo{author}{\bibfnamefont{L.}~\bibnamefont{Lami}}, \bibinfo{journal}{arXiv
  preprint arXiv:1902.02427}  (\bibinfo{year}{2019}).

\bibitem[{\citenamefont{Yadin et~al.}(2016)\citenamefont{Yadin, Ma, Girolami,
  Gu, and Vedral}}]{yadin2016quantum}
\bibinfo{author}{\bibfnamefont{B.}~\bibnamefont{Yadin}},
  \bibinfo{author}{\bibfnamefont{J.}~\bibnamefont{Ma}},
  \bibinfo{author}{\bibfnamefont{D.}~\bibnamefont{Girolami}},
  \bibinfo{author}{\bibfnamefont{M.}~\bibnamefont{Gu}}, \bibnamefont{and}
  \bibinfo{author}{\bibfnamefont{V.}~\bibnamefont{Vedral}},
  \bibinfo{journal}{Physical Review X} \textbf{\bibinfo{volume}{6}},
  \bibinfo{pages}{041028} (\bibinfo{year}{2016}).

\bibitem[{\citenamefont{Barndorff-Nielsen and
  Gill}(2000)}]{barndorff2000fisher}
\bibinfo{author}{\bibfnamefont{O.}~\bibnamefont{Barndorff-Nielsen}}
  \bibnamefont{and} \bibinfo{author}{\bibfnamefont{R.}~\bibnamefont{Gill}},
  \bibinfo{journal}{Journal of Physics A: Mathematical and General}
  \textbf{\bibinfo{volume}{33}}, \bibinfo{pages}{4481} (\bibinfo{year}{2000}).

\bibitem[{\citenamefont{Helstrom}(1969)}]{helstrom1969quantum}
\bibinfo{author}{\bibfnamefont{C.~W.} \bibnamefont{Helstrom}},
  \bibinfo{journal}{Journal of Statistical Physics}
  \textbf{\bibinfo{volume}{1}}, \bibinfo{pages}{231} (\bibinfo{year}{1969}).

\bibitem[{\citenamefont{Gour et~al.}(2018)\citenamefont{Gour, Jennings,
  Buscemi, Duan, and Marvian}}]{gour2018quantum}
\bibinfo{author}{\bibfnamefont{G.}~\bibnamefont{Gour}},
  \bibinfo{author}{\bibfnamefont{D.}~\bibnamefont{Jennings}},
  \bibinfo{author}{\bibfnamefont{F.}~\bibnamefont{Buscemi}},
  \bibinfo{author}{\bibfnamefont{R.}~\bibnamefont{Duan}}, \bibnamefont{and}
  \bibinfo{author}{\bibfnamefont{I.}~\bibnamefont{Marvian}},
  \bibinfo{journal}{Nature communications} \textbf{\bibinfo{volume}{9}},
  \bibinfo{pages}{5352} (\bibinfo{year}{2018}).

\bibitem[{\citenamefont{Konig et~al.}(2009)\citenamefont{Konig, Renner, and
  Schaffner}}]{Konig}
\bibinfo{author}{\bibfnamefont{R.}~\bibnamefont{Konig}},
  \bibinfo{author}{\bibfnamefont{R.}~\bibnamefont{Renner}}, \bibnamefont{and}
  \bibinfo{author}{\bibfnamefont{C.}~\bibnamefont{Schaffner}},
  \bibinfo{journal}{IEEE T. Inform. Theory} \textbf{\bibinfo{volume}{55}},
  \bibinfo{pages}{4337} (\bibinfo{year}{2009}).

\bibitem[{\citenamefont{Cirac et~al.}(1999)\citenamefont{Cirac, Ekert, and
  Macchiavello}}]{cirac1999optimal}
\bibinfo{author}{\bibfnamefont{J.}~\bibnamefont{Cirac}},
  \bibinfo{author}{\bibfnamefont{A.}~\bibnamefont{Ekert}}, \bibnamefont{and}
  \bibinfo{author}{\bibfnamefont{C.}~\bibnamefont{Macchiavello}},
  \bibinfo{journal}{Physical review letters} \textbf{\bibinfo{volume}{82}},
  \bibinfo{pages}{4344} (\bibinfo{year}{1999}).

\bibitem[{\citenamefont{Chiribella and Yang}(2017)}]{chiribella2017optimal}
\bibinfo{author}{\bibfnamefont{G.}~\bibnamefont{Chiribella}} \bibnamefont{and}
  \bibinfo{author}{\bibfnamefont{Y.}~\bibnamefont{Yang}},
  \bibinfo{journal}{Physical Review A} \textbf{\bibinfo{volume}{96}},
  \bibinfo{pages}{022327} (\bibinfo{year}{2017}).

\bibitem[{\citenamefont{Halpern and Renes}(2016)}]{halpern2016beyond}
\bibinfo{author}{\bibfnamefont{N.~Y.} \bibnamefont{Halpern}} \bibnamefont{and}
  \bibinfo{author}{\bibfnamefont{J.~M.} \bibnamefont{Renes}},
  \bibinfo{journal}{Physical Review E} \textbf{\bibinfo{volume}{93}},
  \bibinfo{pages}{022126} (\bibinfo{year}{2016}).

\bibitem[{\citenamefont{Giovannetti et~al.}(2006)\citenamefont{Giovannetti,
  Lloyd, and Maccone}}]{giovannetti2006quantum}
\bibinfo{author}{\bibfnamefont{V.}~\bibnamefont{Giovannetti}},
  \bibinfo{author}{\bibfnamefont{S.}~\bibnamefont{Lloyd}}, \bibnamefont{and}
  \bibinfo{author}{\bibfnamefont{L.}~\bibnamefont{Maccone}},
  \bibinfo{journal}{Physical review letters} \textbf{\bibinfo{volume}{96}},
  \bibinfo{pages}{010401} (\bibinfo{year}{2006}).

\bibitem[{\citenamefont{Giovannetti et~al.}(2011)\citenamefont{Giovannetti,
  Lloyd, and Maccone}}]{giovannetti2011advances}
\bibinfo{author}{\bibfnamefont{V.}~\bibnamefont{Giovannetti}},
  \bibinfo{author}{\bibfnamefont{S.}~\bibnamefont{Lloyd}}, \bibnamefont{and}
  \bibinfo{author}{\bibfnamefont{L.}~\bibnamefont{Maccone}},
  \bibinfo{journal}{Nature Photonics} \textbf{\bibinfo{volume}{5}},
  \bibinfo{pages}{222} (\bibinfo{year}{2011}).

\bibitem[{\citenamefont{Yang et~al.}(2017)\citenamefont{Yang, Chiribella, and
  Hu}}]{yang2017units}
\bibinfo{author}{\bibfnamefont{Y.}~\bibnamefont{Yang}},
  \bibinfo{author}{\bibfnamefont{G.}~\bibnamefont{Chiribella}},
  \bibnamefont{and} \bibinfo{author}{\bibfnamefont{Q.}~\bibnamefont{Hu}},
  \bibinfo{journal}{New Journal of Physics} \textbf{\bibinfo{volume}{19}},
  \bibinfo{pages}{123003} (\bibinfo{year}{2017}).

\bibitem[{\citenamefont{Matsumoto}(2005)}]{matsumoto2005reverse}
\bibinfo{author}{\bibfnamefont{K.}~\bibnamefont{Matsumoto}},
  \bibinfo{journal}{arXiv preprint quant-ph/0511170}  (\bibinfo{year}{2005}).

\bibitem[{\citenamefont{Aberg}(2006)}]{aberg}
\bibinfo{author}{\bibfnamefont{J.}~\bibnamefont{Aberg}},
  \bibinfo{journal}{arXiv preprint quant-ph/0612146}  (\bibinfo{year}{2006}).

\bibitem[{\citenamefont{Liu et~al.}(2017)\citenamefont{Liu, Hu, and
  Lloyd}}]{liu2017resource}
\bibinfo{author}{\bibfnamefont{Z.-W.} \bibnamefont{Liu}},
  \bibinfo{author}{\bibfnamefont{X.}~\bibnamefont{Hu}}, \bibnamefont{and}
  \bibinfo{author}{\bibfnamefont{S.}~\bibnamefont{Lloyd}},
  \bibinfo{journal}{Physical review letters} \textbf{\bibinfo{volume}{118}},
  \bibinfo{pages}{060502} (\bibinfo{year}{2017}).

\bibitem[{\citenamefont{Rudolph et~al.}(2003)\citenamefont{Rudolph, Spekkens,
  and Turner}}]{rudolph2003unambiguous}
\bibinfo{author}{\bibfnamefont{T.}~\bibnamefont{Rudolph}},
  \bibinfo{author}{\bibfnamefont{R.~W.} \bibnamefont{Spekkens}},
  \bibnamefont{and} \bibinfo{author}{\bibfnamefont{P.~S.}
  \bibnamefont{Turner}}, \bibinfo{journal}{Physical Review A}
  \textbf{\bibinfo{volume}{68}}, \bibinfo{pages}{010301}
  (\bibinfo{year}{2003}).

\bibitem[{\citenamefont{Berry}(1941)}]{berry1941accuracy}
\bibinfo{author}{\bibfnamefont{A.~C.} \bibnamefont{Berry}},
  \bibinfo{journal}{Transactions of the american mathematical society}
  \textbf{\bibinfo{volume}{49}}, \bibinfo{pages}{122} (\bibinfo{year}{1941}).

\bibitem[{\citenamefont{Durrett}(2019)}]{durrett2019probability}
\bibinfo{author}{\bibfnamefont{R.}~\bibnamefont{Durrett}},
  \emph{\bibinfo{title}{Probability: theory and examples}},
  vol.~\bibinfo{volume}{49} (\bibinfo{publisher}{Cambridge university press},
  \bibinfo{year}{2019}).

\bibitem[{\citenamefont{{S.L. Braunstein and C.M.
  Caves}}(1994)}]{Braunstein:94}
\bibinfo{author}{\bibnamefont{{S.L. Braunstein and C.M. Caves}}},
  \bibinfo{journal}{Phys. Rev. Lett.} \textbf{\bibinfo{volume}{72}},
  \bibinfo{pages}{3439} (\bibinfo{year}{1994}).

\bibitem[{\citenamefont{Gour et~al.}(2017)\citenamefont{Gour, Jennings,
  Buscemi, Duan, and Marvian}}]{gour2017quantum}
\bibinfo{author}{\bibfnamefont{G.}~\bibnamefont{Gour}},
  \bibinfo{author}{\bibfnamefont{D.}~\bibnamefont{Jennings}},
  \bibinfo{author}{\bibfnamefont{F.}~\bibnamefont{Buscemi}},
  \bibinfo{author}{\bibfnamefont{R.}~\bibnamefont{Duan}}, \bibnamefont{and}
  \bibinfo{author}{\bibfnamefont{I.}~\bibnamefont{Marvian}},
  \bibinfo{journal}{arXiv preprint arXiv:1708.04302}  (\bibinfo{year}{2017}).

\bibitem[{\citenamefont{Renner}(2008)}]{renner2008security}
\bibinfo{author}{\bibfnamefont{R.}~\bibnamefont{Renner}},
  \bibinfo{journal}{International Journal of Quantum Information}
  \textbf{\bibinfo{volume}{6}}, \bibinfo{pages}{1} (\bibinfo{year}{2008}).

\end{thebibliography}

\end{document}